\documentclass{article}       

\usepackage[affil-it]{authblk}

\usepackage{graphicx}
\usepackage{mathptmx}      

\usepackage{hyperref}
\usepackage{amssymb}
\usepackage{amssymb}
\usepackage{geometry} 
\usepackage{float}
\usepackage{booktabs}
\usepackage{algorithm,algorithmic}
\usepackage{amsmath}
\usepackage{pdflscape}
\usepackage{url}
\usepackage{natbib}
\usepackage{graphicx,color}
\usepackage[utf8]{inputenc}
\usepackage{listings} 
\usepackage{verbatim}
\usepackage{subcaption}

\begin{document}

\title{BROUTE: a benchmark suite for the implementation of standard
  vehicle routing algorithms}


\author{Fabien Tricoire}

\affil{Vienna University of Economics and Business,
              Welthandelsplatz 1, 1020 Wien, Austria \\
              fabien.tricoire@wu.ac.at}

\date{\today}

\maketitle

\begin{abstract}
  We introduce BROUTE, a benchmark suite for vehicle routing optimization
  algorithms. We define a selection of algorithms traditionally used
  in vehicle routing optimization. They capture essential features
  that are also relevant in optimization algorithms for different
  application domains, like local search move evaluation, memory
  allocation, dynamic programming, or insertion and deletion from a
  list. Each algorithm is deterministic. We implement these benchmark
  algorithms using a selection of programming languages and different
  data structures.
  BROUTE is free, open-source, and can be used to inform early decisions in
  projects that involve programming, such as which language to use.
\end{abstract}

\section{Introduction}
\label{sec:intro}
When it comes to implementing optimization algorithms and routing
algorithms, some important design decisions such
as the choice of the programming language must be made early. While
most programming languages are Turing-complete, thus allowing to run any
deterministic algorithm, they vary in many aspects including 
conciseness, ease of writing, ease of reading, ease of debugging and
runtime performance. Many of these aspects are subjective in
nature, but runtime performance can be measured objectively. We
introduce a benchmark suite for vehicle routing optimization
algorithms called
BROUTE, available at \url{https://github.com/fa-bien/broute}, which
focuses on CPU effort. CPU effort is especially relevant when running
the final version of a program, as it is a direct measure of
performance. However it also matters while developing and debugging,
as these tasks typically involve repeatedly running the program and waiting
for the runs to complete.

The goal of this suite is to inform early design decisions when
programming optimization algorithms, such as the choice of the
programming language, or how to represent a distance matrix. Readers
may already know a few programming languages and know both (i) which
one they appreciate using the most and (ii) which one offers the best
performance, but they do not necessarily have a quantified
assessment of the trade-off between performance and convenience,
i.e. of the performance cost of convenience. Other
readers may need to decide which programming language their new
student should learn and use for the implementation of their
research work. Others may elect to program using Julia because it
makes them more productive, but still want to know what kind of
performance to expect when compared to C++. Some will want to
find out, given the hardware and software they have access
to, which language is the most efficient for which task. These
decisions can all be informed by using BROUTE.

\subsection{Scope of this study}
We provide the means to compare the reasonably efficient
implementation of various algorithms used in combinatorial
optimization, within a given experimental setting. In particular, we
look at five algorithms that are commonly used in vehicle routing
optimization. However these algorithms capture features that are also
common in other domains of optimization, such as modifying
subsequences, inserting and removing elements in a sequence, or other
operations on graphs.
The five algorithms are implemented using different programming
languages.
We put ourselves in the situation of a
programmer with limited experience, for example a student, having to
implement these algorithms. For the time being all
benchmarks are implemented in C++, Python, Java, JavaScript, Rust and
Julia. We also assess the impact on performance of some implementation
decisions, e.g. using a flat matrix representation.

Our goal is not to provide the most efficient implementation of each
algorithm for a set of languages, or even to use the most efficient
algorithm to solve the given problems. In fact, the value of the run
time performance comparisons is independent from the efficiency of the
algorithms, as long as all implementations are of the same
algorithms and use comparable data structures. Nonetheless, we
believe the algorithms used (i)
capture an interesting range of features encountered in many routing
algorithms and (ii) are reasonably good implementations of reasonably
efficient algorithms. Additionally, we welcome the addition of
new benchmark algorithms to BROUTE in the future, which might then
allow a comparison of the efficiency of different algorithms. However,
this falls outside the scope of this contribution.

Drawing general conclusions about which programming
language is the best, or yields the most productivity, or even runs
the fastest, is of limited interest: software evolves fast, hardware
architectures evolve as well and compilers have to adapt. Some outcomes
observed in the example experimental study in Section~\ref{sec:exp}
may have been different a few years ago, and might be different on a
different platform. However, even though benchmark results depend on
constant software and hardware evolution, the tools to provide these
results are independent from it and can be used in any experimental
setting, now or in the future. BROUTE is such a tool.

Readers looking for state-of-the-art implementations of algorithms for
vehicle routing, including efficient data structures, may find these
contributions interesting: LKH~\citep{LKH,LKH-2,LKH-POP} is likely the
most efficient implementation of local search for the TSP currently in
existence, and tackles instances with several million nodes efficiently; a linear
time Split algorithm, which is employed in some of the best known
metaheuristics for VRPs, is provided by~\cite{Vidal_split}; a generic
state-of-the-art branch-and-price code is provided by~\cite{bapcod}.

\subsection{What is measured}
In all proposed benchmarks, the CPU time of the
benchmarked algorithm is measured. Every other task, such
as reading input data or constructing an initial solution for local search
benchmarks, is not measured. If an auxilliary graph needs to
be constructed every time the algorithm is run, then we also measure
the construction of said auxilliary graph. This is the case for
benchmarks \emph{maxflow} and \emph{espprc}. We do not, however,
measure the time required to allocate this auxiliary graph in memory:
this typically happens once when a program starts, then the same
memory can be used many times.

In general, compilation time is not measured. However, some
implementations do just-in-time (JIT) compilation, for example Julia,
JavaScript, Pypy, Numba and Java. In such cases, this JIT compilation time is
included in the running time of the algorithm. In general we measure
\emph{clock} time, i.e. the CPU budget used by the algorithm. In the
case of JavaScript we measure \emph{wall} time as we are not aware of
any way to measure clock time in JavaScript.

In order to represent a tour in various heuristics, we use variable-size
vectors. Dynamic size containers are intuitive to use in algorithms for 
vehicle routing, for instance to insert or delete a vertex from a
tour. Such containers can also be used to represent solutions in
other domains, e.g. scheduling or any application that requires the
representation of sequences of variable length.
Inserting and deleting elements in a vector has linear worst-case time
complexity. However, for heuristics it can still be a good idea to
use vectors rather than, say, linked lists, since neighborhood
exploration typically requires to evaluate many moves then perform
only one of these moves. The complexity of performing a move is thus
a secondary concern and unlikely to be a performance
bottleneck. Additionally, performing certain moves can actually be
more efficient using a vector than using a linked list, e.g. inverting
a subsequence in the context of 2-opt.
There exist a variety of data structures for vehicle routing, for
example one can use arrays of successors~\citep{pvrp-dam}; their size
is dynamic but they do not require further
memory allocation and allow insertion and deletion in
$\mathcal{O}(1)$.
Efficient data structures for TSPs are discussed by~\cite{LKH}. One
conclusion is that below 1000 cities, some sophisticated data
structures are outperformed by arrays and lists, due to their time
overhead. Additionally, their implementation is not simple.

\subsection{Limitations}
Benchmarks are by definition limited in scope. Any given benchmark
measurement is only valid within a certain environment. In our case
such environmental factors include the choice of compilers, compiler
versions, operating system and CPU family. Therefore all results are
to be taken as indicators but not absolute truths. The intent of this
study is not to declare a language the winner because it is
consistently faster than others by a few percents, but rather to
assess what kind of performance losses can be expected by, for
example, using Python instead of C++ on a given computing
cluster. Another example is to assess what kind of performance gains
can be achieved by using a JIT compiler for Python. One could also
determine differences in performance between different C++ compilers.

Additionally, there are some implicit conditions for the good
functioning of interpreters and compilers. For instance Pypy is
advertised to work better when things are ``kept simple'', which is
generally good advice regardless of the compiler. However, this
typically requires discipline.

\section{The benchmarks}
\label{sec:benchmark}
We consider five different benchmark algorithms that perform tasks
commonly encountered in transport optimization. Together they capture
key aspects of computationally intensive tasks in heuristics and
exact methods for transport optimization, although they do not
necessarily use the most efficient algorithm to solve the problem they
are tackling. Additionally, we believe
that they also capture such aspects for other fields of application
of operations research, such as e.g. scheduling. For the sake of
simplicity, we consider the symmetric travelling salesperson problem
(TSP) as a base problem, i.e. input data are in the form of a
symmetric distance matrix while a solution is a permutation. However
some benchmark algorithms solve different problems than the TSP based on these
TSP data, as described below.
Abstract algorithms are providex in Appendix~\ref{app:abstract}. The
implementations are all available at
\url{https://github.com/fa-bien/broute}.

All five benchmarking algorithms described below are
deterministic. In cases where we want to capture the features of a
stochastic algorithm, we simulate randomness in a deterministic
fashion for the sake of reproducibility. 
Additionally, the input provided to them is also deterministic,
i.e. each implementation receives the exact same inputs and performs
the same operations from these inputs. For that purpose, we generate
various instance files, each with a given size $n$, which represents
the number of vertices in the graph, and a number $p$ of
permutations. A permutation is a solution to the TSP.
Vertex coordinates are generated randomly in $[0, 100)$. Let $d_{ij}$
be the Euclidean distance between $i$ and $j$, then the distance
considered is $c_{ij} = \lfloor 100 d_{ij} \rfloor$. Since it is an
integer number, it is sufficient to use integer number representation
to compute the cost of a TSP solution; however certain benchmarks use
floating-point number representation, as explained below.
Each instance
file contains one $n \times n$ distance matrix as well as $p$ randomly
generated permutations, which are used as starting \emph{seeds} for
the algorithms (e.g. as starting solution for 2-opt).

Since the same instance files are given to 
each implementation, all implementations perform the same operations
and return the same result. In order to control result integrity we use a
mechanism similar to \emph{checksum}. The checksum calculation for
applying a certain algorithm using a given permutation as starting seed differs
based on the algorithm and is explained separately for each algorithm
below. The checksum calculation for an instance file is the sum of
checksum values over all permutations in that instance file.

Performances are measured per instance file, i.e. each time reported
is for running an algorithm $p$ times, using one distance matrix with the $p$
different seed permutations. 

\subsection{2-opt}
The 2-opt heuristic was first described by~\cite{2-opt}.
One of the most commonly used heuristics in vehicle routing, 2-opt
improves a tour by performing 2-exchanges. A 2-exchange consists in
removing two edges from a tour and reconnecting the tour with two
other edges. It is equivalent to inverting a sub-sequence of the
tour, and can be performed in place using an array or vector solution
representation. If the distances are symmetric, then each move can be
evaluated in constant time. 

The checksum for a given permutation seed is the number of
improvements found while applying first-improvement 2-opt until no
improving move exists, using this permutation as starting solution.

\subsection{Or-opt}
Or-opt is a heuristic that was first introduced by~\cite{Or-opt}.
It relies on the exploration of a neighborhood that is a subset of all
3-exchanges. A 3-exchange consists in removing 3 edges from a tour and
reconnecting the tour in a different way. Of all these moves, Or-opt
only considers those that shift a sequence of 1, 2 or 3 vertices to a
different position in the tour. Each move can be evaluated in constant time. 

The checksum for a given permutation seed is the number of
improvements found while applying first-improvement Or-opt until no
improving move exists, using this permutation as starting solution.

\subsection{lns: Large neighbourhood search}
Large neighborhood search (LNS) was originally introduced in
conjunction with constraint programming to solve the vehicle routing
problem with time windows (VRPTW)~\citep{lns}. Since it has been
a very popular method for tackling a wide variety of vehicle routing
problems, see e.g.~\cite{lnschapter}.

LNS iteratively (i) copies the \emph{incumbent} solution, (ii)
destroys the copy by removing some vertices from it, (iii) repairs the
partial solution by re-inserting previously removed customers in it
and (iv) determines whether this newly produced solution becomes the
new incumbent solution or not. The method relies, among other
things, on randomness. This is an issue since we want each benchmark
to be deterministic. To remedy this issue, we design deterministic
schemes for both destroying and repairing stages. For each starting
seed, we apply a certain number of LNS iterations. At each iteration,
the vertices at even indices in the tour are removed. These removed
vertices are stored in a vector in the order in which they were
removed, i.e. from smallest to largest index. Then the cost of each
possible insertion is calculated and the cheapest insertion is
performed. Because removed vertices are stored in a vector, the
algorithm's behaviour in case of same-cost insertions is
deterministic. 

The checksum for a given permutation seed is the total
cost of insertions performed while applying 10 iterations of LNS using
that permutation as starting solution.

\subsection{espprc and espprc-index: Dynamic programming for column
  generation}
\label{sec:espprc}
Many state-of-the-art exact methods for vehicle routing problems
(VRPs) rely on column generation (see
e.g.~\cite{Baldacci:2012survey}). In such methods, a significant
amount of CPU effort is spent solving the \emph{pricing subproblem},
which differs depending on the problem at hand but is nonetheless
often an elementary shortest path problem with resource constraints
(ESPPRC). While it can be solved heuristically, in order to establish
optimality it eventually needs to be solved exactly as well. For this
benchmark we implement a dynamic programming algorithm similar to the
one by~\cite{feillet:2004exact} to solve the ESPPRC, albeit without
time windows. One difficulty in this problem stems from the existence
of negative-cost cycles in the graph. 

In order to simulate an environment similar to the one encountered in
column generation for VRPs, we derive an ESPPRC instance from each TSP
seed permutation by using an auxilliary distance matrix $c'$ from the
distance matrix $c$, as follows:
\begin{enumerate}
\item For each vertex $j$, let $\pi_j$ be the cost of arc $(i, j)$,
  where $i$ is the predecessor of $j$ in the seed permutation.
\item For each arc $(i, j)$, set $c'_{ij}$ to $c_{ij} - d_j$.
\end{enumerate}
Additionally, we generate resource consumption as follows: we consider
6 resources numbered from 0 to 5, and any vertex $i$ consumes resource
$r$ iff the $r^{th}$ bit of $i$'s binary representation is 1
(considering that the first bit is bit 0). That is, in C syntax, iff
\verb|i & (1 << r) > 0|. Capacity is set to 1 for each resource.
The algorithm then computes the elementary shortest path with resource
constraints on $c'$ from 0 to 0 (1 to 1 in Julia).

For this benchmark, floating-point numbers are used to represent cost,
even though the values are integer. This is to emulate the usual
setting when solving such problems in the context of column
generation.

Each label needs (i) a reference to its predecessor, so that
paths can be reconstructed and (ii) a collection of references to
its successors, in order to recursively propagate dominance and later
ignore labels that are successors of an already proved-to-be-dominated
label. Rust does not allow that, at least not in the intuitive way
where a label is encapsulated in a \verb|struct|, including references
to predecessor and successors. This is because Rust is focused on
security, while this type of construct can lead to unsafe operations if
not used properly. This means that this benchmark cannot be safely
implemented with Rust, even though it is consistent and reliable in all
other implementations. In order to remediate this issue, we produce a
second implementation of the same algorithm, although
in a less intuitive way: all labels are stored in a collection, and
the indices in this collection are used in other collections to store all
predecessors and successors. This second implementation represents a
different benchmark, which we call \emph{espprc-index}, whereas the
base benchmark is called \emph{espprc}. The version with indices is more
tedious to implement, but it may also be more efficient, so we also
implement it in languages other than Rust.

The checksum for a given permutation seed is the cost of the shortest
elementary path obtained when using that permutation seed to generate the
ESPPRC instance, truncated to its integer part.

\subsection{maxflow: Maximum flow problem}
Another popular framework for the exact solution of routing
problems is \emph{branch and cut}. Branch and cut is typically applied
to solve mixed-integer linear programs; it is akin to branch and bound
but only considers a subset of all constraints explicitly. Other
constraints are generated dynamically when they are found to be
violated while exploring the search tree. These constraints are said
to be \emph{separated}. One early success story of branch and cut is
actually in application to the symmetric TSP~\citep{padberg:532}, by
separating subtour elimination constraints. The separation procedure
involves finding a minimum capacity cut, which itself involves solving
a maximum flow problem. For this reason the maximum flow problem is
important in vehicle routing, and we dedicate one benchmark to solving
it. The \emph{maxflow} benchmark implements the Edmonds-Karp
algorithm~\citep{edmonds-karp}, which is a specific version of the
Ford-Fulkerson~\citep{ford-fulkerson} augmenting path method. The
specificity resides in how the augmenting paths are generated.

We derive a capacity graph $C$ from the distance matrix $c$ and a
given TSP solution as follows: 
\begin{enumerate}
\item For each vertex $j$, let $t_j$ be the cost of arc $(i, j)$,
  where $i$ is the predecessor of $j$ in the TSP solution.
\item For each arc $(i, j)$, set $C_{ij}$ to $\frac{c_{ij}}{1000}$ if $c_{ij}
  > t_j$, otherwise set it to 0.
\end{enumerate}

For any starting seed, the capacity graph is generated then the
Edmonds-Karp algorithm is applied with 0 as \emph{source} and every other
node as \emph{sink}. The checksum for that permutation seed is the sum
of all maximum flow values thus obtained, truncated to its integer part.

\section{Languages and implementations considered}
We consider six different programming languages: C++, Python, Java,
Julia, Rust and Javascript. Any rule of selection
is arbitrary in nature; however we indicate below, for each of these
languages, our perceived advantages of these languages, which justify
their selection. It is possible to add other languages to this list in
the future, and keep the benchmark collection alive in an online form.
For a given language several implementations are possible, e.g. one
with nested distance matrices and one with flat distance matrices.
As a convention, implementations are named in lower-case letters, for
example \emph{python-flat-matrix}.
We also use the simplest names for
the best implementations, i.e. \emph{python} uses a nested distance
matrix because it performs better than \emph{python-flat-matrix}, but
\emph{pypy} uses a flat matrix because it performs better than
\emph{pypy-nested-matrix}. Language versions and options are
summarised in Appendix~\ref{app:lang}.

For each language, we may consider multiple \emph{implementations},
which differ by which mechanisms or library they use. For
instance with Python we can compare an implementation using native
lists against an implementation using Numpy arrays. Implementations are
detailed individually below for each language. The code for all
implementations can be found at
\url{https://github.com/fa-bien/broute}.

It is difficult to properly measure the popularity of a programming
language. There exist rankings, for example based on online
searches~\citep{TIOBE} or on GitHub contributions~\citep{octoverse}. All
these approaches have flaws. Basing an index on online searches is
subject to all the biases and exploits that target online
search. Measuring GitHub contributions is not clear, as
there is no trivial way to decide what should count among
repositories, number of lines written, number of commits,
etc. Ultimately, language popularity for its own sake does not need to
be a crucial factor in the choice of a programming language, especially for
academic research. These aspects seem more important to us:
\begin{itemize}
\item The programs written in a given programming language should be
  easily run on any computer. For that reason, we only look at
  languages that are free and work under multiple operating
  systems. The ability to run under Linux is especially important,
  since it is typically the operating system of high-performance
  computing (HPC) clusters~\citep{top500}.
\item The language should be well documented, but also have online
  resources for support. This in fact also depends on popularity.
\end{itemize}

\subsection{C++}
C++ is one of the most popular programming languages in the World, and
has been so for decades. Among its many features there is good
runtime performance. In fact the performance is good enough that whole
operating systems can be written in C++, for example Microsoft
Windows. C++ has many features and it is often possible to
implement the same algorithm using different subsets of the language;
one criticism is that people and teams evolve into using
only their own subset of the language.

We develop two C++ implementations of every benchmark, using the C++98
and C++14 standards, which we call \emph{c++98} and \emph{c++14}. they
both use a flat distance matrix. With C++98 we use references to
objects, while with C++14 we use smart pointers such as
\verb|shared_ptr|. Comparing these two implementations will allow us
to evaluate the cost of smart pointers.

For each benchmark the reference CPU time is provided by the c++98
implementation. We also develop a variant of c++14 using a nested
distance matrix, and another using static arrays.

\subsection{Python}
Python is another of the most popular programming languages in the
World, especially for tasks of machine learning in recent years. The
base Python implementation is called \emph{python} and uses a nested
distance matrix. There is also an implementation using a flat distance
matrix, called \emph{python-flat-matrix}, and one using a nested
distance matrix wrapped in function calls, called
\emph{python-nested-matrix-function}.

One issue with Python is that much of the code is not compiled, rather
directly interpreted, which can lead to poor performance. One way to
address that issue is to rely on libraries that are implemented in C,
however that is not always possible. We consider alternative
implementations in Python, all bringing different solutions to this
issue.

The simplest way to speed up the run time of a Python program is
to run this program with Pypy~\citep{pypy} instead of the
standard CPython interpreter. Pypy uses JIT compilation to speed
things up. There is no need to modify the Python program, which is a
strong advantage compared to library-based solutions or
annotation-based solutions. Pypy's website claims that it is on
average 4.2 times faster than CPython~\citep{pypy}, however this is
just an average and actual numbers may vary a lot depending on the
benchmark. We use both the python and python-flat-matrix
implementations with Pypy, resulting in \emph{pypy-nested-matrix} and
\emph{pypy} implementations, respectively.

We also implement the benchmarks using Numpy arrays instead
of standard Python lists. Numpy is one of the most popular external
libraries for Python; it is coded in C for runtime
performance~\citep{numpy}. Numpy performs especially well when using
\emph{array operations}, for instance matrix multiplication. However
the benchmarks that we consider do not take advantage of Numpy
functions, only of its array data structure. Since Numpy provides
multi-dimensional arrays, we use a two-dimensional array to represent
the distancematrix. Additionally, we consider a flat matrix in
implementation \emph{numpy-flat-matrix}.

Numpy itself does not provide any kind of JIT compilation, but another
Python package, Numba, performs JIT compilation when using Numpy
arrays. Numba requires code annotation, therefore we annotate the
Numpy implementations to take advantage of them with Numba.

\subsection{Java}
Java is a very popular programming language of the past two
decades. Java code is compiled to \emph{bytecode}, which is then run
by a virtual machine (VM). The bytecode compiled on one computer can
be run by a VM on another computer. Nonetheless, JIT compilation to
native code happens dynamically at run time. The syntax is based on
C/C++ syntax. The base Java implementation, \emph{java}, uses a flat
distance matrix and a \verb|ArrayList<Integer>| object to represent a
tour. We also implement a version with nested distance matrix,
\emph{java-nested-matrix}, as well as a version with static arrays to
represent a tour, \emph{java-static-arrays}.

\subsection{Julia}
Julia is a relatively recent programming language. It is a
general-purpose language but it is also particularly aimed at
computationally intensive tasks such as numerical
analysis~\citep{Julia-2017}. Version 1.0 was 
released in 2018 and the language is well-documented and has a strong
community.

Compilation happens in a JIT fashion. Like Numpy but
unlike all other programming languages considered, Julia supports
multi-dimensional arrays, therefore a native two-dimensional array can be
used to represent a distance matrix. We also develop an
implementation using flat matrix representation in Julia, called
\emph{julia-flat-matrix}; we will determine which version is better
through experiment.

One valuable feature of Julia is the fact that projects have an
environment, including package dependencies with specific
versions. The language makes it easy to \emph{instantiate} the
project environment, i.e. recreate the same environment in which the
project was developed, with the same packages in the same
versions. This provides safety from backward-compatibility-breaking
library updates.

One notable issue with Julia is the inability to generate a
compiled binary executable file. In general, in order to run a Julia
program, one must install the Julia ecosystem. There are active
projects to remedy this, but we are not aware of any easy one-command
solution.

\subsection{Rust}
Rust is also relatively recent, with version 1.0 released in 2015. It
is a general-purpose language with an emphasis on memory safety and
on performance. Notable software using Rust include Firefox, Dropbox
and Cloudflare~\citep{rust-lang}.

Programs in Rust are pre-compiled, like in C++ for
instance. Like Julia, the language is well documented and has a strong
community. Also like Julia, projects have an environment, and it is
easy to recreate the environment for a given project, including
dependencies and their specific version, which again provides safety
from backward-compatibility-breaking library updates. The \emph{rust}
implementation uses a flat matrix representation.

\subsection{JavaScript}
JavaScript is likely one of the most popular programming languages in
the World, as it is used on countless websites as well as mobile
applications based on web technology. JavaScript is not expected to
perform as well as, say, C++. However there are several reasons why it
can be appealing to implement routing algorithms in JavaScript:
\begin{itemize}
\item Any program written in JavaScript can run on virtually any
  computer with a modern web browser, including smartphones.
\item Integrating JavaScript with a web interface is especially easy.
\item JavaScript engines have received considerable attention from
  major companies and been the subject of fierce competition in
  relation with web browsers engines. As a result they have seen vast
  performance improvements over the years and the trend is likely to continue.
\end{itemize}

We use Node.js, which allows to run JavaScript programs from the
command line. This means that there is in fact not even the need for a
web browser. Node.js currently uses Google Chrome v8's JavaScript
engine~\citep{nodejs}.

The \emph{javascript} implementation uses a flat matrix representation
but there is also an implementation using nested matrix representation,
called \emph{javascript-nested-matrix}.

\subsection{Considerations on how to implement a distance matrix}
Perhaps the most straightforward way to implement a distance matrix in
a number of languages is to use nested arrays, i.e. each element of the
main array is an array representing a row of the distance matrix. This
is typically done in C++, Java, Python (although the structure is
officially called a \emph{list} and not an array), Javascript. C and
C++ use pointers to achieve that effect. All these languages use the
same C syntax for looking up values in the distance matrix: the
distance between vertices $i$ and $j$ using distance matrix $d$ is
written \lstinline{d[i][j]}.

Another easy way to implement a distance matrix is to use what we
call from here on a \emph{flat} distance matrix representation, which
is a single-dimensional array containing all distance values. Assuming
indices start at 0, the distance from $i$ to $j$ in array $d$ can be
coded as \lstinline{d[i*n+j]}, where $n$ is the total number of
vertices. This representation guarantees that the whole distance
matrix is stored in contiguous memory. Additionally, looking up in
arrays also takes time, so one lookup is better than two. The drawback
is that we have to pay a multiplication and an addition for each
lookup, and that lookups have to be wrapped in a function call for the
sake of readability. Such function calls can usually be inlined,
i.e. the function content is substituted to the function call, so
there should be no runtime penalty from using a function call. In
general we expect a speedup from using a flat representation. We
implemented some of the benchmarks with both flat and nested matrix
representation in order to determine if there are significant
performance differences.

There are a few cases where the above considerations on using a flat
representation do not apply: Python does not allow inlining, while
Numpy and Julia provide multi-dimensional arrays. Nonetheless we also
wrote a flat matrix version of these implementations, in order
to elicit any difference in performance. 

\subsection{Other general cross-language considerations}
Here we discuss considerations that are subjective in nature, as they
relate to individual experience implementing all benchmarks in all
languages.
The author's experience with all considered programming languages is
heterogeneous, which can cause bias in the perception of the ease of
implementation in a given language.
To add to the subjectivity, each benchmark is first implemented in
Python then in other languages, with Rust typically being the last one
to be implemented. Therefore the languages implemented last represent
supposedly easier tasks, due to increased familiarity with the
algorithm. For all reasons mentioned above, measuring the
time spent implementing each benchmark in each language would be
meaningless, therefore we limit ourselves to subjective opinion here.
Keeping this in mind, we see a subset of perceived
\emph{easy} languages, as in ``easy to program in and get things running
with the expected outcome'', which consists of Python, Julia and
JavaScript. These 
easy languages typically allow to program the same algorithms faster
than the other ones with the same results, performance considerations
aside.
On the other hand, we found the process of getting benchmarks to work
with Rust to be much more tedious than with any other language,
although this is certainly due to lack of experience. Nonetheless, the
JavaScript implementation was typically easier than the C++ or Java ones
despite having more experience with C++ and Java, so
experience does not explain everything.
Ultimately, there is little doubt that any experienced programmer in
any of these languages would have no trouble implementing any of these
benchmarks. Nonetheless, this programmer with experience in mostly C++
and Python implemented every benchmark faster in Python, JavaScript or Julia
than in C++.

We also want to underline that no matter the language, it is possible
to write both efficient and inefficient code. Familiarity and
experience are beneficial; beginners in a language are more likely to
write inefficient code. This being said, not every language is equal
when it comes to documenting how to write efficient code. We note that
the official Julia documentation has a whole section dedicated to that
topic~\citep{julia-performance}. There also exists an automated tool to
detect inefficient code patterns, based on the official
documentation~\citep{julia-traceur}.

\section{Implementation-specific notes}
\subsection{Python}
CPython, Python's official interpreter, does not allow to inline
functions. Since using a flat matrix requires to wrap matrix value
lookup in a function, using a flat distance
matrix in Python is not beneficial. It would of course be possible to
inline by hand, i.e. formulate the correct 1-dimensional index in the
matrix for every lookup, but that would be impractical and defeat the
purposes and philosophy stated in Section~\ref{sec:intro}.
Running Python code that uses a flat representation with a JIT compiler
(e.g. Pypy, Numba) introduces automatic inlining and is likely to
remedy this.

On the ESPPRC benchmark, a label needs to store information about
which customers have already been visited in the partial path it
represents. In Python, preliminary testing reveals that using a set of
integers for that purpose is twice as fast as using a vector of
booleans. Similar preliminary testing in C++ reveals the opposite:
using an array of booleans is faster. This is the behaviour we would
expect in general, the observation on Python is the surprising one
here. There are two reasons for this expectation: (i) copying an array,
represented in contiguous memory, can in general be performed faster
than copying a set structure and (ii) lookup in an array is performed
in $\mathcal{O}(1)$ while lookup in a set is performed in
$\mathcal{O}(log(n))$.

\subsection{Julia}
Julia has native multi-dimensional arrays, which may be more efficient
than a flat representation. We will determine which version to use
based on experiment. Since Julia's arrays start with index 1, the code
for flat representation lookup of the distance between vertices $i$
and $j$ in matrix $d$ is \lstinline{d[i*n-n+j]}. It is worth noting
that it involves one subtraction on top of the addition and
multiplication used for 0-indexed languages.

\subsection{JavaScript}
All benchmarks are repetitive in nature, as they successively perform similar
operations with different input data. In JavaScript, when our
implementation loads 40 instances and performs the same
benchmarking operations on all 40 them, we observe a significant performance
hit after a few instances (2-5 times slower). In order to remedy this,
we re-start the program individually for each instance. Advanced
knowledge of this specific JavaScript engine might allow to remove that
performance hit, but this is clearly outside the scope of this
study. However it is worth noting that this can be a concern in general.

\subsection{Numba}
Numba is not yet feature-complete for Numpy. A direct consequence is that some
benchmarks cannot be implemented for Numba. For example
\verb|numpy.insert()| is not implemented, but this function is
needed for the $lns$ benchmark, therefore the $lns$ benchmark cannot be
implemented with Numba. Additionally, we were not able to implement
the ESPPRC benchmark in a satisfying manner. In our experience, Numba
error messages are sometimes unrelated to the issue causing them, or
too obscure to make sense, keeping in mind the premises of this
paper. It would likely be possible to implement a dynamic
programming algorithm for the espprc benchmark using Numba's JIT
compilation features, however our conclusion is that it would not be
possible to do so without advanced knowledge of Numba.

Still, Numba is relevant enough to be
benchmarked and its current performance is an indication of what to
expect in years to come, as the project matures.

\subsection{Java}
The ecosystem of Java implementations and versions can be hard to
navigate. There exist multiple compilers, multiple virtual
machines, and there is a new Java version every 6 months, sometimes
breaking backwards compatibility. The official reference
implementation is OpenJDK since version 7~\citep{openjdk-reference}.
We use OpenJDK version 11, which is the current long-term support version as
of writing this article.

\subsection{Rust}
As mentioned in Section~\ref{sec:espprc}, Rust does not allow a given
\verb|struct| to have multiple references to the same \verb|struct|
type. For this reason, a different implementation of the espprc
benchmark is needed. In general Rust puts restrictions on what
constitutes valid code, for the sake of safety. This is a valuable
feature in systems programming, which is the main application that Rust was
designed for. This also means that writing code that compiles and runs
is actually a more tedious task, but that once the code compiles,
there is also a higher chance that it will run according to
expectations, like with the Ada programming language for instance.

In the context of this study, these benefits did not appear relevant
however: each benchmark was programmed in all other languages as well,
providing the same results as with the Rust implementation, while no
evidence of memory leak was observed. As mentioned above, The Rust
implementation was typically way more time-consuming than other
implementations, although one reason for that is lack of
experience. Nonetheless, in the context of this study, the generally
assumed benefits of using Rust are not apparent, with the notable
exception of good runtime performance.

\section{Experiments}
\label{sec:exp}
In this section, we provide examples of some possibilities offered
by BROUTE and perform topic-specific comparisons for a given hardware
and software setup.
The computer is a cluster where each node has an Intel
Xeon E5-2650 v3 CPU, 20 cores per CPU, 3 GHz, running Linux. The
cluster is shared and it is not possible to reserve an entire
node. The compiler and interpreter versions used are described in
Appendix~\ref{app:lang}. For every benchmark and every
instance, CPU times are reported as a ratio of a \emph{reference time} for that
benchmark and instance. In practice, for each benchmark and instance,
the reference time is the time required by the c++14 implementation
for that benchmark and that instance. Therefore the c++14
implementation always has a CPU ratio of 1.

\subsection{Benchmark data}
Each instance file contains one $n \times n$ distance matrix as well
as $p$ randomly generated permutations, which are used as starting
\emph{seeds} for the algorithms (e.g. as starting solution for 2-opt).
For $n$ taking each value of $\{20, 40, 60\}$ and fixing $p = 1000$ we
generate 40 instance files, for a total of 120 instance files. Each
instance file contains 1000 permutations, therefore involves running
each benchmark 1000 times. As mentioned in
Section~\ref{sec:benchmark}, distances are Euclidean distances multiplied
by 100 and truncated to their integer part. In the following, each data
point represents the total CPU time for running one benchmark implementation
for each of the 1000 permutations of one given instance file.

These data points are represented in two types of charts: box plots and
performance profiles. Box plots contain different boxes for different
values of $n$, and these boxes are always displayed, from left to
right, in the order in which they
appear in the legend from top to bottom. Performance profiles
aggregate all data points for a
given implementation into one profile, i.e. one curve. A performance
profile is a chart that indicates, for a given level of performance,
the ratio of data points produced with that performance level or a
better one. In our case, the performance level is the CPU effort
required to produce a data point, as a ratio of the CPU time taken by
the fastest implementation over all implementations compared in the
same chart, for the same input file. For more information,
see~\cite{Dolan2002}. Box plots are produced with the \emph{ggplot2}
package~\citep{ggplot2} in R~\citep{R}. Performance profiles are
produced using a homemade script in Python, relying on
matplotlib~\citep{matplotlib}.

\subsection{A comparison of Python implementations}
We first compare different implementations based on Python:
\begin{itemize}
\item Implementation 'python' uses CPython, the official Python interpreter
\item Implementation 'numpy' uses Numpy, a Python library for
  numerical operations that relies on code implemented in C. We also
  call it from CPython.
\item Implementation 'numba' used Numba, a Python library that
  performs JIT compilation. Numba is advertised to work well in
  conjunction with Numpy, i.e. the functions from Numpy are compiled
  by Numba. Numba is under development. It is also called from CPython.
\item Implementation 'pypy' uses Pypy, a different Python interpreter
  that also performs JIT compilation. The code is exactly the same as
  with implementation 'python'.
\end{itemize}

\begin{figure}
  \begin{subfigure}{0.5\textwidth}
    \centering
    \includegraphics[scale=.4]{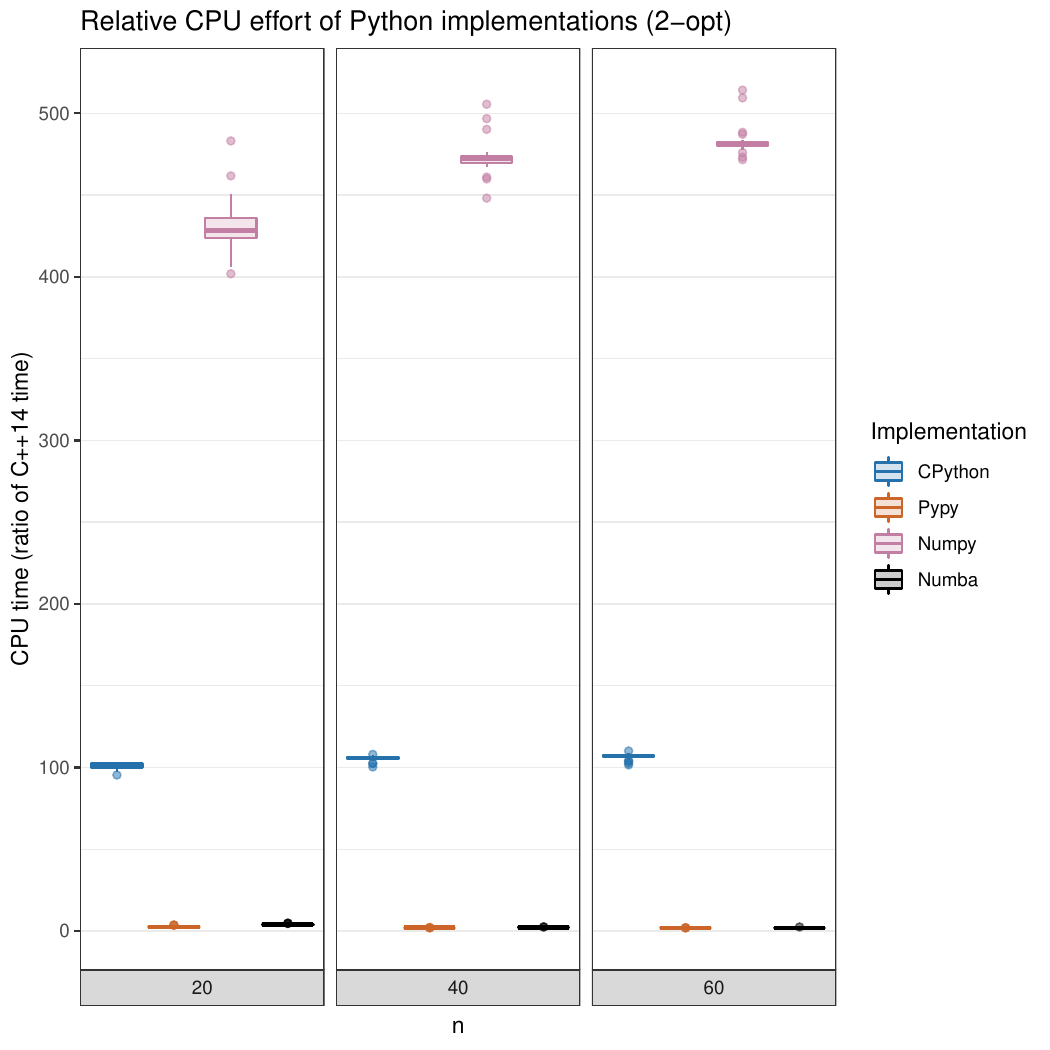}
    \includegraphics[scale=.3]{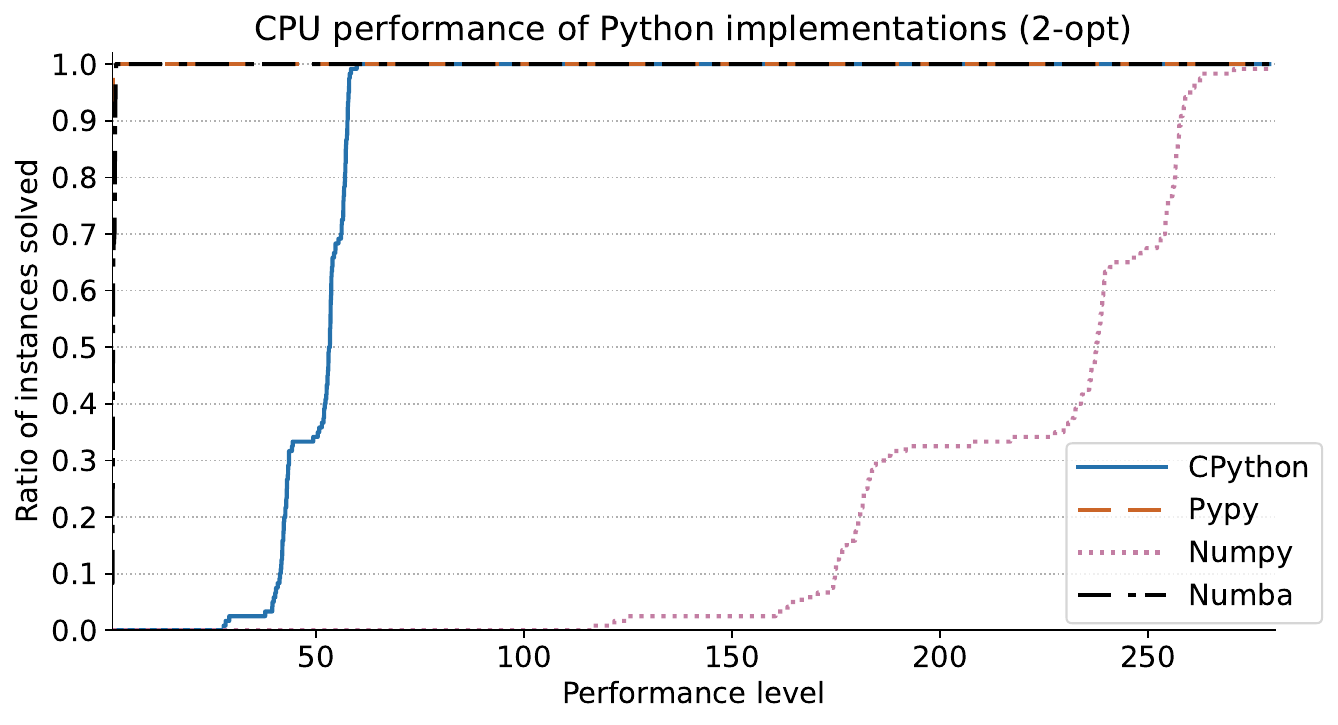}
    \caption{2-opt benchmark}
  \end{subfigure}
  \begin{subfigure}{0.5\textwidth}
    \centering
    \includegraphics[scale=.4]{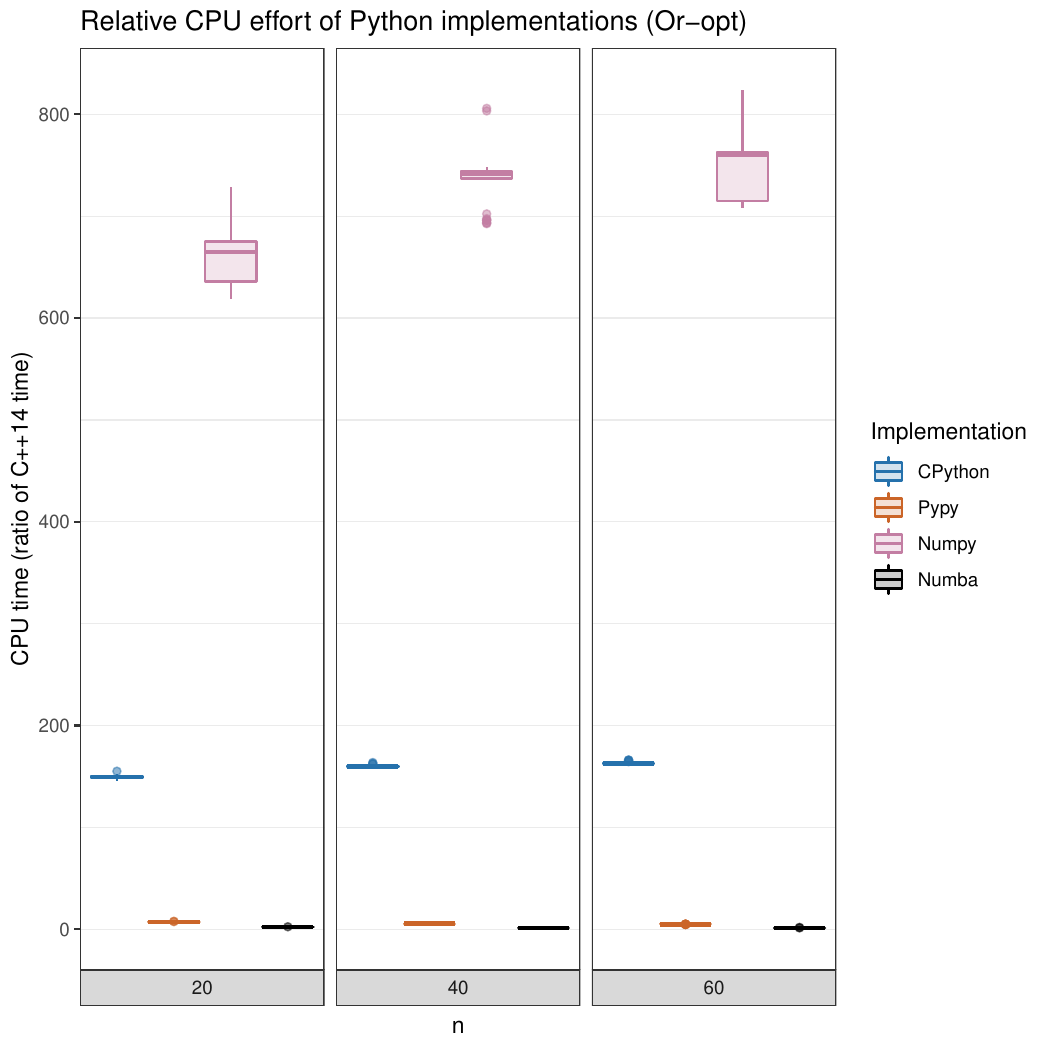}
    \includegraphics[scale=.3]{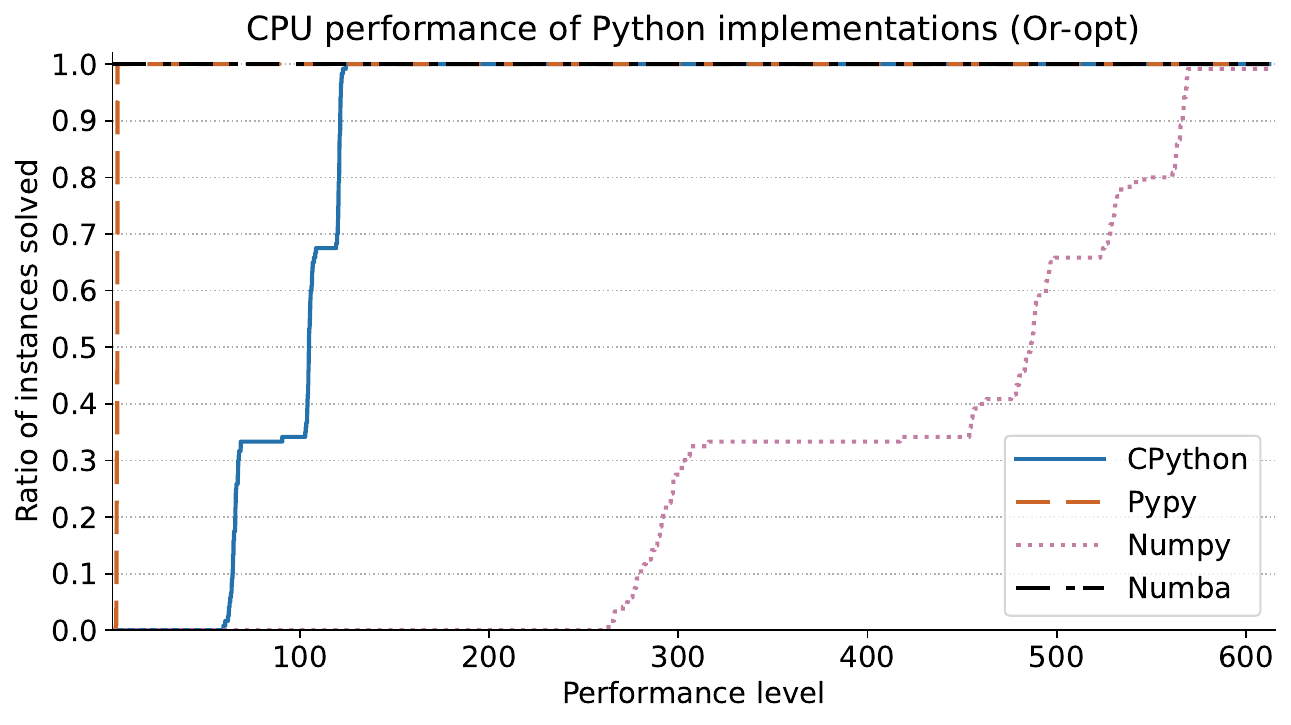}
    \caption{Or-opt benchmark}
  \end{subfigure}
  \caption{Python implementations: box plots and performance profiles}
  \label{fig:python-implementations}
\end{figure}

We first compare the performance of these four implementations on
benchmarks 2-opt and Or-opt. The results for this comparison are
summarised in Figure~\ref{fig:python-implementations}.
The box plots indicate that the performance of the numpy
implementation is always more than 400 times slower than the c++14
implementation. The performance profiles show that it is also up to
600 times slower than the fastest Python implementation in some
cases. Overall, it is clear that the Numpy implementation has the
worst performance. This is easily explained: every time a function
from Numpy is called, it is wrapped in order to call the C code for
that function. At the same time, none of the benchmarks actually use
functions which would make this beneficial, i.e. functions that would
perform a lot of calculations written in C then return the result of
this calculation (e.g. matrix multiplication). Such operations are
simply not needed in these benchmarks. Therefore the cost of calling C
code is paid often, but the benefits are never reaped. Additionally, it
is clear that the base CPython implementation is also very slow,
although not as much as the Numpy one. However, the Numpy structures
combined with JIT compilation from Numba look promising.

\begin{figure}
  \begin{subfigure}{0.33\textwidth}
    \centering
    \includegraphics[scale=.28]{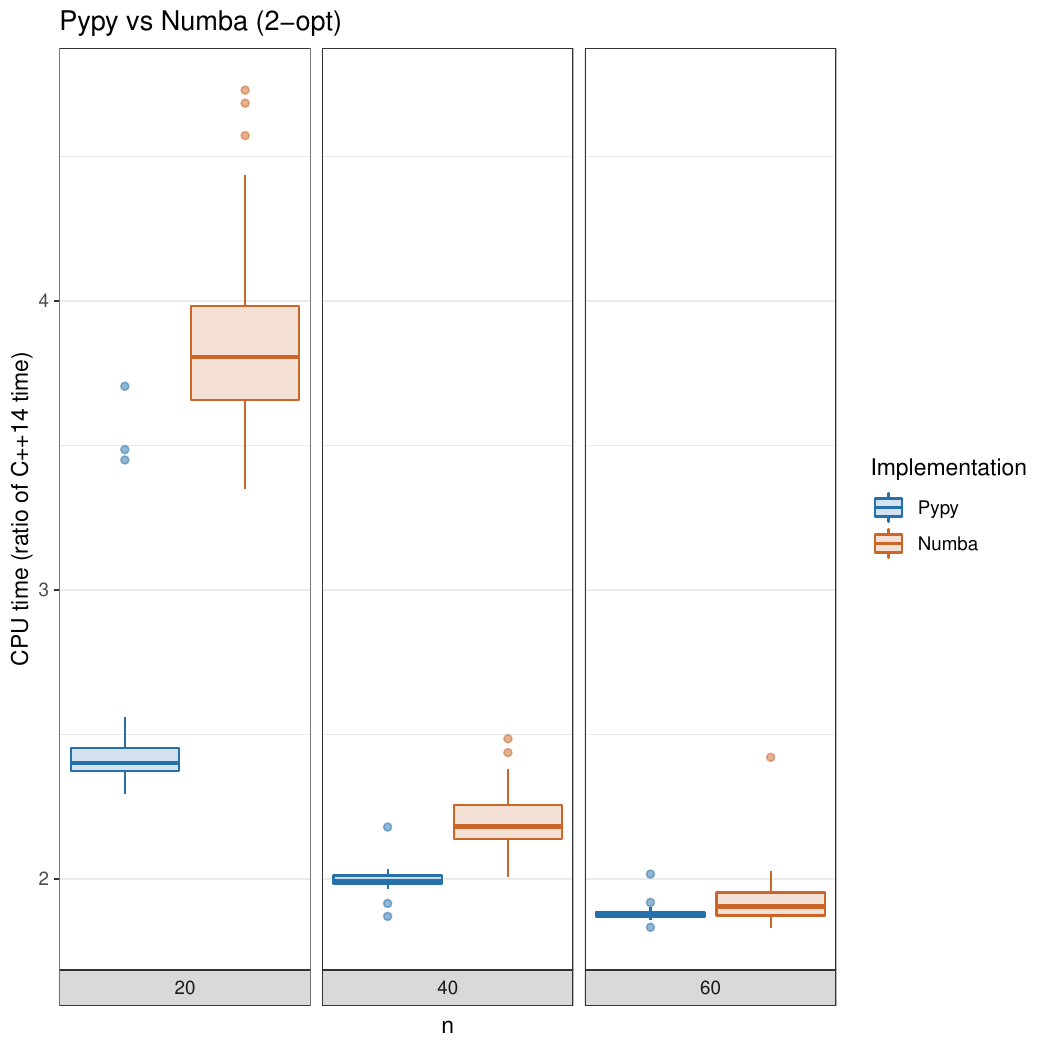}
    \includegraphics[scale=.22]{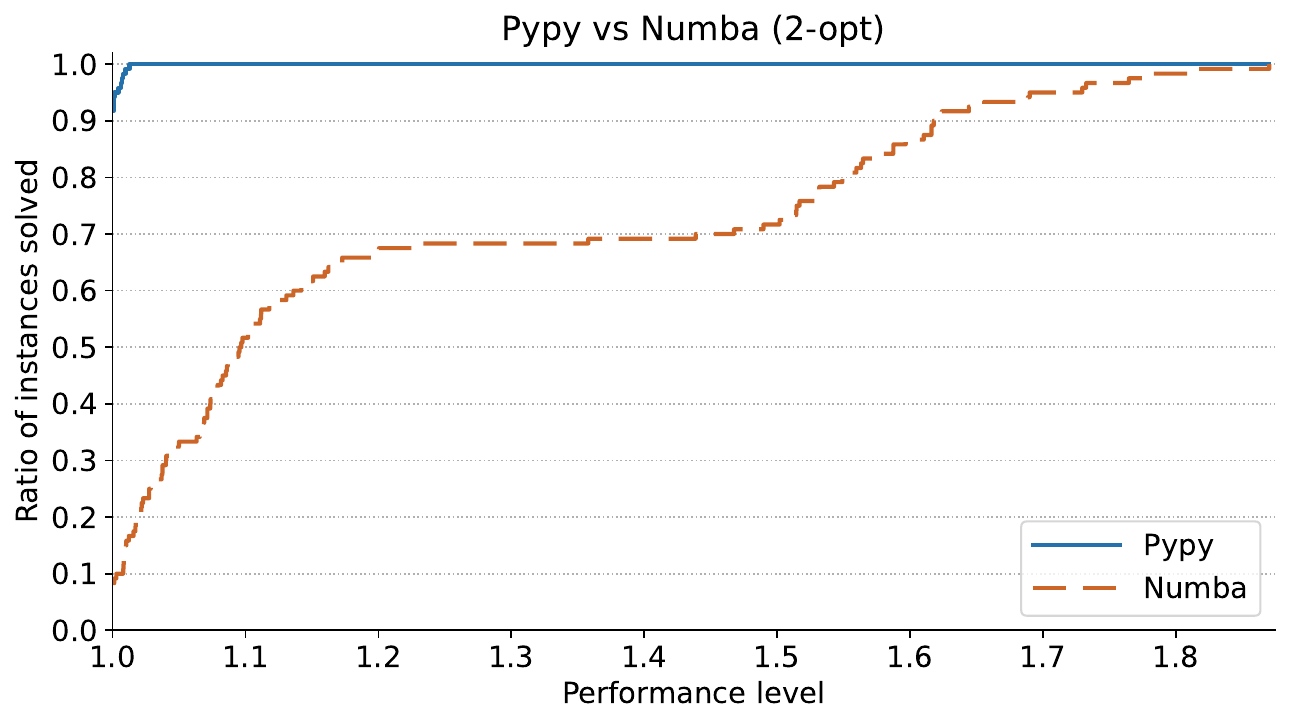}
    \caption{2-opt benchmark}
  \end{subfigure}
  \begin{subfigure}{0.33\textwidth}
    \centering
    \includegraphics[scale=.28]{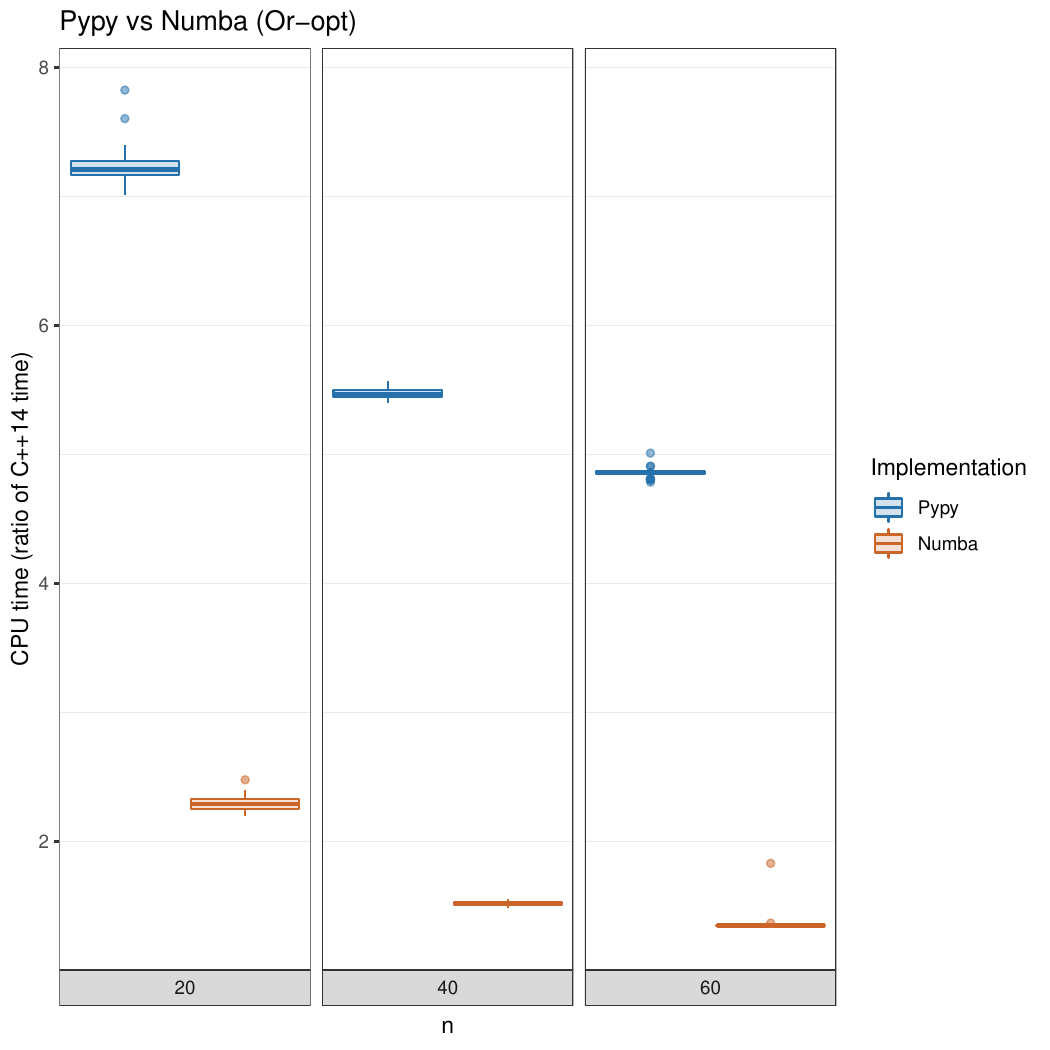}
    \includegraphics[scale=.22]{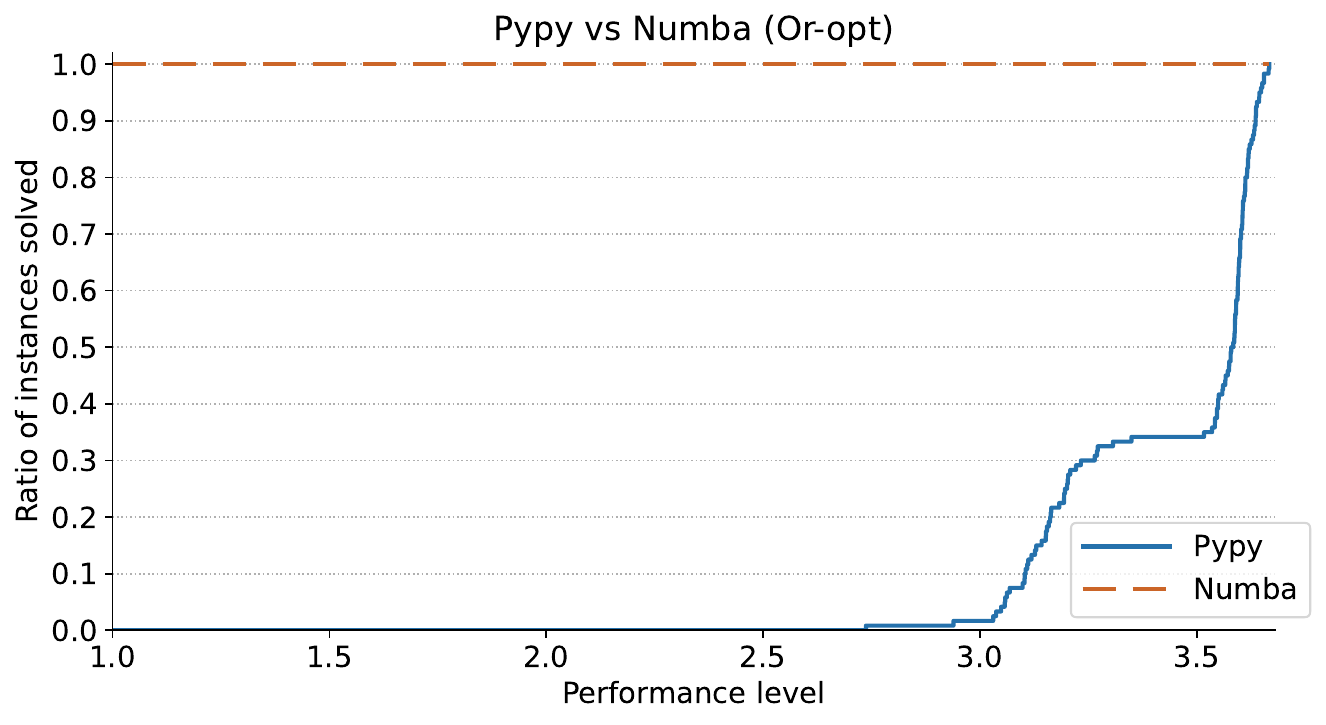}
    \caption{Or-opt benchmark}
  \end{subfigure}
  \begin{subfigure}{0.33\textwidth}
    \centering
    \includegraphics[scale=.28]{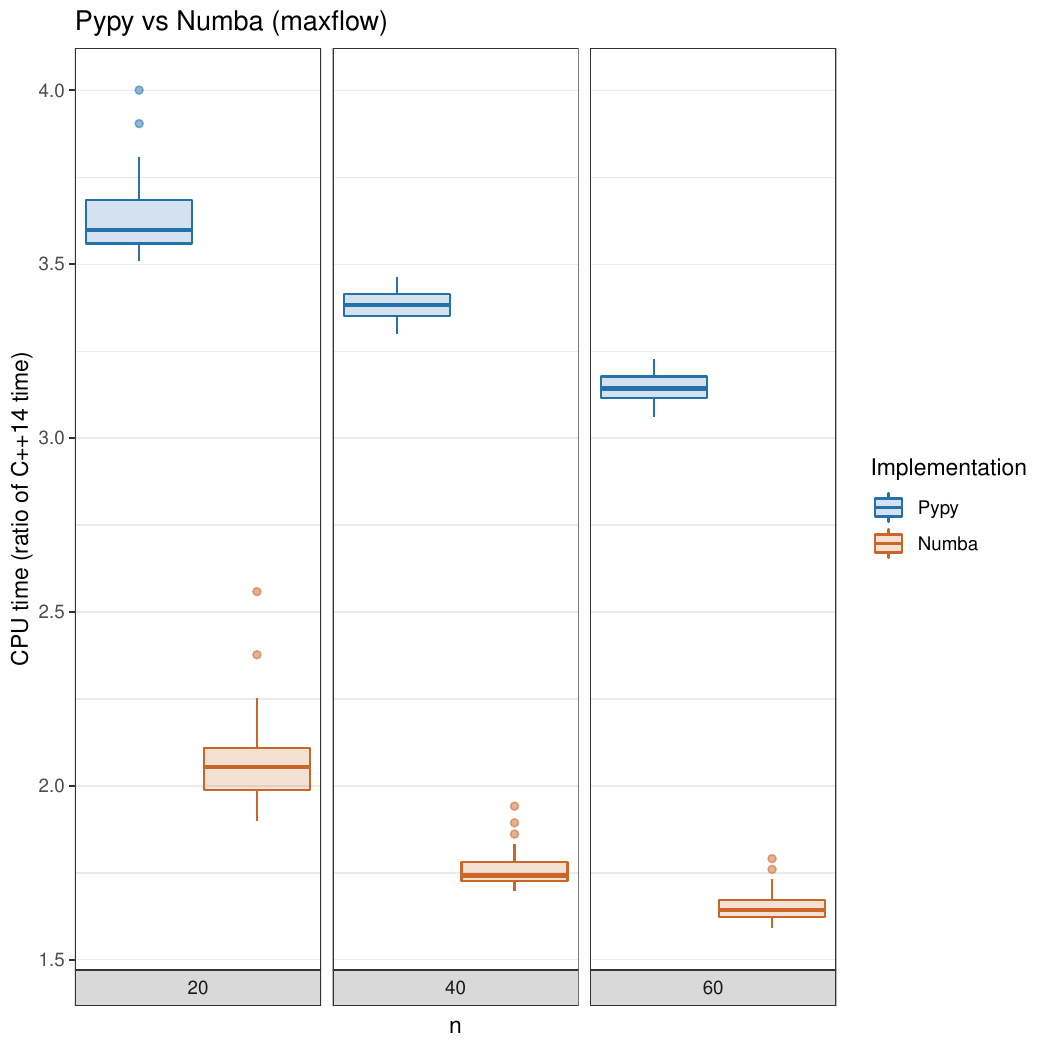}
    \includegraphics[scale=.22]{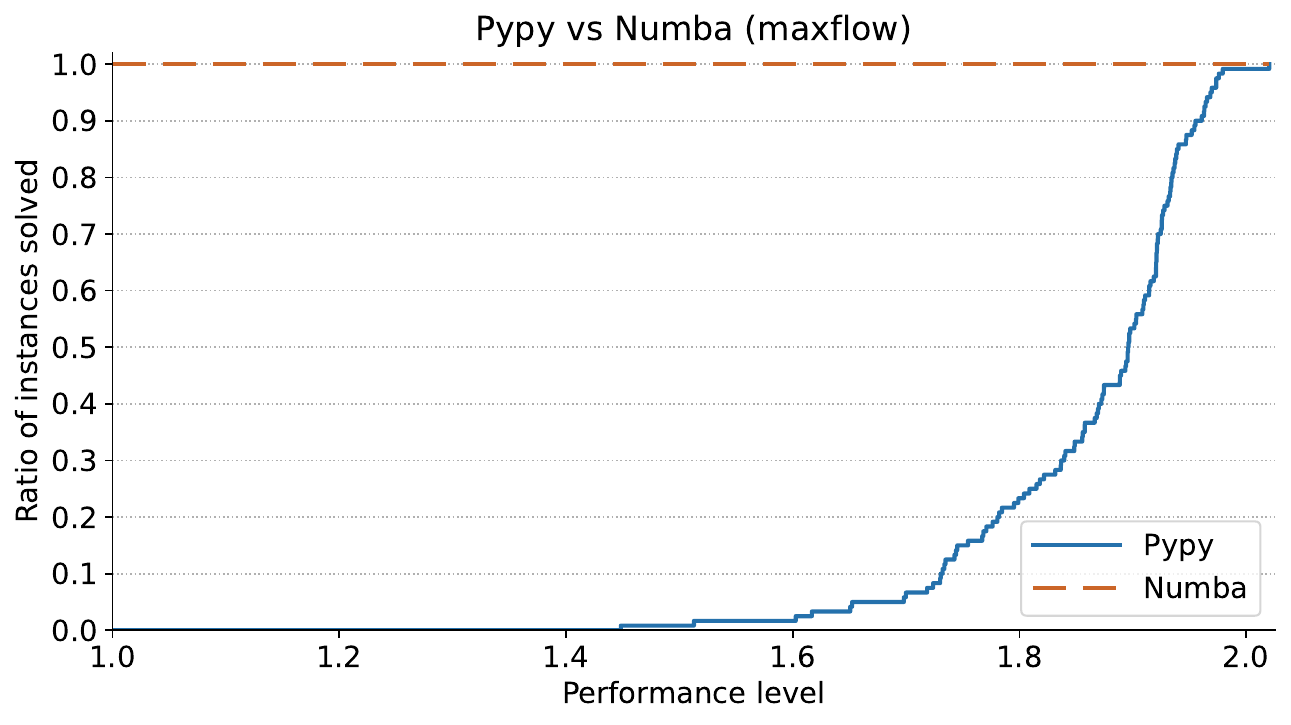}
    \caption{maxflow benchmark}
  \end{subfigure}
  \caption{Pypy vs Numba: box plots and performance profiles}
  \label{fig:pypy-vs-numba}
\end{figure}

We now compare Pypy and Numba only, on all benchmarks except for lns and
espprc, since they are not implemented with Numba as explained
above. These results are presented in Figure~\ref{fig:pypy-vs-numba}.
Pypy is up to almost twice as fast as Numba on the 2-opt benchmark,
while Numba is more than three times as fast in a majority of cases for
the Or-opt benchmark and up to twice as fast for the maxflow
benchmark. We also note that for the 2-opt benchmark, the performance
gap reduces when instance size increases, keeping in mind that the
2-opt benchmark is the fastest of the lot. This may simply mean that
Numba has a higher initial cost for JIT compilation. Overall it seems
that Numba is faster, except for very quick computations. However we
still see two reasons to use Pypy: it does not require any
modification to the Python code, and it is feature-complete, thus
allowing to implement all benchmarks. In fact the code for the CPython
and Pypy implementations of each benchmark is fully identical.

\subsection{Impact of using a flat matrix}
\begin{figure}
  \begin{subfigure}{0.33\textwidth}
    \centering
    \includegraphics[scale=.28]{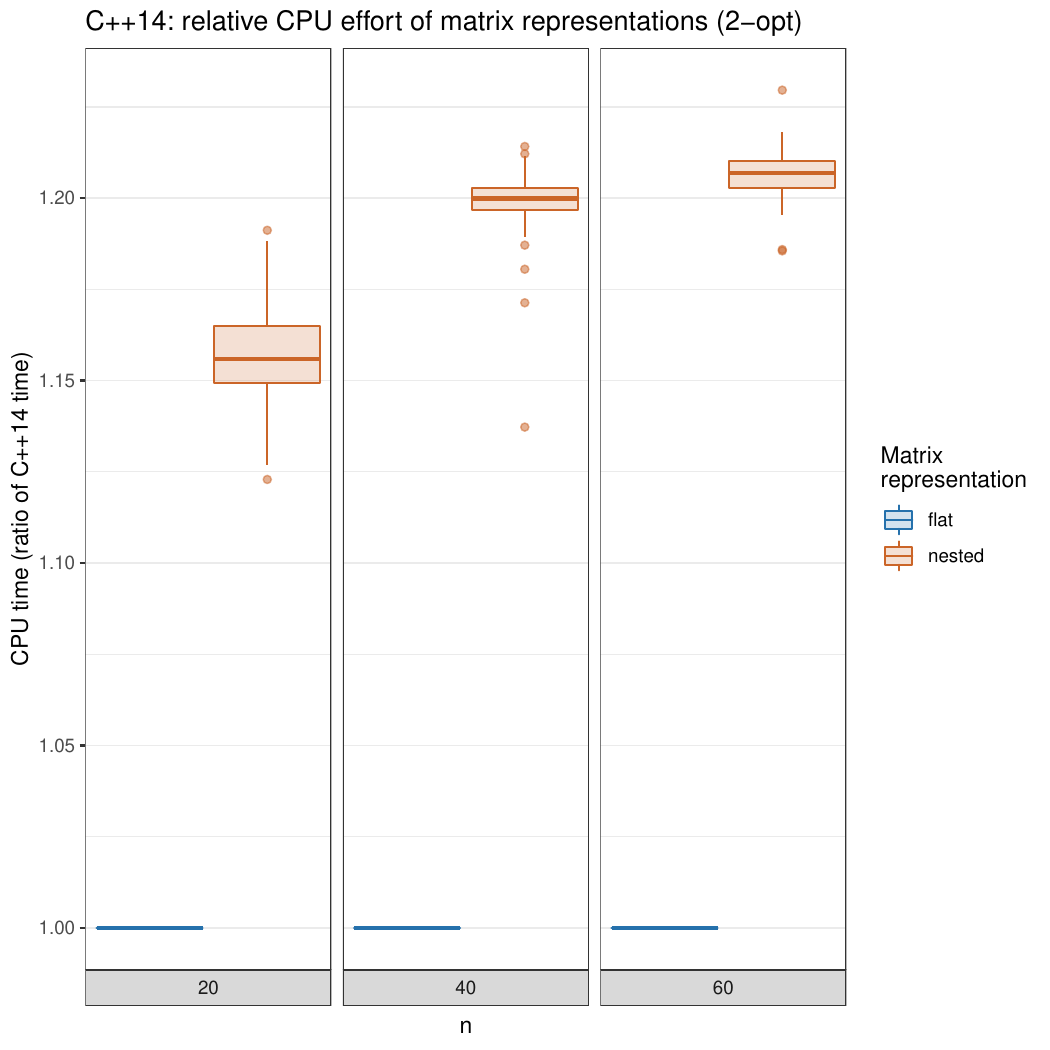}
    \includegraphics[scale=.22]{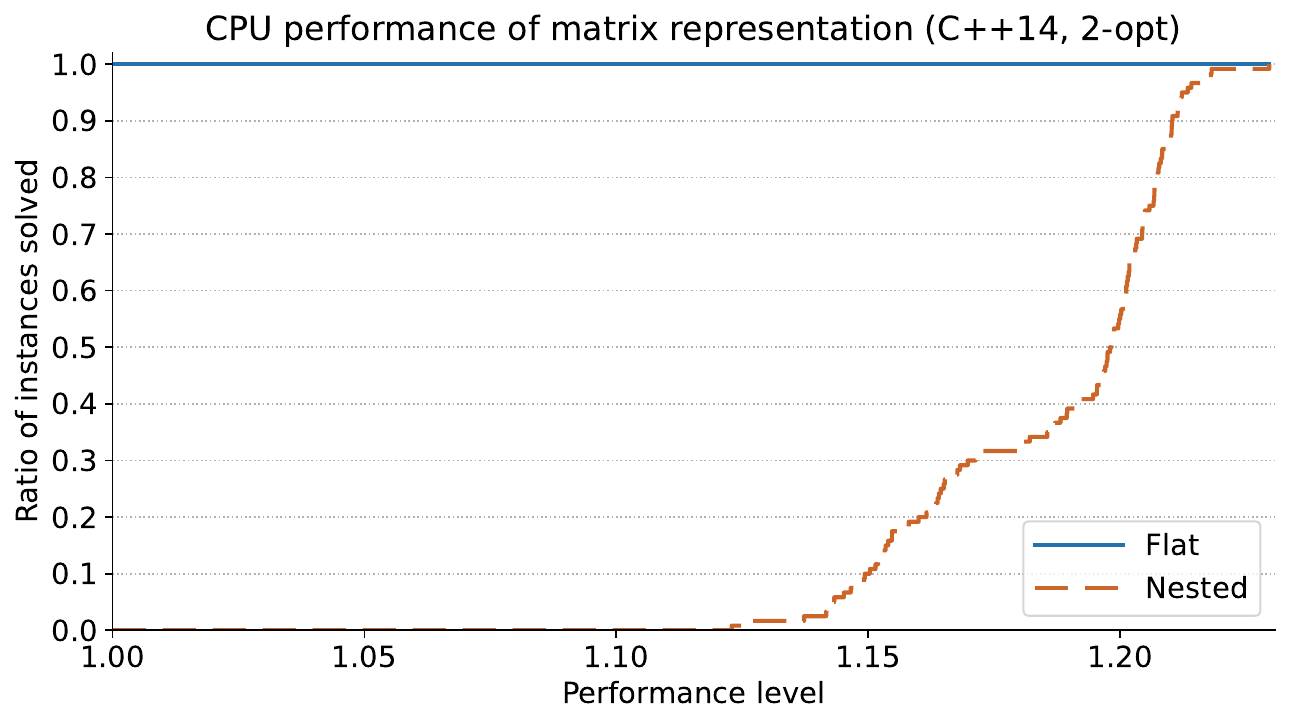}
    \caption{2-opt benchmark}
  \end{subfigure}
  \begin{subfigure}{0.33\textwidth}
    \centering
    \includegraphics[scale=.28]{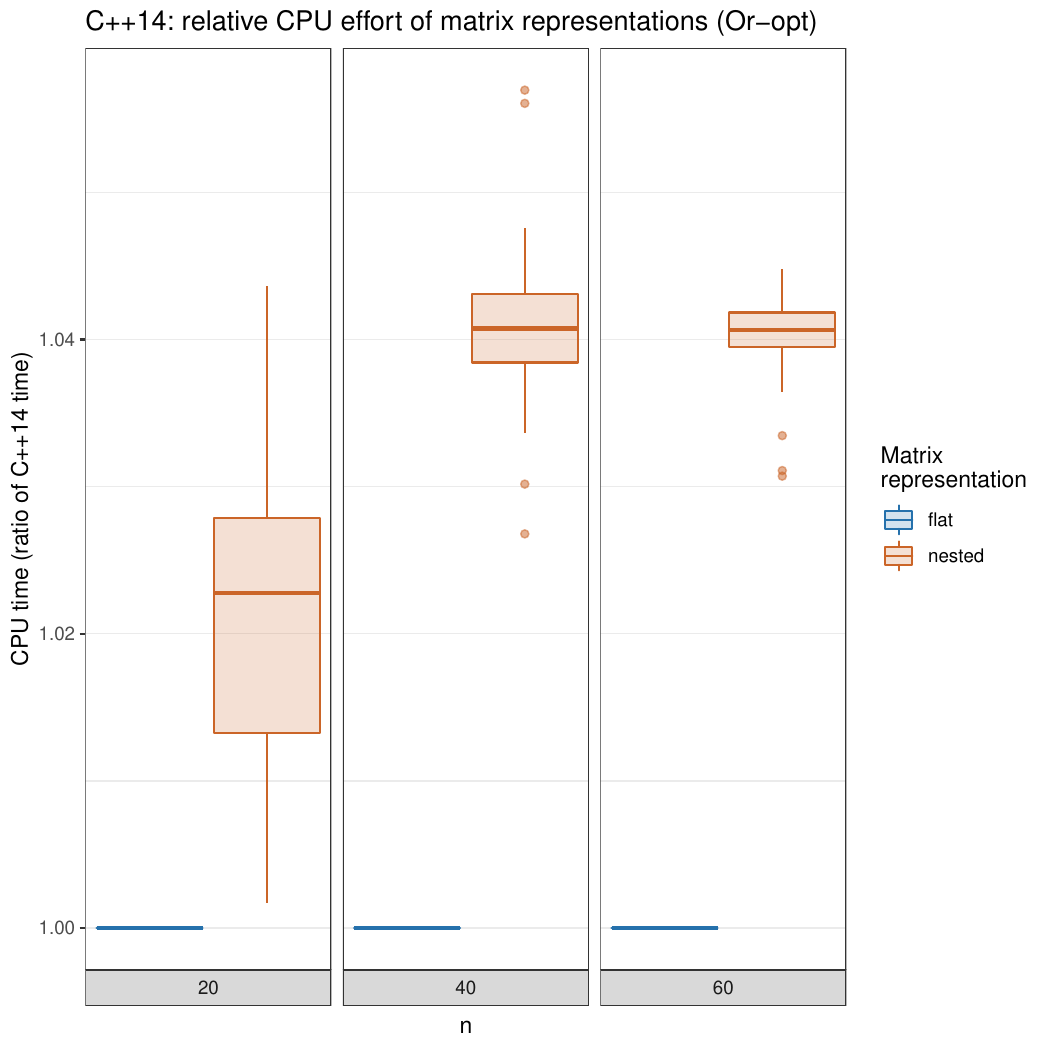}
    \includegraphics[scale=.22]{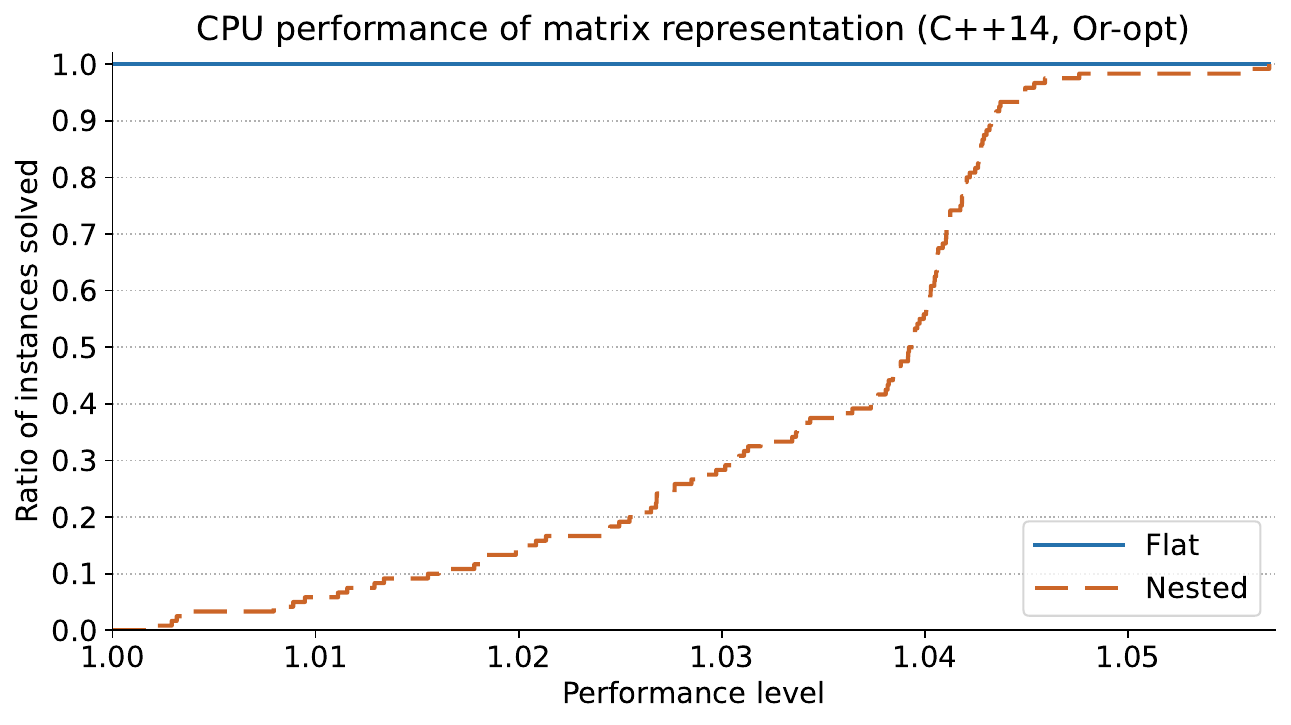}
    \caption{Or-opt benchmark}
  \end{subfigure}
  \begin{subfigure}{0.33\textwidth}
    \centering
    \includegraphics[scale=.28]{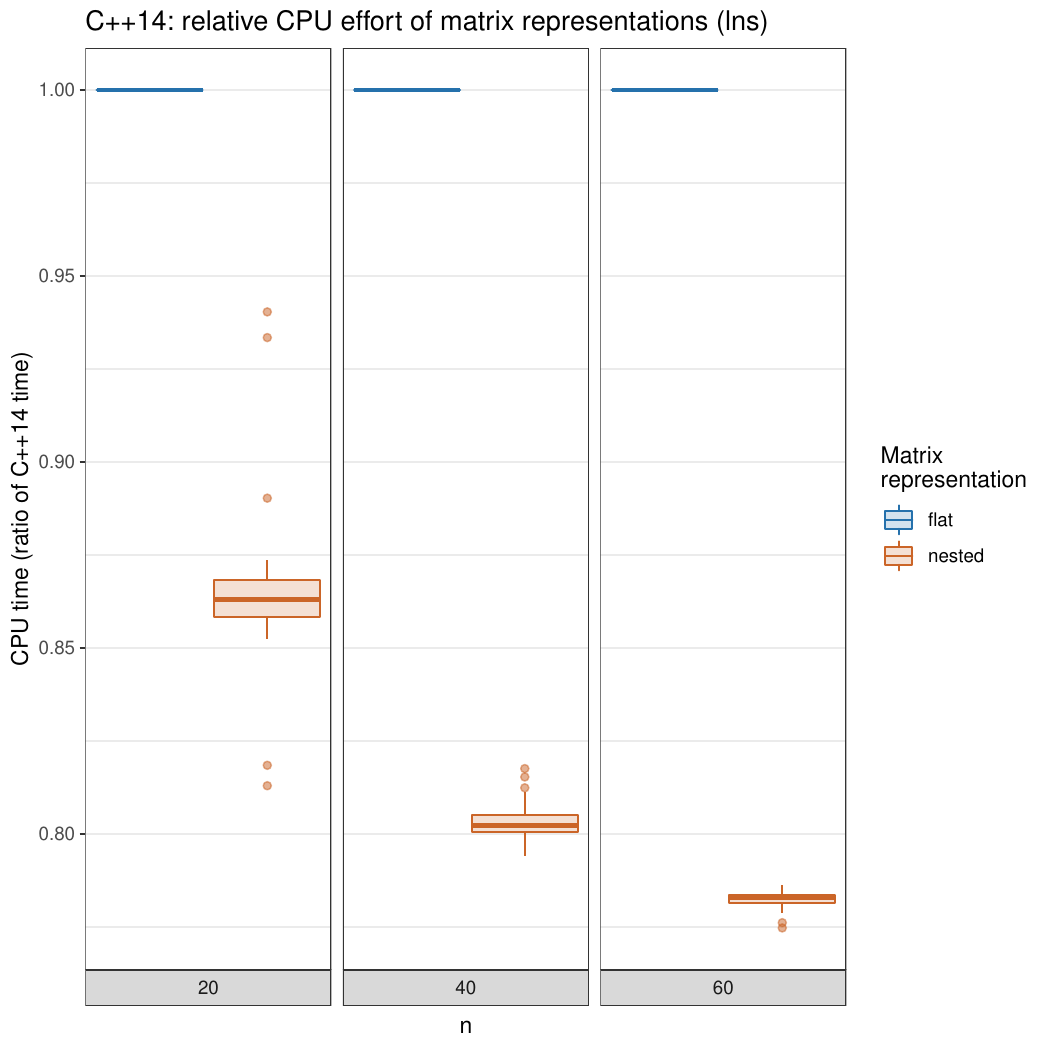}
    \includegraphics[scale=.22]{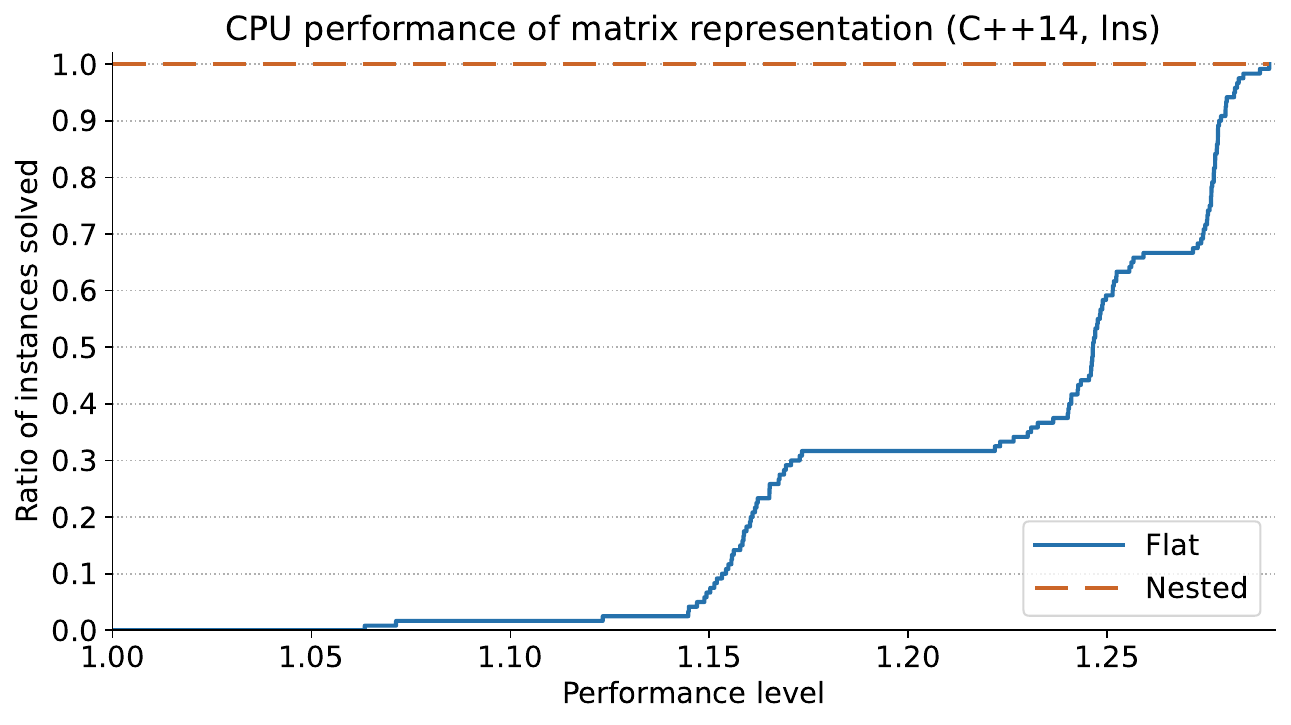}
    \caption{lns benchmark}
  \end{subfigure}
  \caption{C++14: comparison of matrix representations}
  \label{fig:c++14-matrix}
\end{figure}

We now look at the impact of using a flat distance matrix
representation, across multiple languages. The expectation is that it
should generally be at least as fast, as long as function calls are
inlined; however, in the case of languages
that have multi-dimensional arrays, it is harder to predict. We first
look at such a comparison for the c++14 
implementation in Figure~\ref{fig:c++14-matrix}, considering
benchmarks 2-opt and Or-Opt. Benchmarks maxflow and espprc use their
own matrix structure, therefore we do not include them in this comparison.
Although not depicted here, similar observations can be made with most
languages, with the notable exception of base Python (i.e. using
CPython without Numba). Since CPython does not do any kind of function
inlining, the performance cost of calling functions for each distance
lookup outweighs the performance benefit of using a flat, contiguous
matrix representation.

\begin{figure}
  \begin{subfigure}{0.33\textwidth}
    \centering
    \includegraphics[scale=.28]{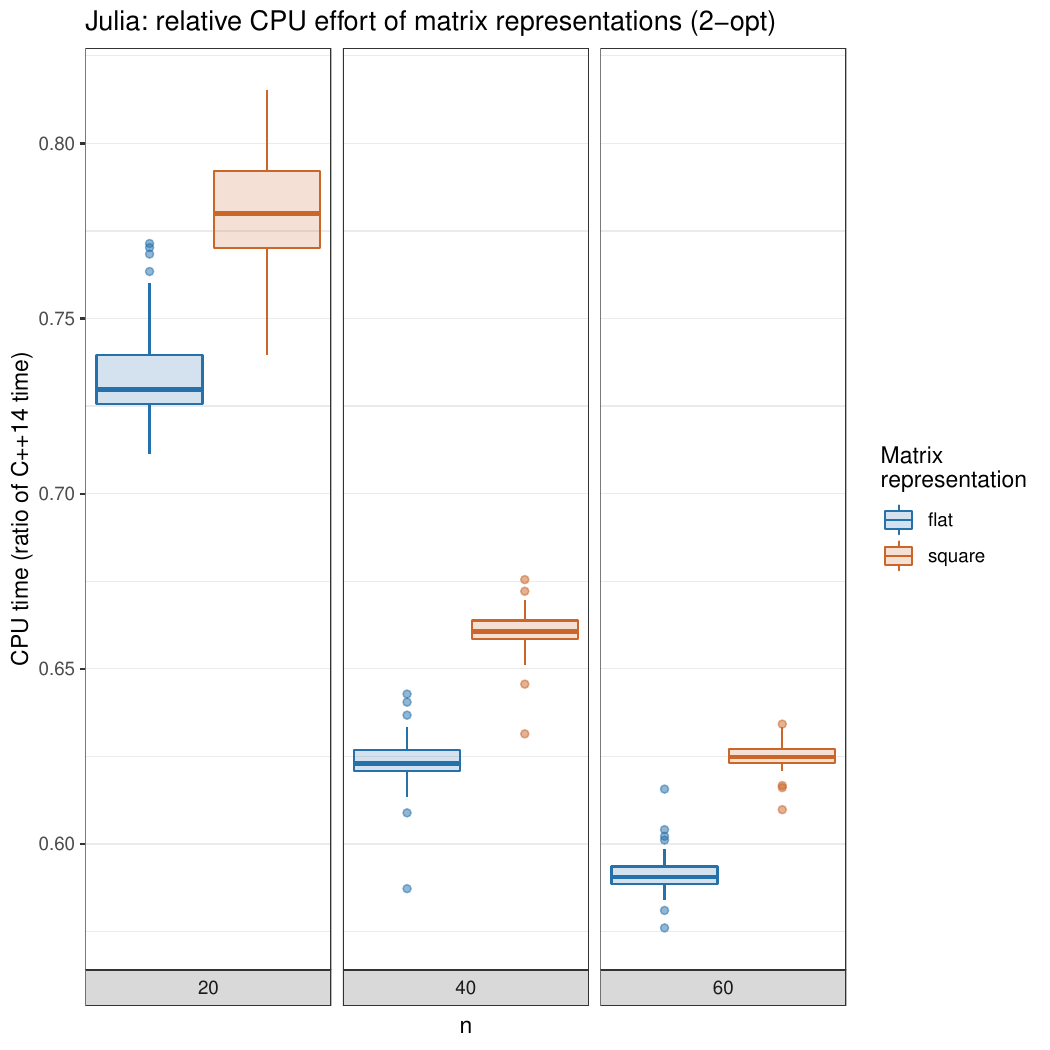}
    \includegraphics[scale=.22]{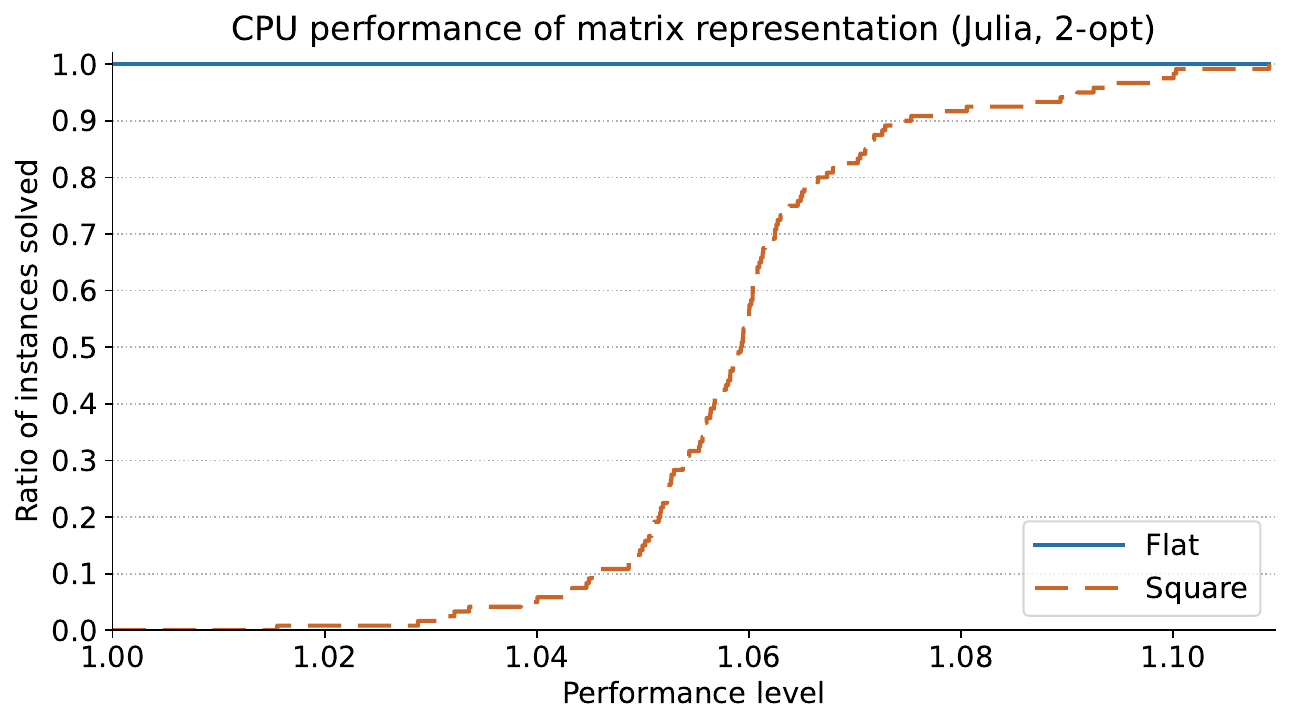}
    \caption{2-opt benchmark}
  \end{subfigure}
  \begin{subfigure}{0.33\textwidth}
    \centering
    \includegraphics[scale=.28]{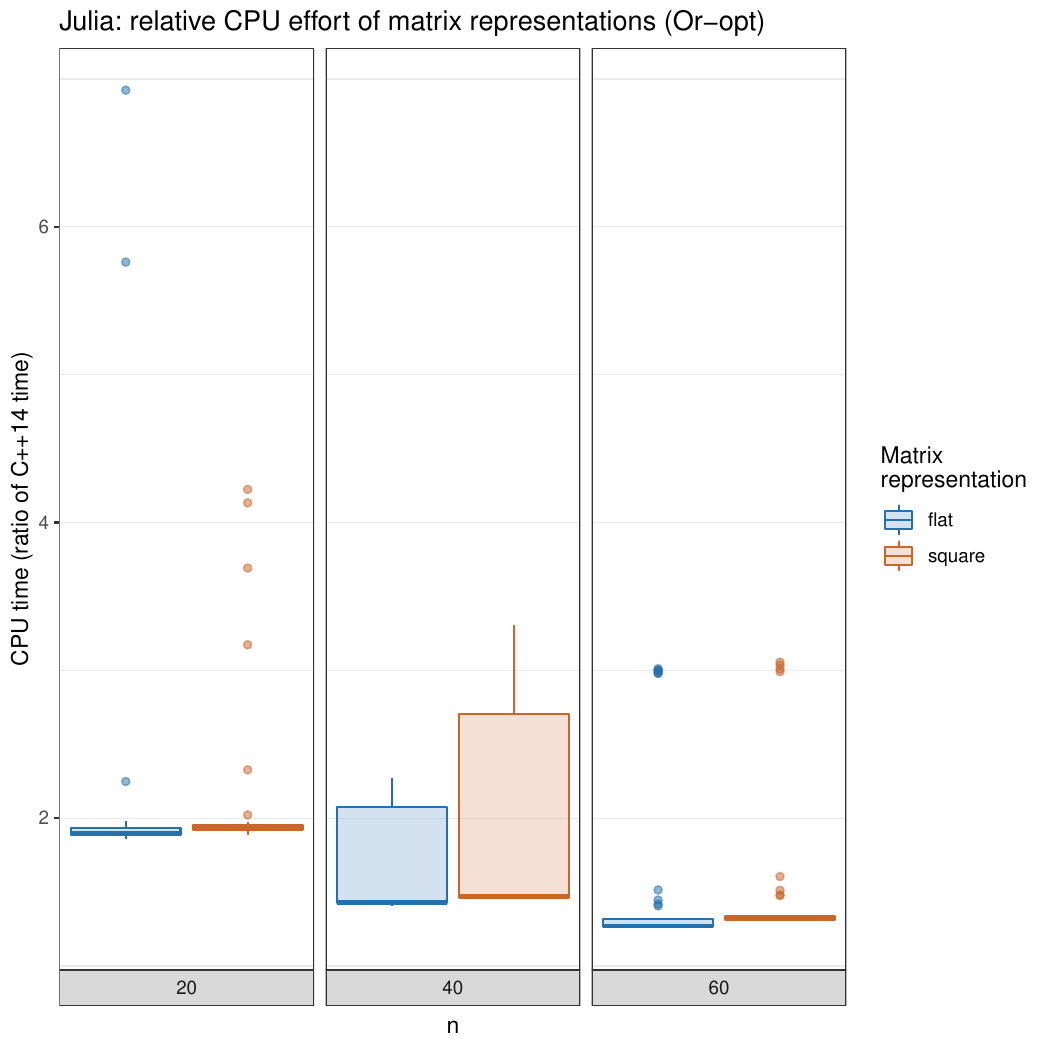}
    \includegraphics[scale=.22]{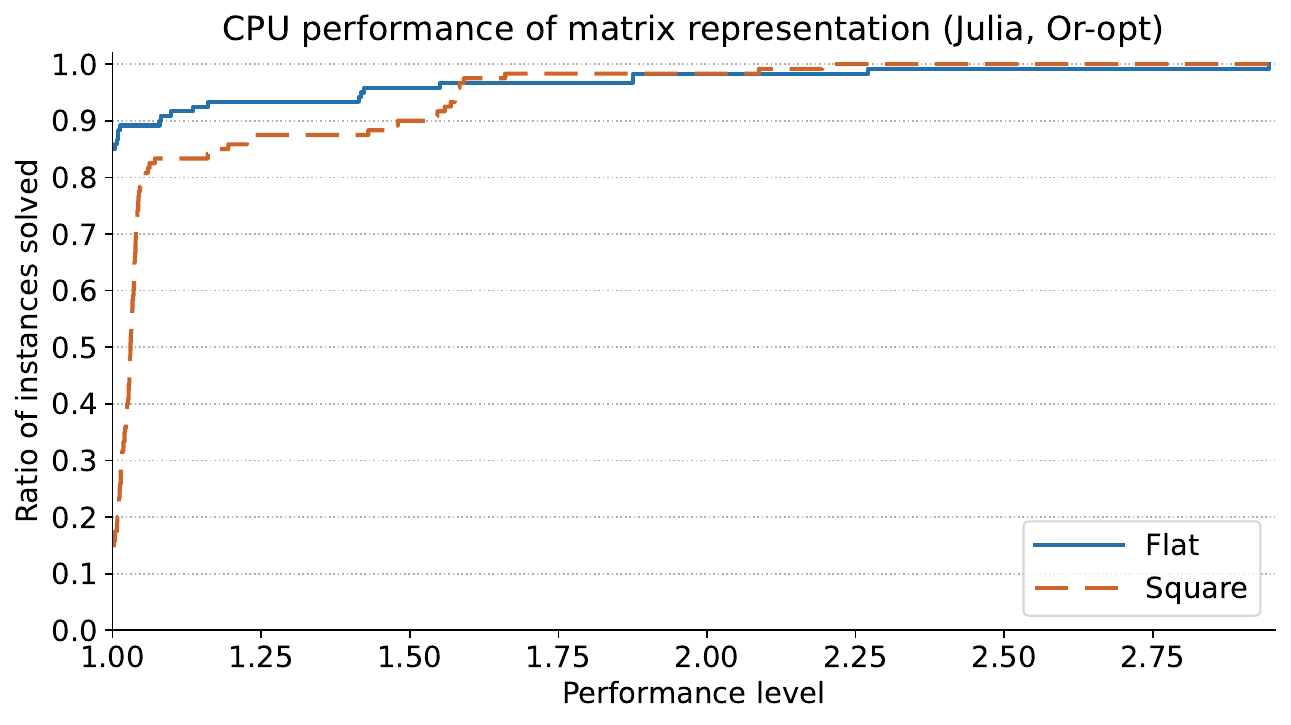}
    \caption{Or-opt benchmark}
  \end{subfigure}
  \begin{subfigure}{0.33\textwidth}
    \centering
    \includegraphics[scale=.28]{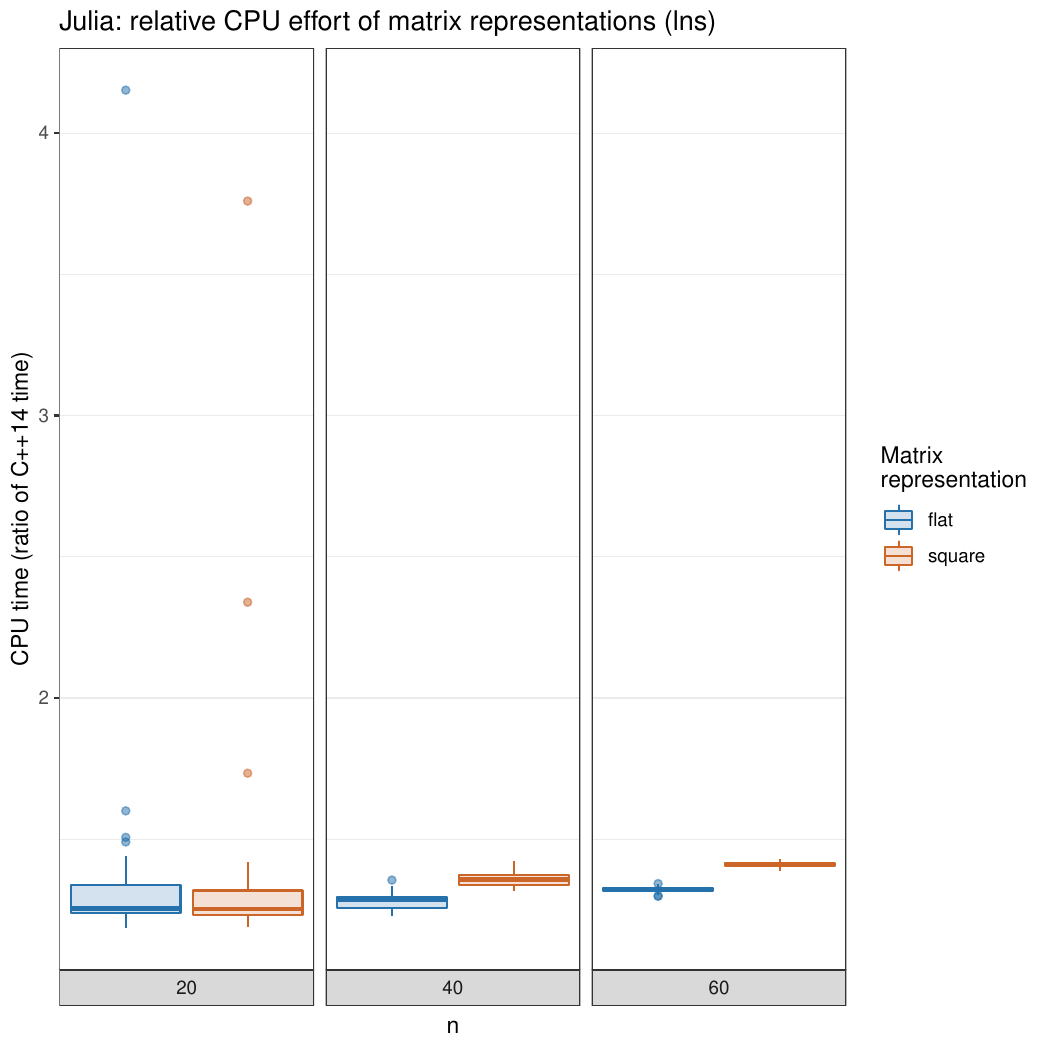}
    \includegraphics[scale=.22]{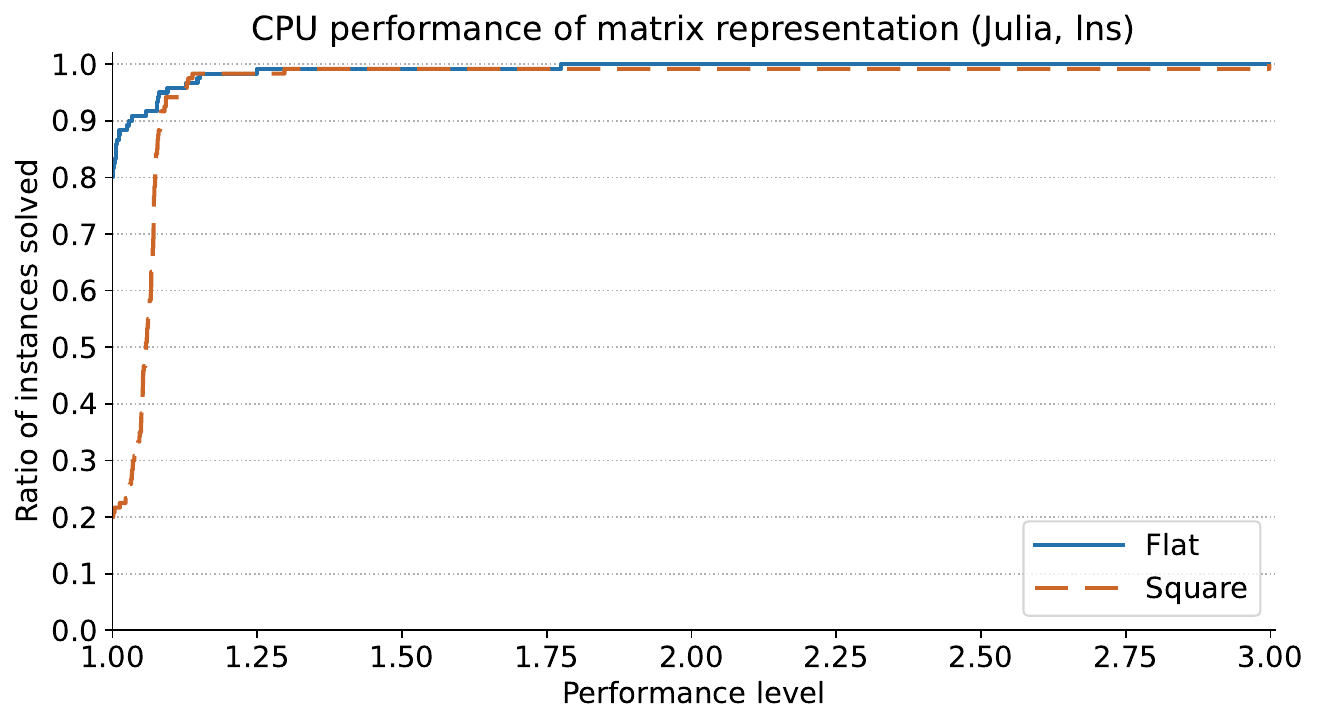}
    \caption{lns benchmark}
  \end{subfigure}
  \caption{Julia: comparison of matrix representations}
  \label{fig:julia-matrix}
\end{figure}

We now compare a flat matrix representation in Julia against the
``natural'' representation using Julia's native multi-dimensional
arrays. This comparison is presented in Figure~\ref{fig:julia-matrix}.
While it is not overwhelming, there can still be a benefit in using a
flat matrix representation.

Overall, it appears that it is never a significantly bad idea to use a flat
matrix representation in our setup, while it can bring
benefits. Therefore, from this point on, flat matrix representation is
used by default.

\subsection{Impact of using static arrays}
\begin{figure}
  \begin{subfigure}{0.5\textwidth}
    \centering
    \includegraphics[scale=.4]{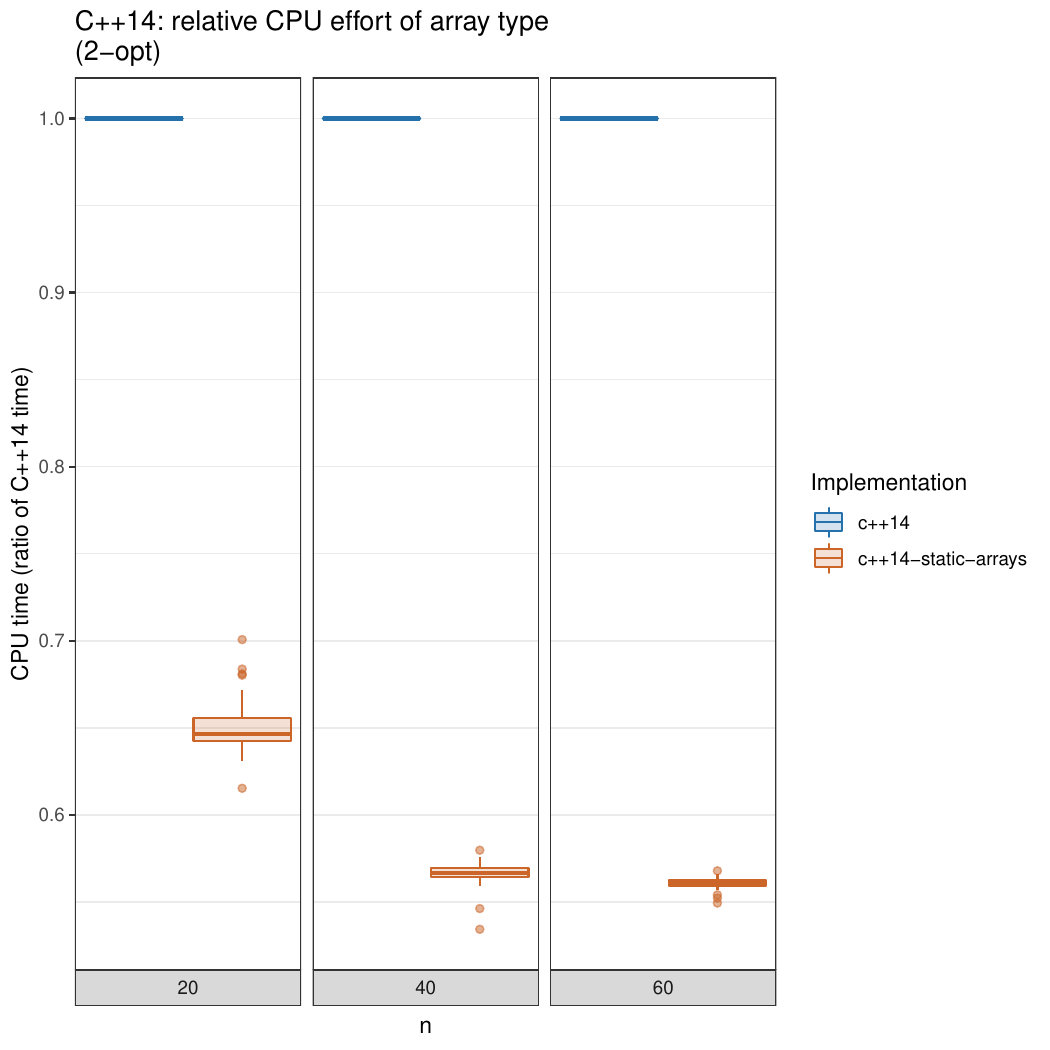}
    \includegraphics[scale=.3]{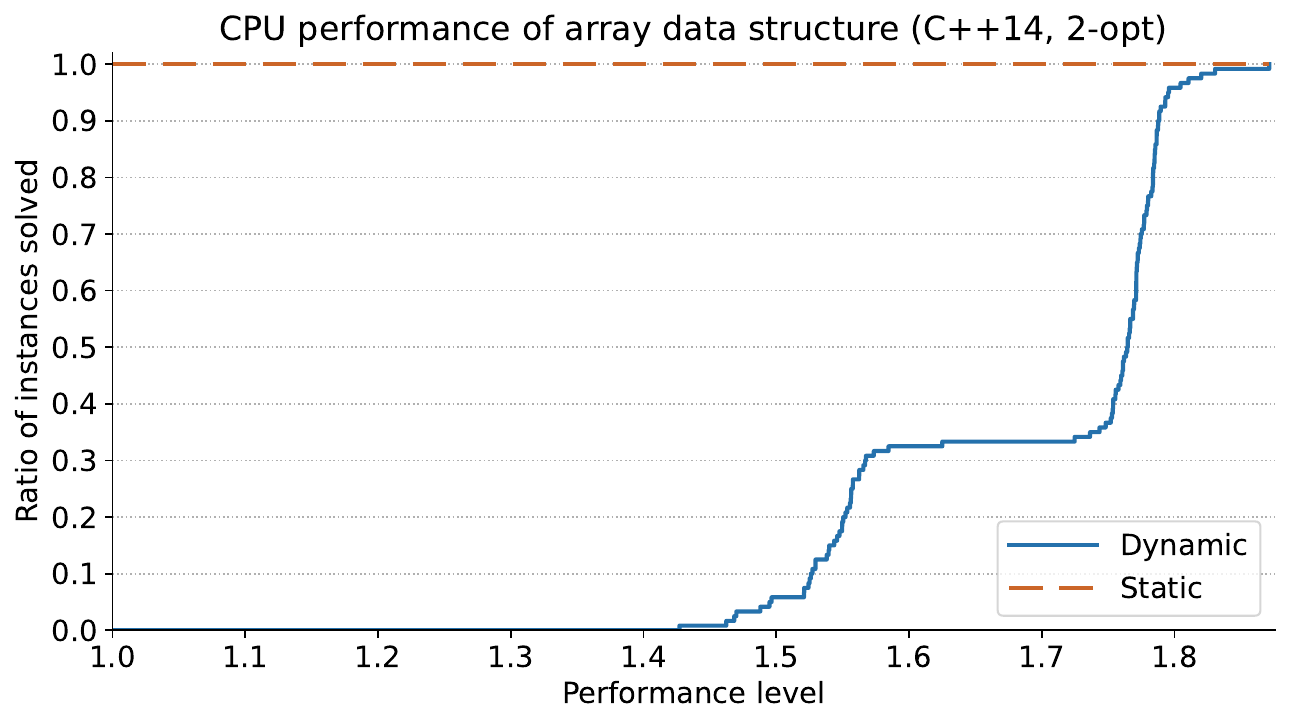}
    \caption{C++14, 2-opt benchmark}
  \end{subfigure}
  \begin{subfigure}{0.5\textwidth}
    \centering
    \includegraphics[scale=.4]{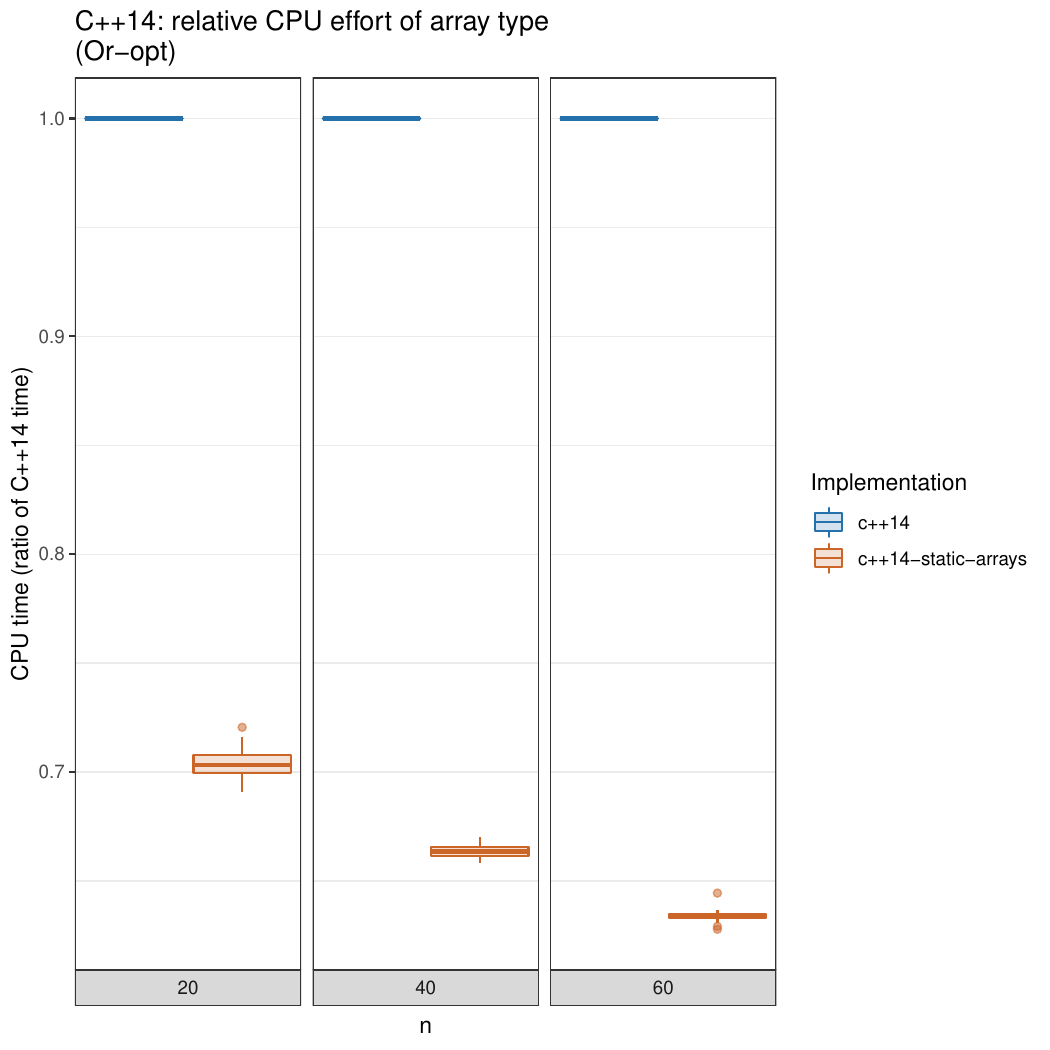}
    \includegraphics[scale=.3]{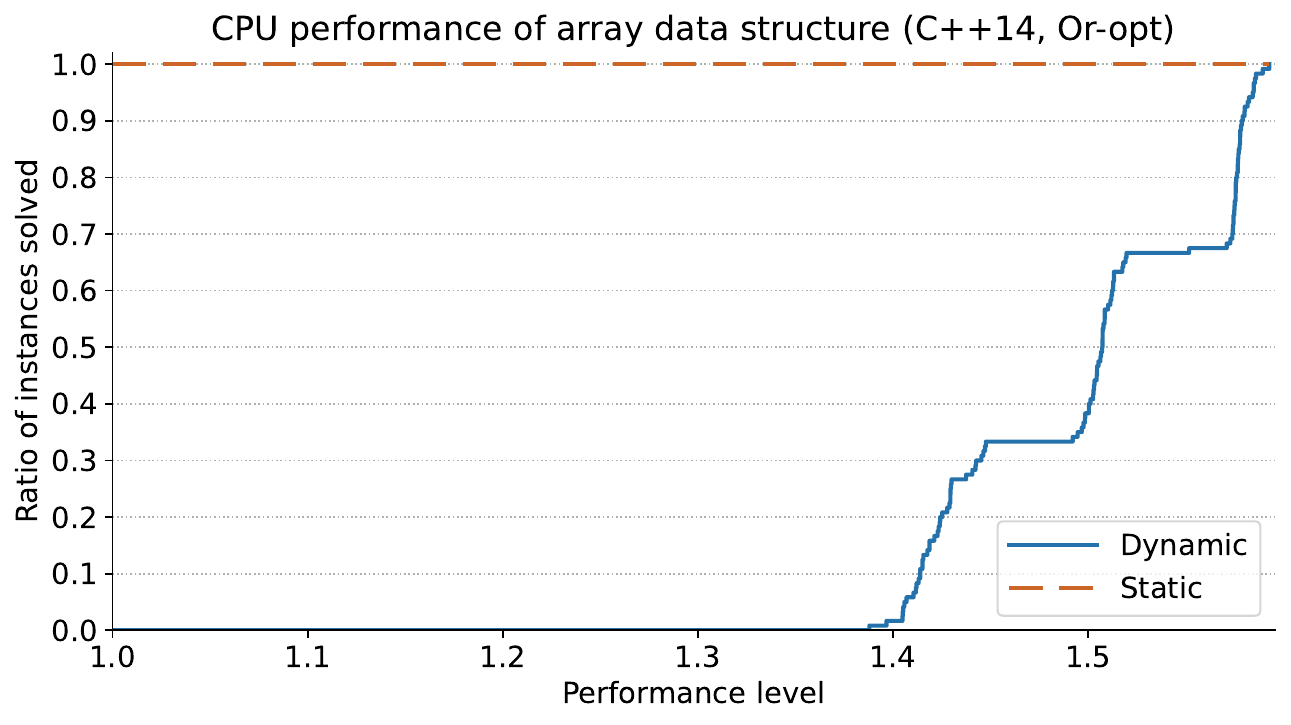}
    \caption{C++14, Or-opt benchmark}
  \end{subfigure}
  \begin{subfigure}{0.5\textwidth}
    \centering
    \includegraphics[scale=.4]{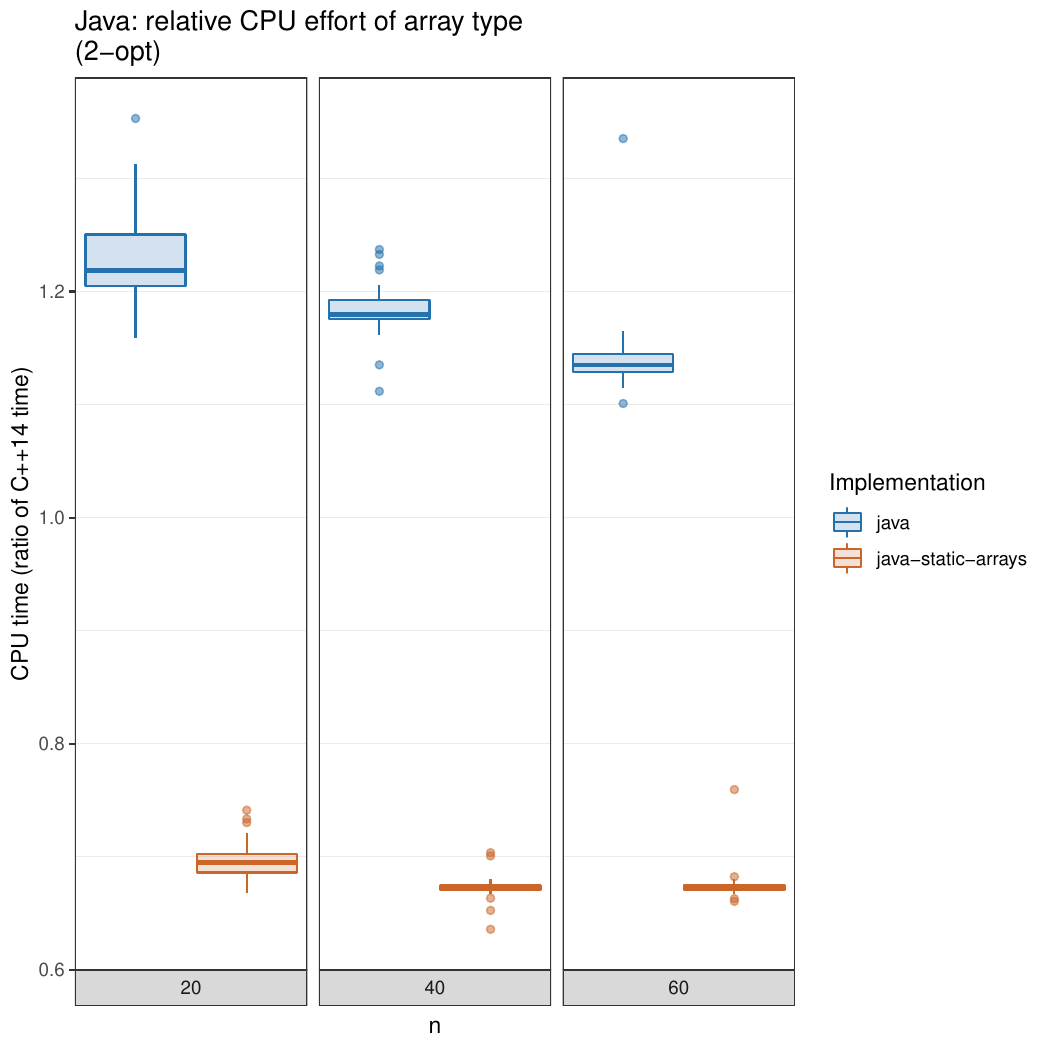}
    \includegraphics[scale=.3]{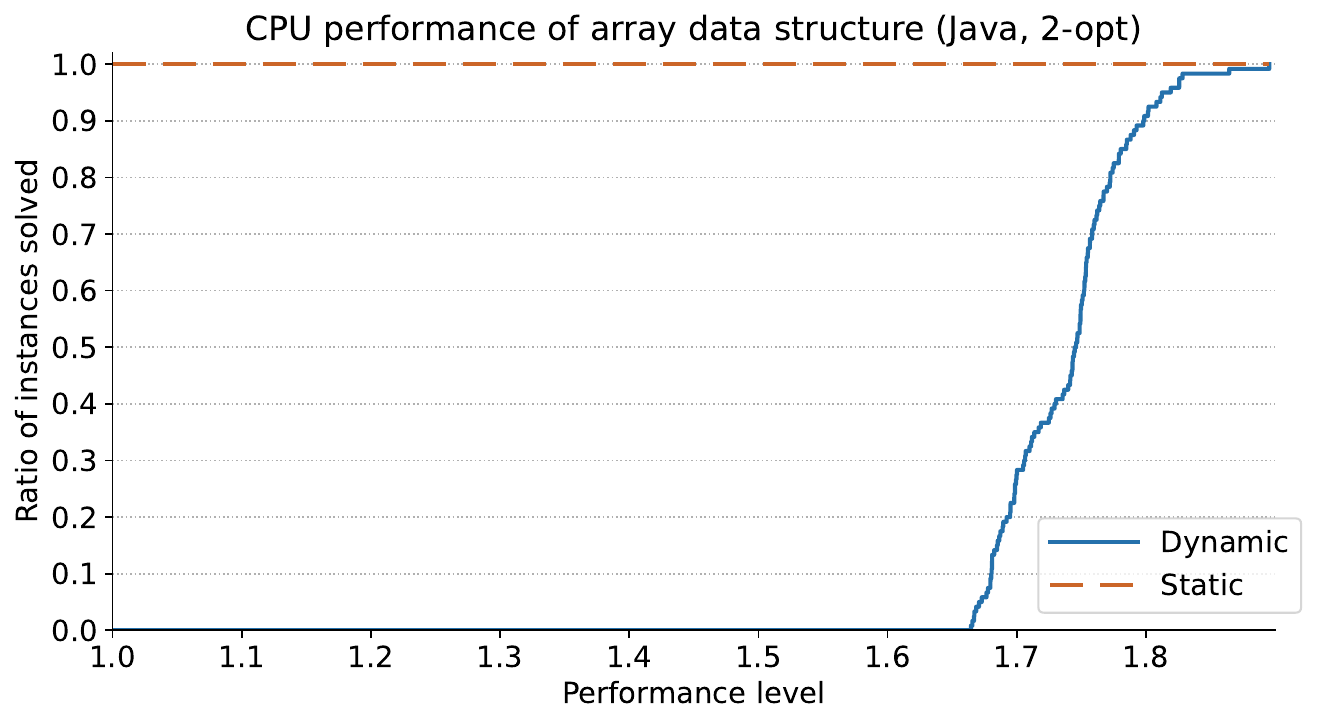}
    \caption{Java, 2-opt benchmark}
  \end{subfigure}
  \begin{subfigure}{0.5\textwidth}
    \centering
    \includegraphics[scale=.4]{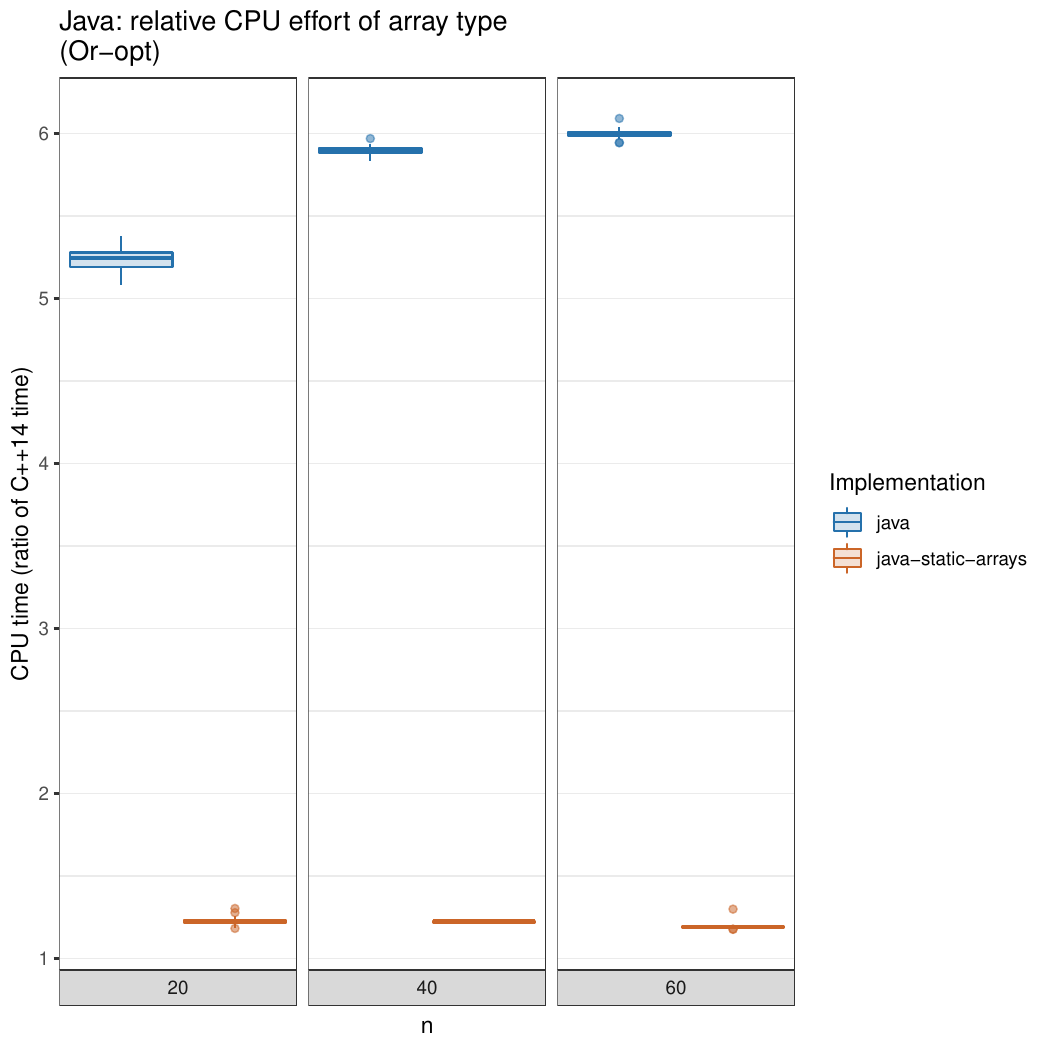}
    \includegraphics[scale=.3]{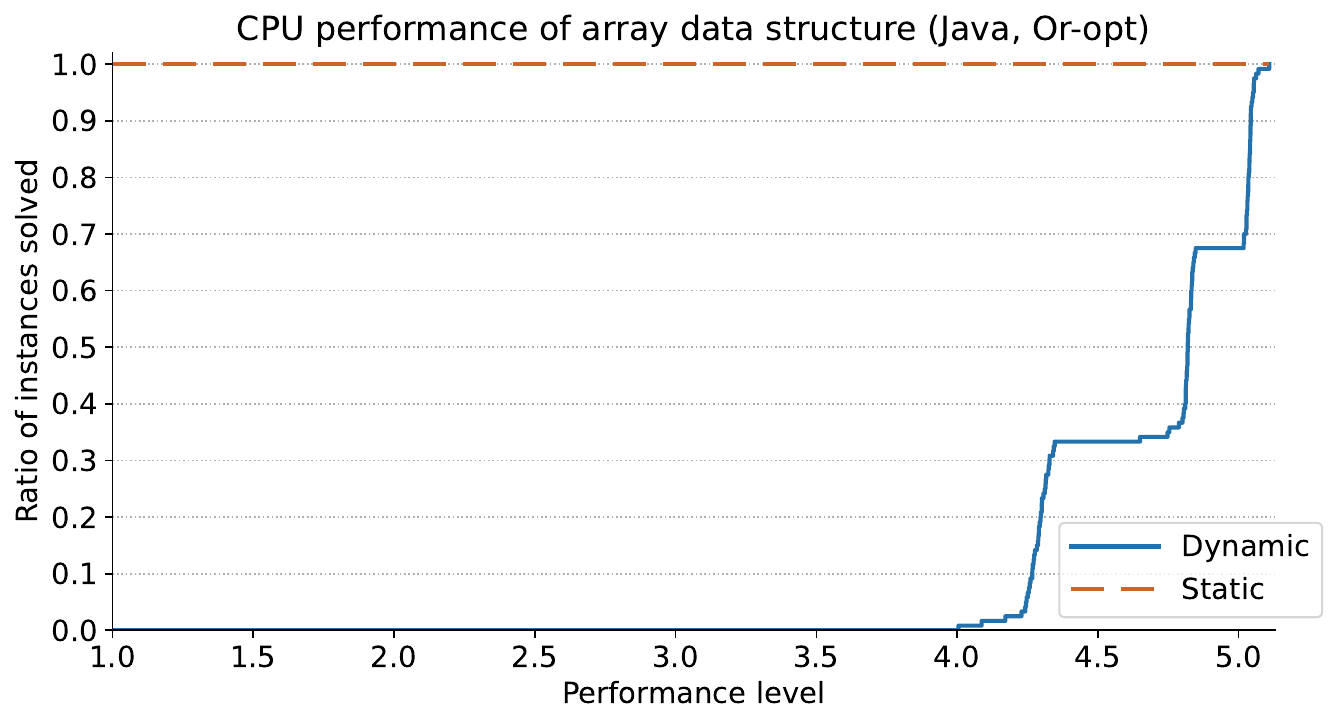}
    \caption{Java, Or-opt benchmark}
  \end{subfigure}
  \caption{Performance of static arrays versus variable-size vectors.}
  \label{fig:static-arrays}
\end{figure}

In some cases, variable-size arrays are not necessary. This is the
case for example when implementing a 2-opt procedure for a TSP. In
these cases, fixed-size arrays are enough and their performance have
to be at least as good as the performance of variable-size arrays
(like C++'s vectors). We run this comparison for benchmarks 2-opt and
Or-opt, which can be both implemented easily using the fixed-size native
arrays of C++, which are C native arrays, and which work in a similar
fashion in Java. We run this comparison for both C++14 and Java, the
results are summarized in Figure~\ref{fig:static-arrays}.
Using static arrays in C++ can be up to almost twice as fast, while in
Java it can be up to five times as fast. One possible explanation why
the gap is larger in Java lies in the nature of the Java containers
used: we use \verb|ArrayList| objects, which take as generic parameter
the class of the objects stored in the collection,
e.g. \verb|ArrayList<Integer>|. This parameter can only be a class,
not a native type, which matters in Java. This means that it is not
possible to have a collection of \verb|int| variables, only a
collection of \verb|Integer| objects, each wrapping an
\verb|int|.

\subsection{A comparison of C++ implementations}
\begin{figure}
  \begin{subfigure}{0.33\textwidth}
    \centering
    \includegraphics[scale=.28]{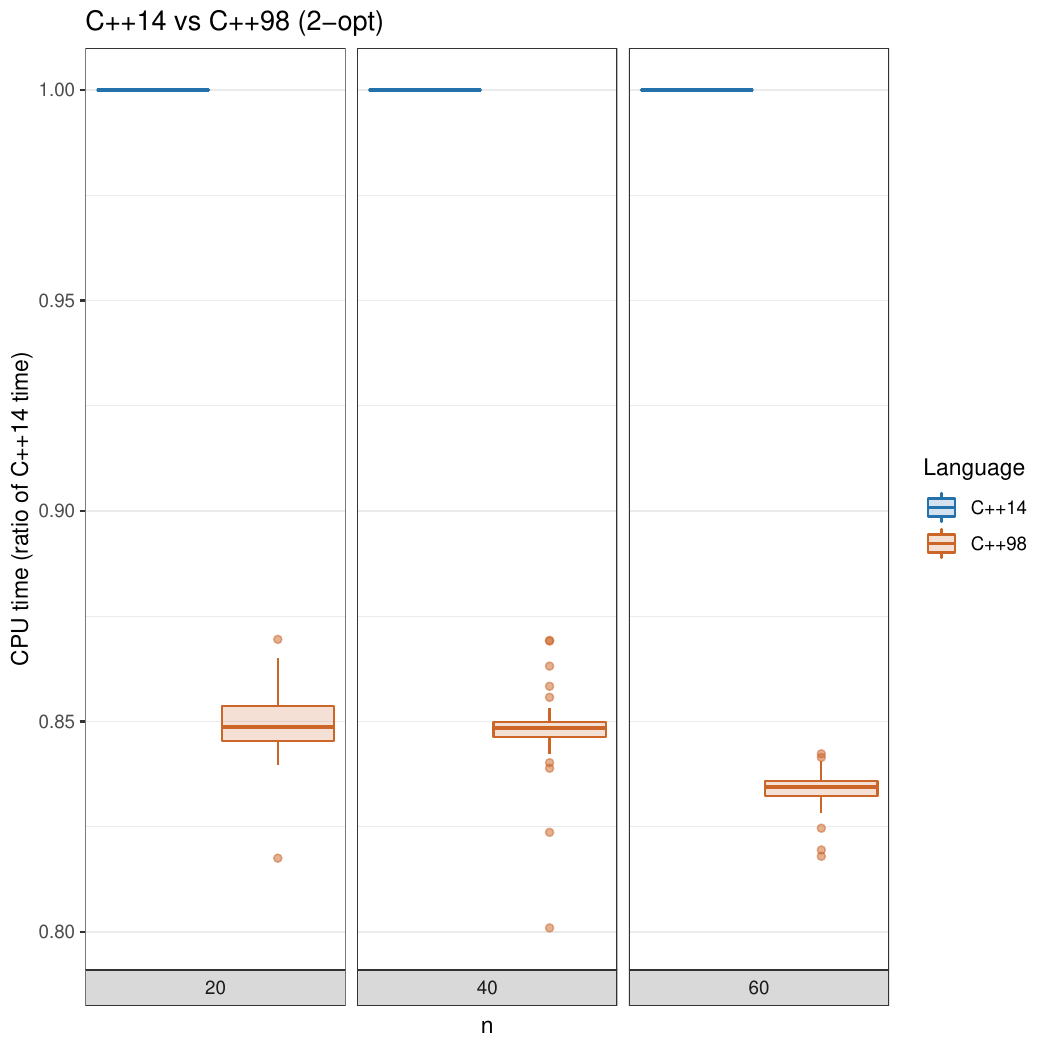}
    \includegraphics[scale=.22]{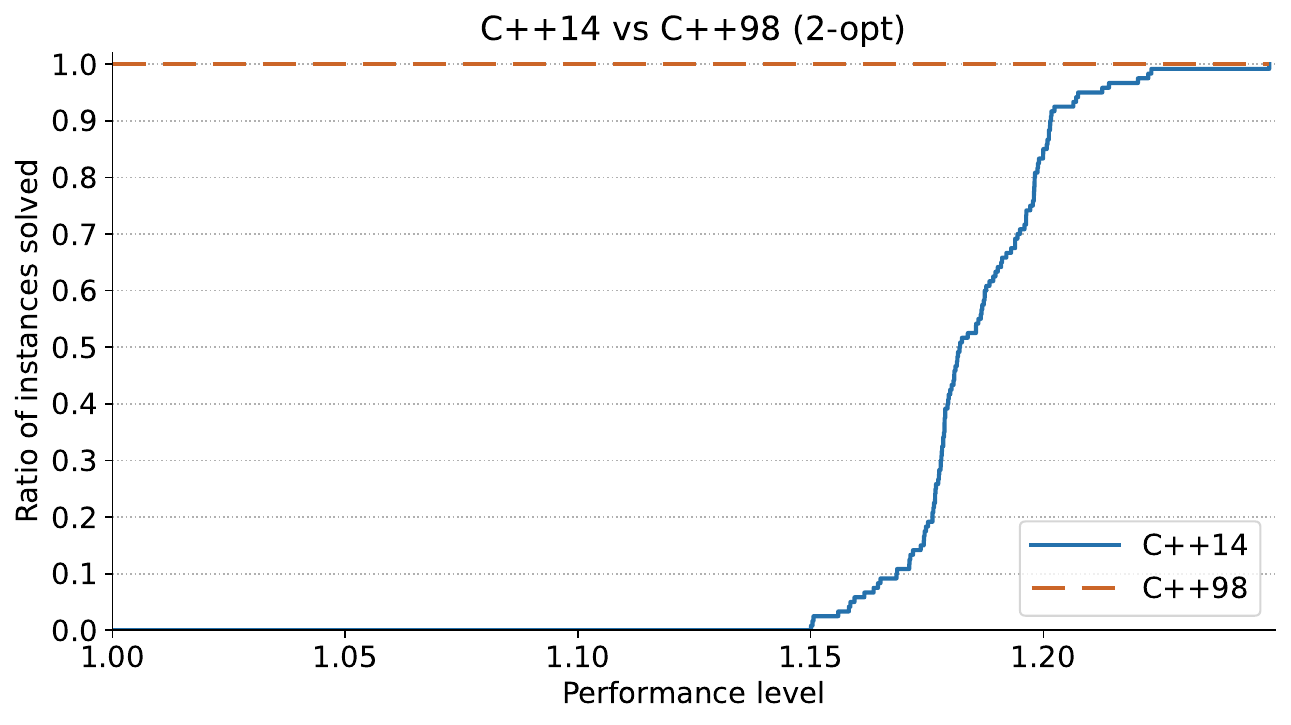}
    \caption{2-opt benchmark}
  \end{subfigure}
  \begin{subfigure}{0.33\textwidth}
    \centering
    \includegraphics[scale=.28]{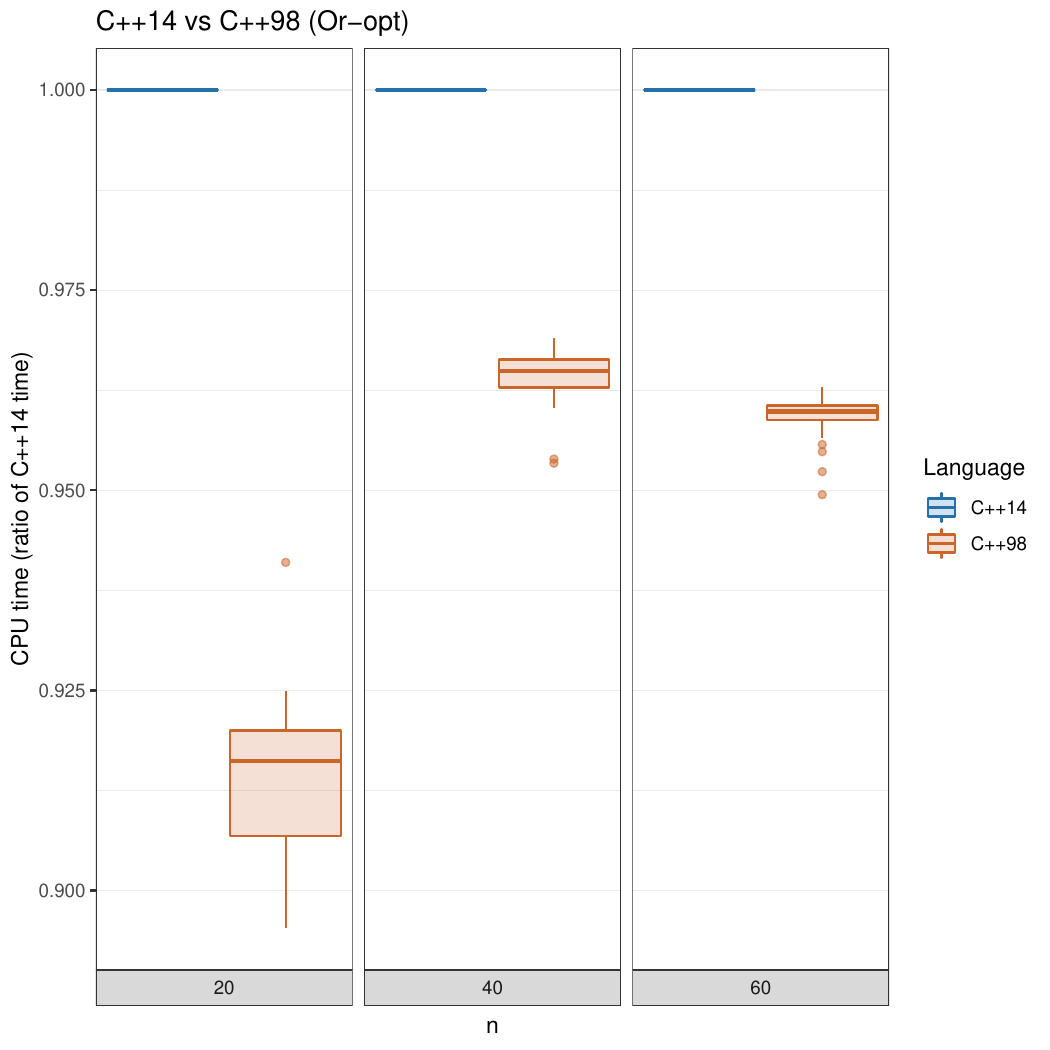}
    \includegraphics[scale=.22]{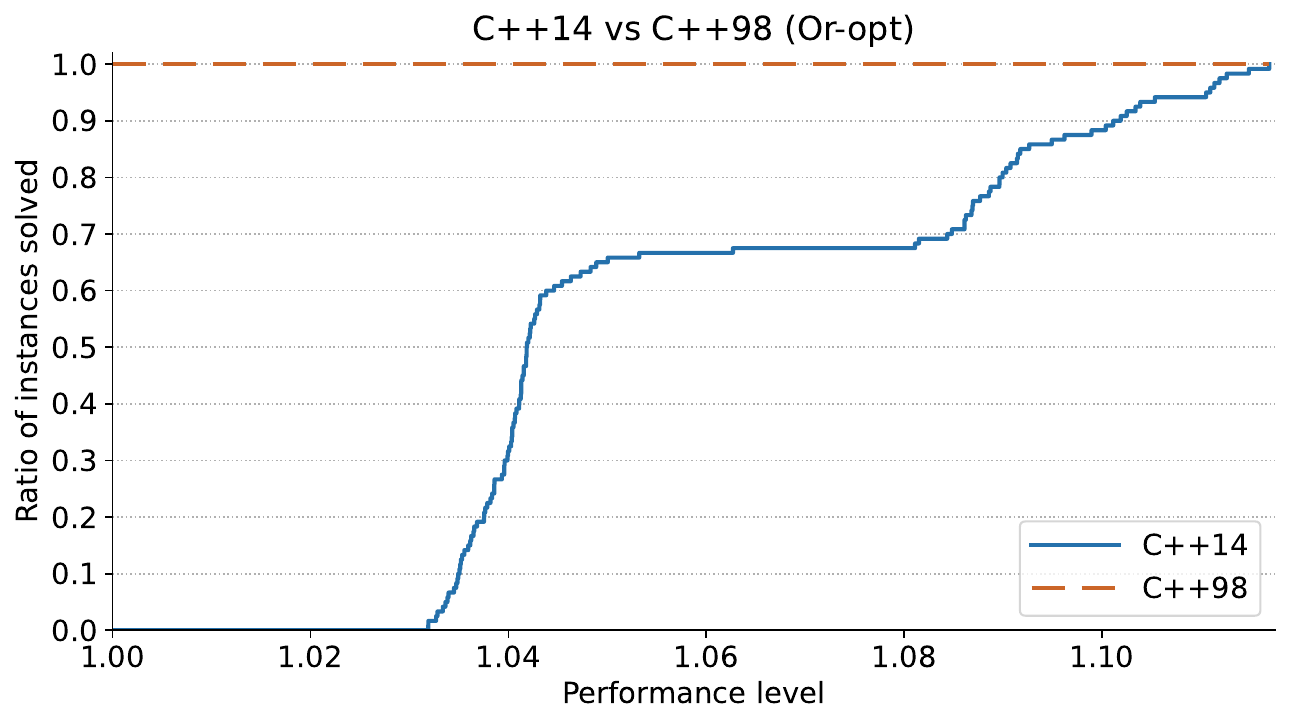}
    \caption{Or-opt benchmark}
  \end{subfigure}
  \begin{subfigure}{0.33\textwidth}
    \centering
    \includegraphics[scale=.28]{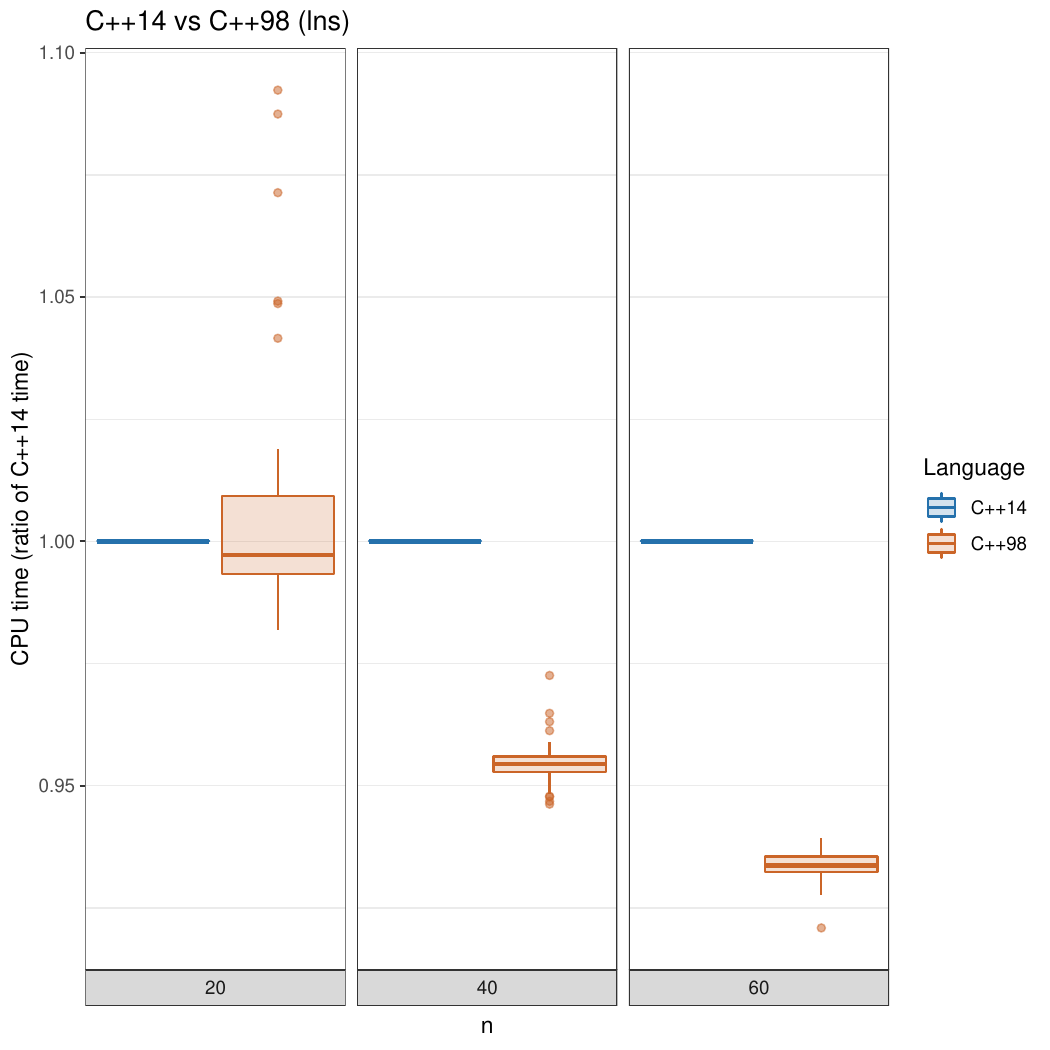}
    \includegraphics[scale=.22]{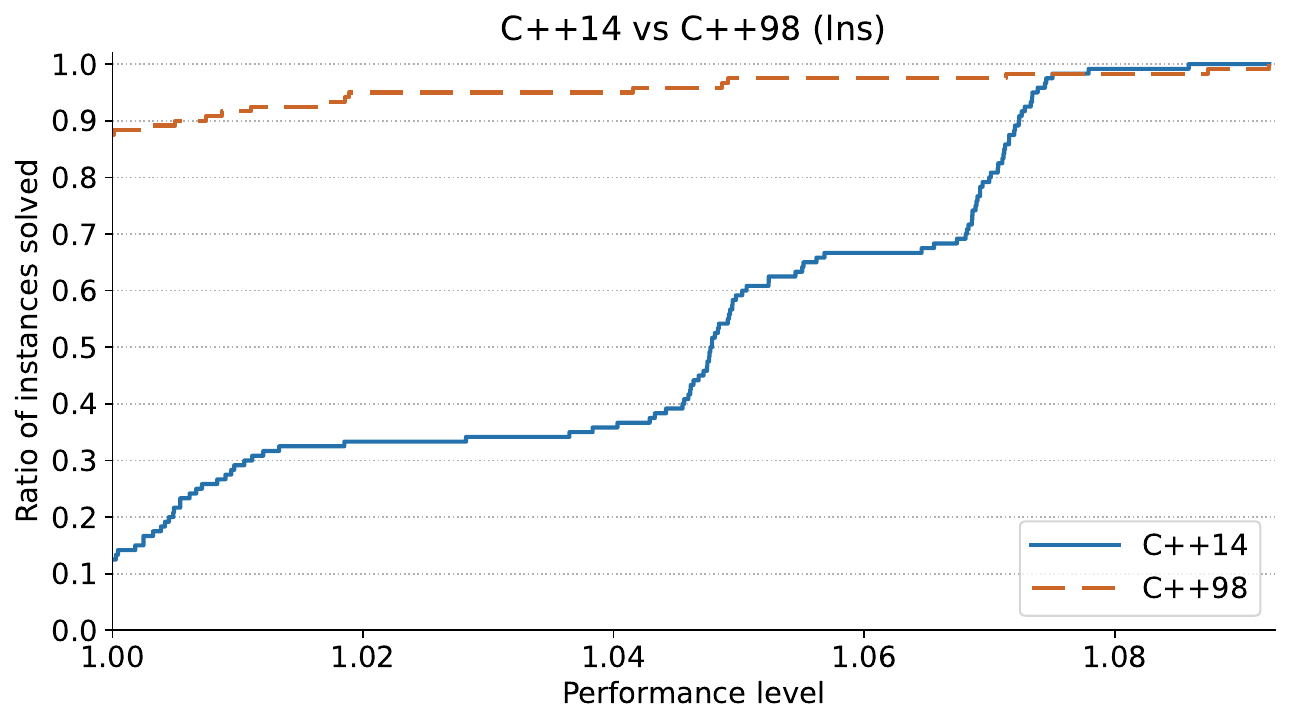}
    \caption{lns benchmark}
  \end{subfigure}
  \begin{subfigure}{0.33\textwidth}
    \centering
    \includegraphics[scale=.28]{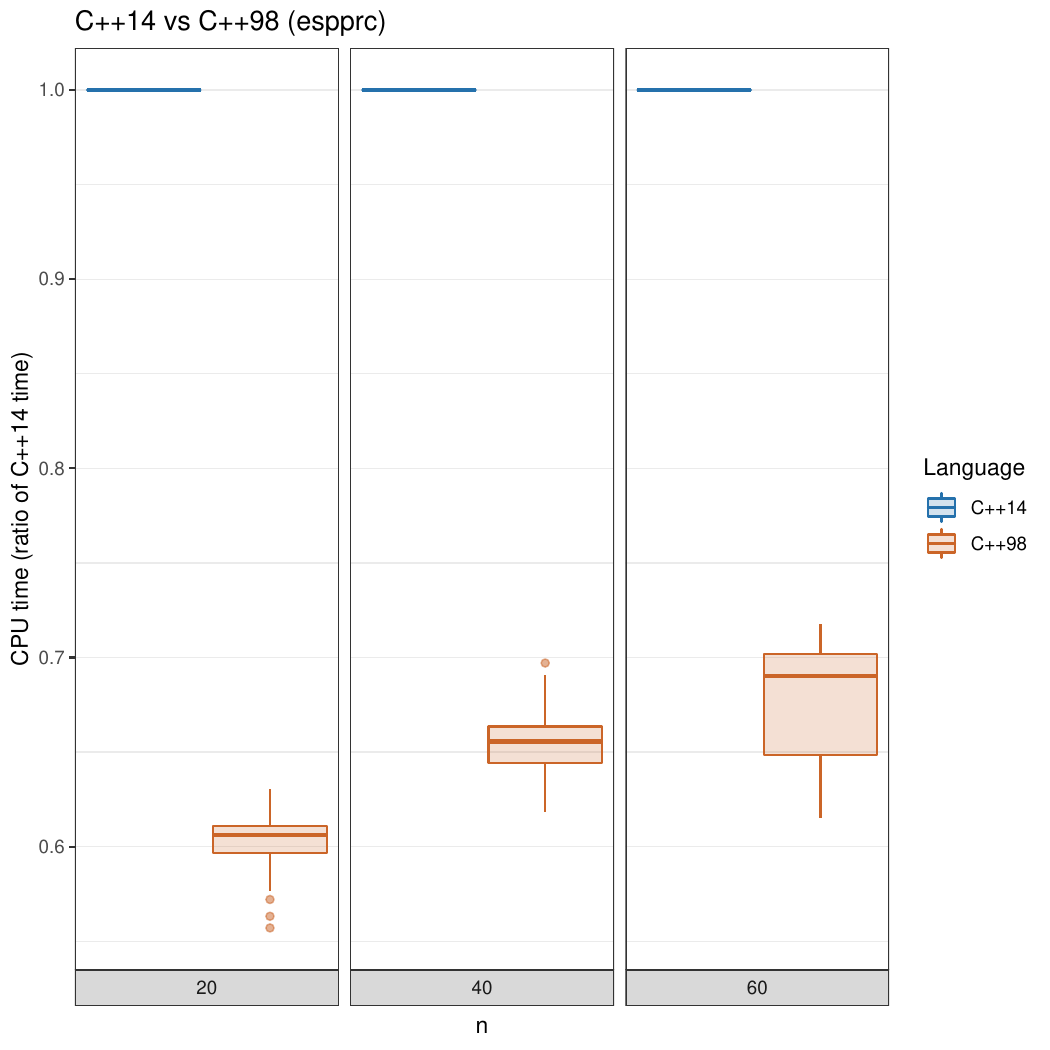}
    \includegraphics[scale=.22]{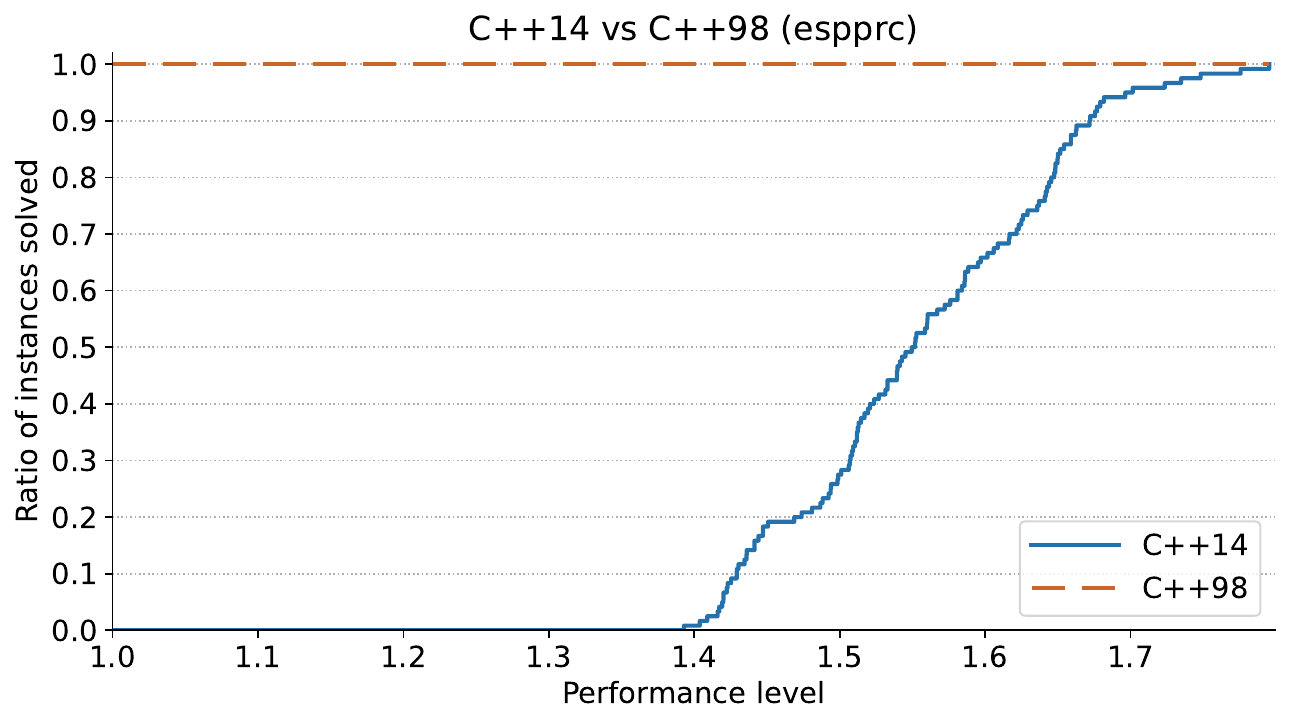}
    \caption{espprc benchmark}
  \end{subfigure}
  \begin{subfigure}{0.33\textwidth}
    \centering
    \includegraphics[scale=.28]{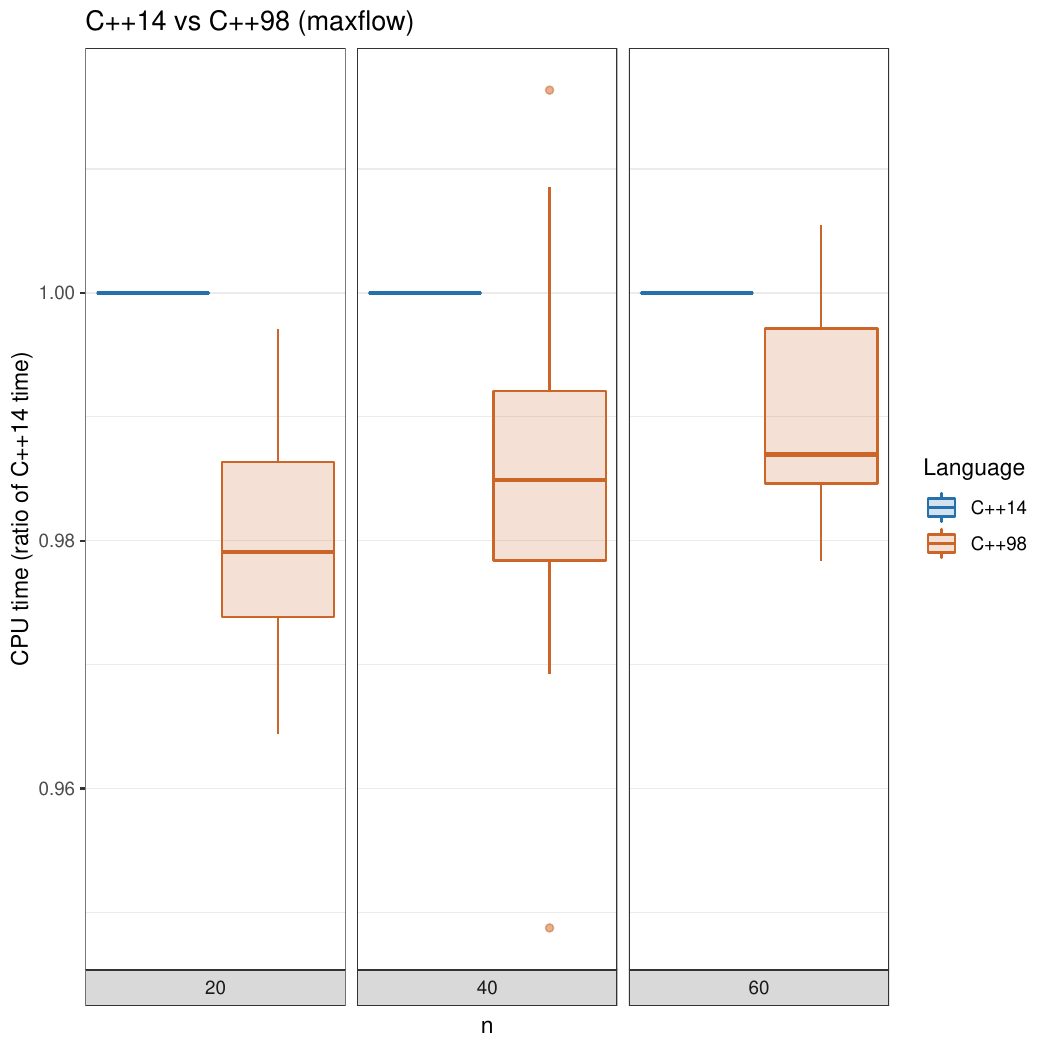}
    \includegraphics[scale=.22]{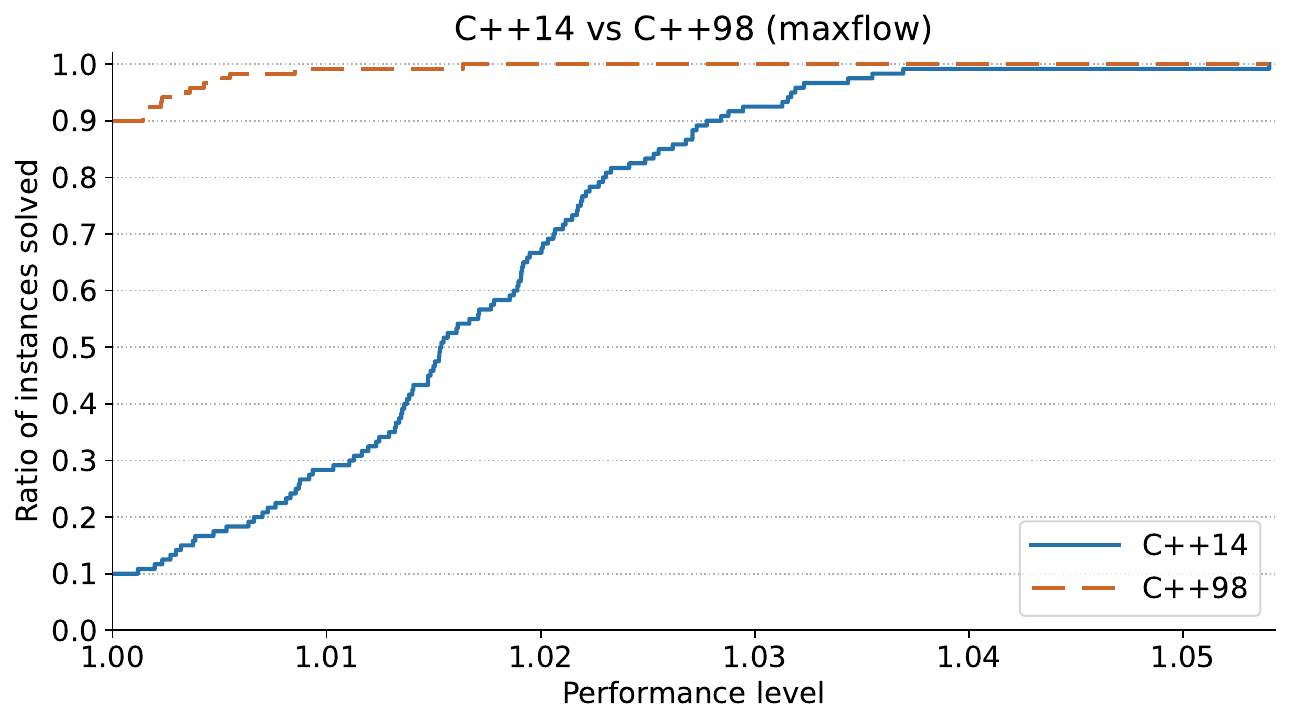}
    \caption{maxflow benchmark}
  \end{subfigure}
  \caption{Performance of C++14 versus C++98.}
  \label{fig:c++-implementations}
\end{figure}

We investigate the difference in performance between using C++14's
smart pointers and older traditional C++ coding with
references and/or C pointers. For that purpose, we compare the C++14
and C++98 implementations of each
benchmark. Figure~\ref{fig:c++-implementations} summarizes this comparison.
The C++98 implementation is faster in all five benchmarks, although not always
by much. We conclude that if something can be implemented in our setup
without using smart pointers and without extra effort, then it is a good idea
to do so.

\subsection{General cross-language comparison}
We now compare the performance of various programming
languages when performing the same tasks. We compare the following
languages: C++, Java, JavaScript, Julia, 
Python and Rust. For that purpose, we use what seems to be the best
choice for each language based on the above experiments, meaning that
we use the C++98 implementation for C++, and the Pypy
interpreter when running a Python program. There is a natural
separation between compiled ``fast'' languages (C++, Java, Julia,
Rust) on one side and ``interpreted'' languages (Python, JavaScript)
on the other side. No language is exclusively interpreted, as Pypy and
Node use JIT compiling; this being said, there is a performance gap
between fast and interpreted languages, and in order to produce more
readable charts we compare them separately.

\subsubsection{Cross-language comparison: ``fast'' languages}
\begin{figure}
  \begin{subfigure}{0.33\textwidth}
    \centering
    \includegraphics[scale=.28]{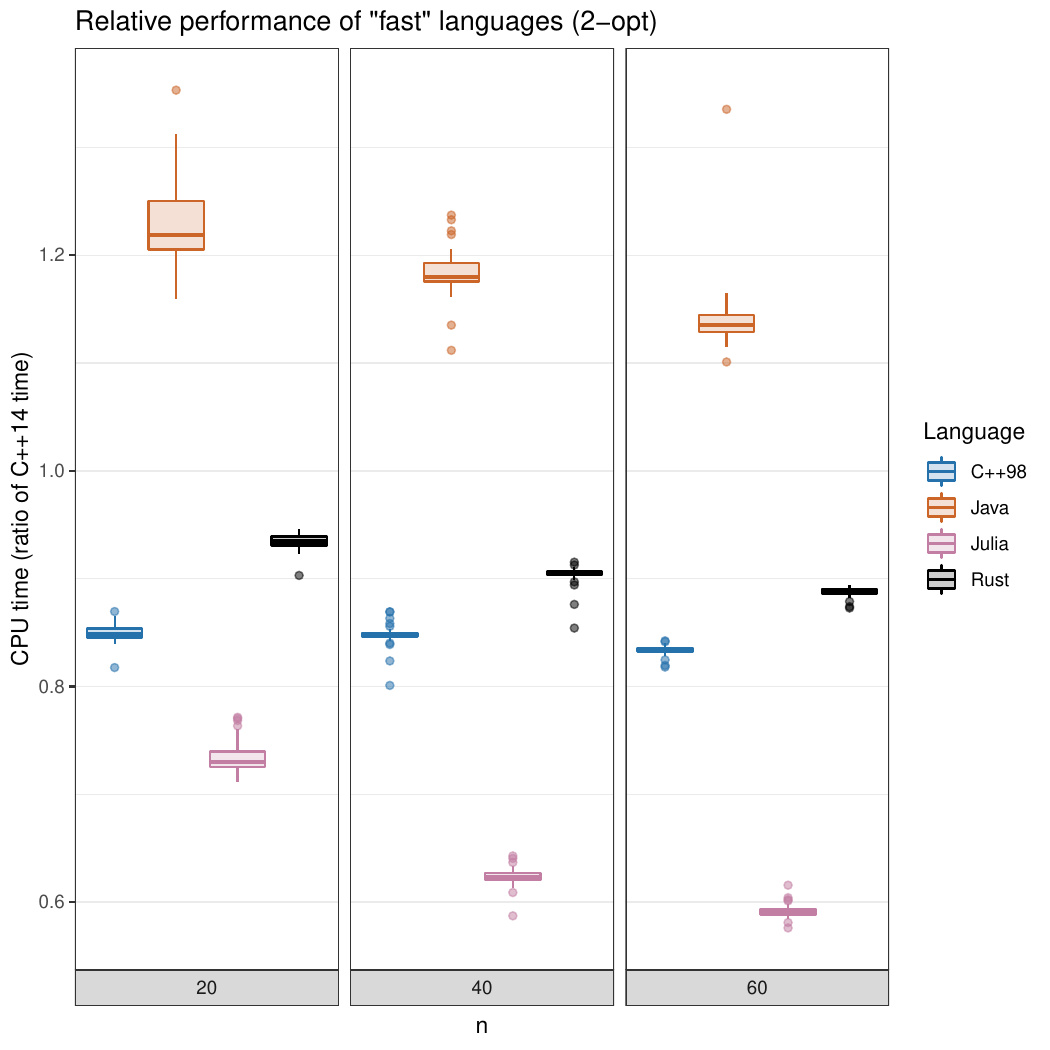}
    \includegraphics[scale=.22]{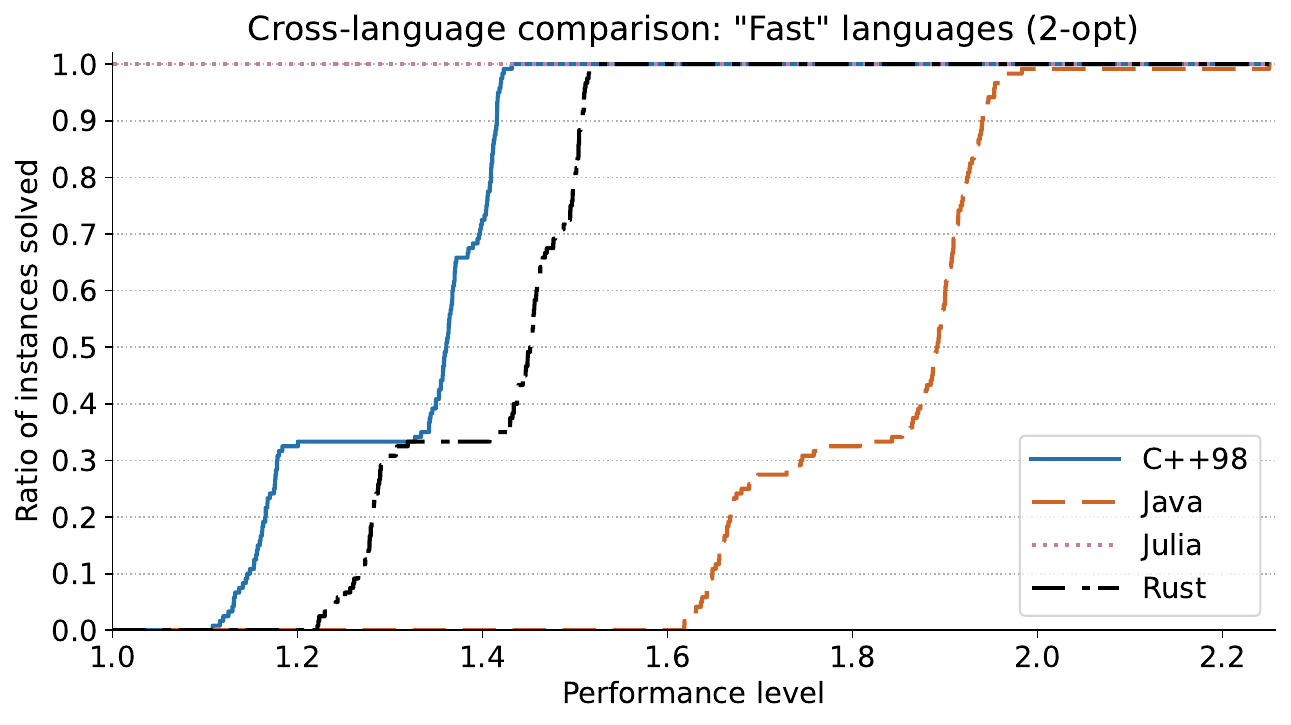}
    \caption{2-opt benchmark}
  \end{subfigure}
  \begin{subfigure}{0.33\textwidth}
    \centering
    \includegraphics[scale=.28]{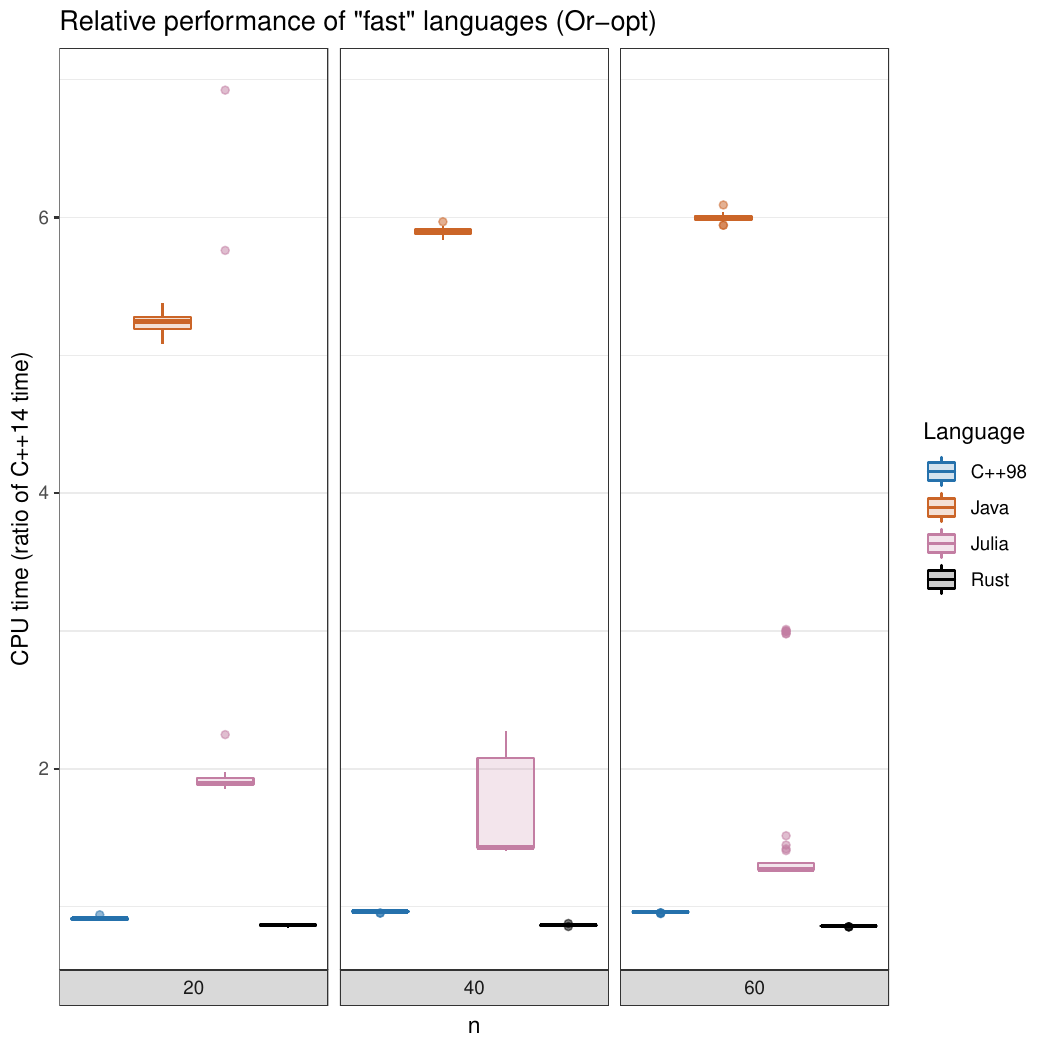}
    \includegraphics[scale=.22]{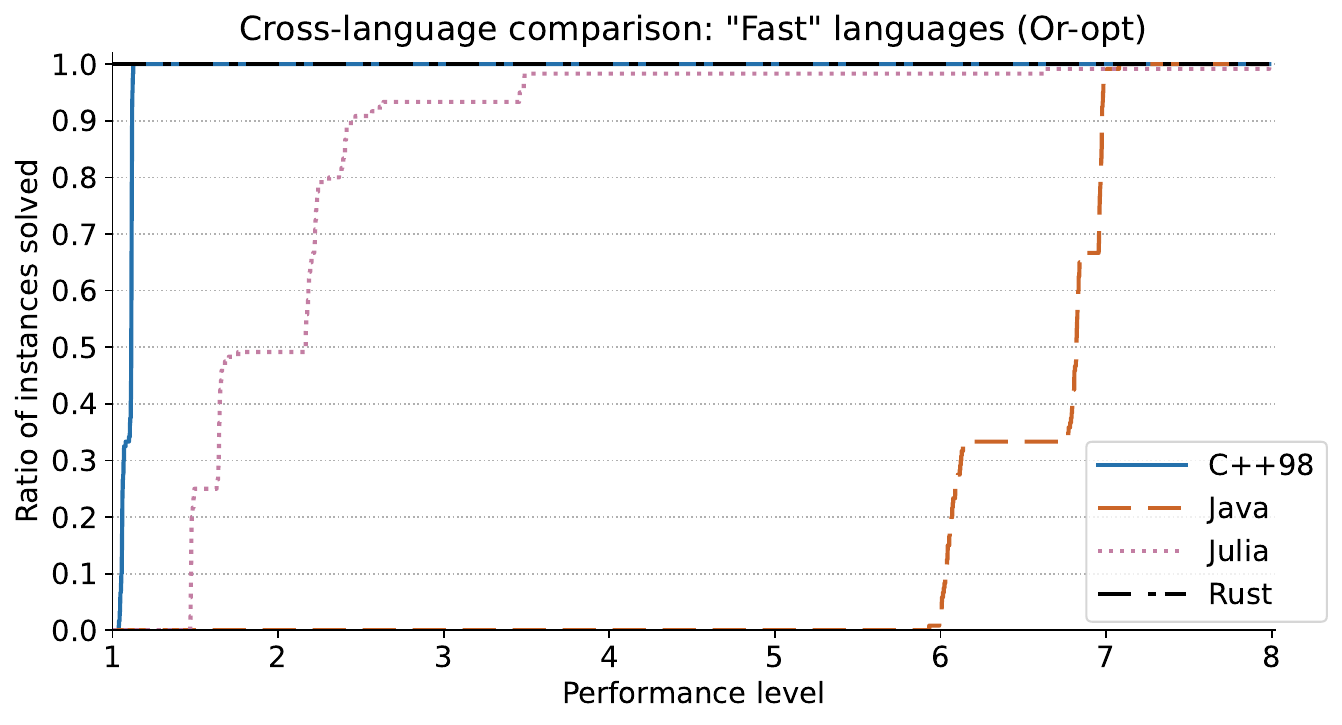}
    \caption{Or-opt benchmark}
  \end{subfigure}
  \begin{subfigure}{0.33\textwidth}
    \centering
    \includegraphics[scale=.28]{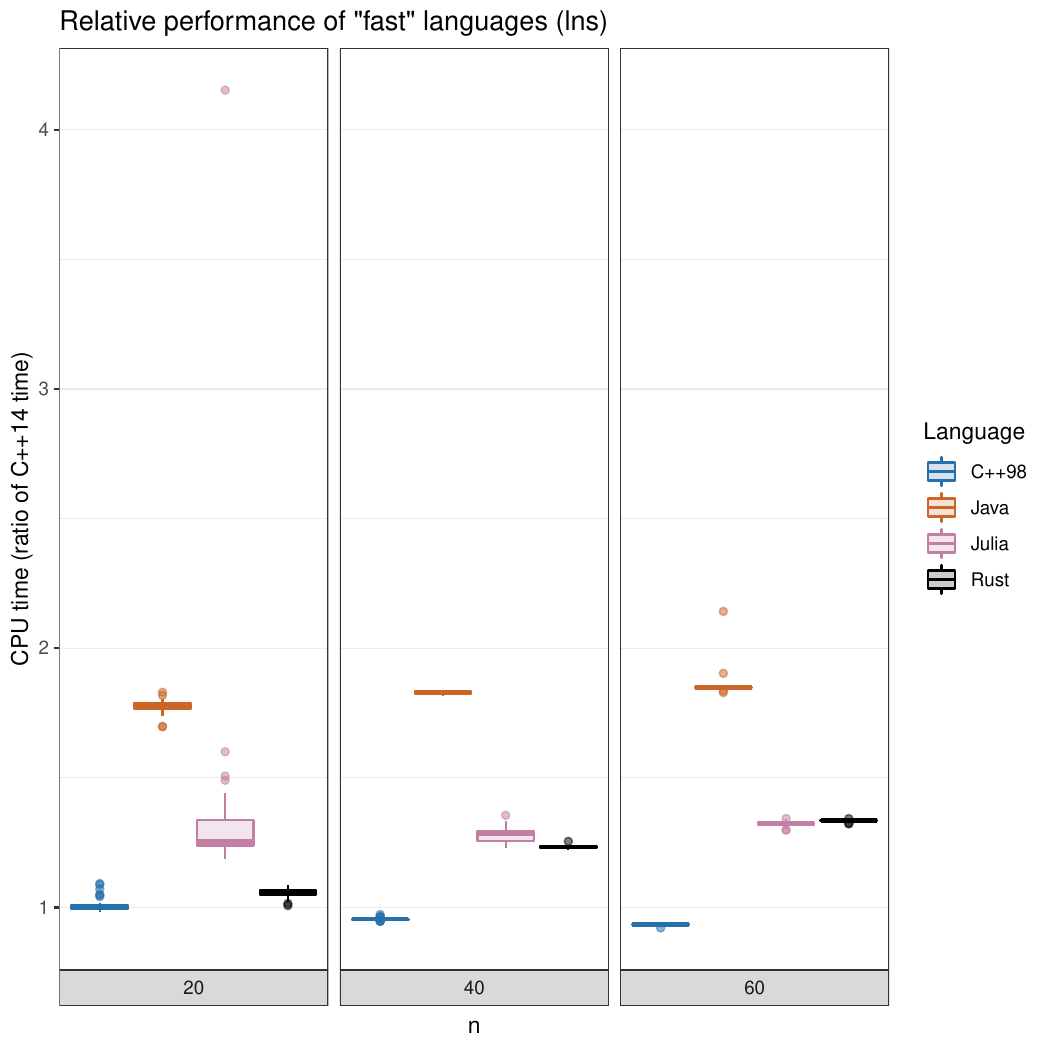}
    \includegraphics[scale=.22]{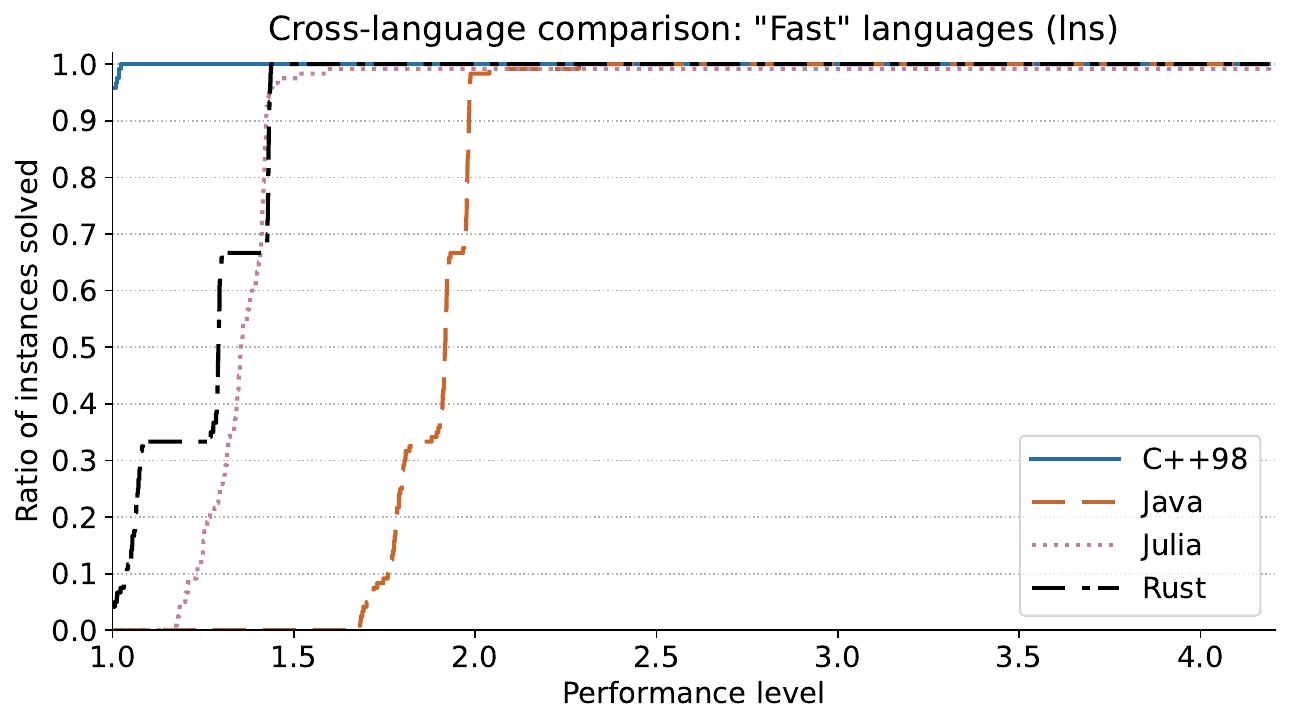}
    \caption{lns benchmark}
  \end{subfigure}
  \begin{subfigure}{0.33\textwidth}
    \centering
    \includegraphics[scale=.28]{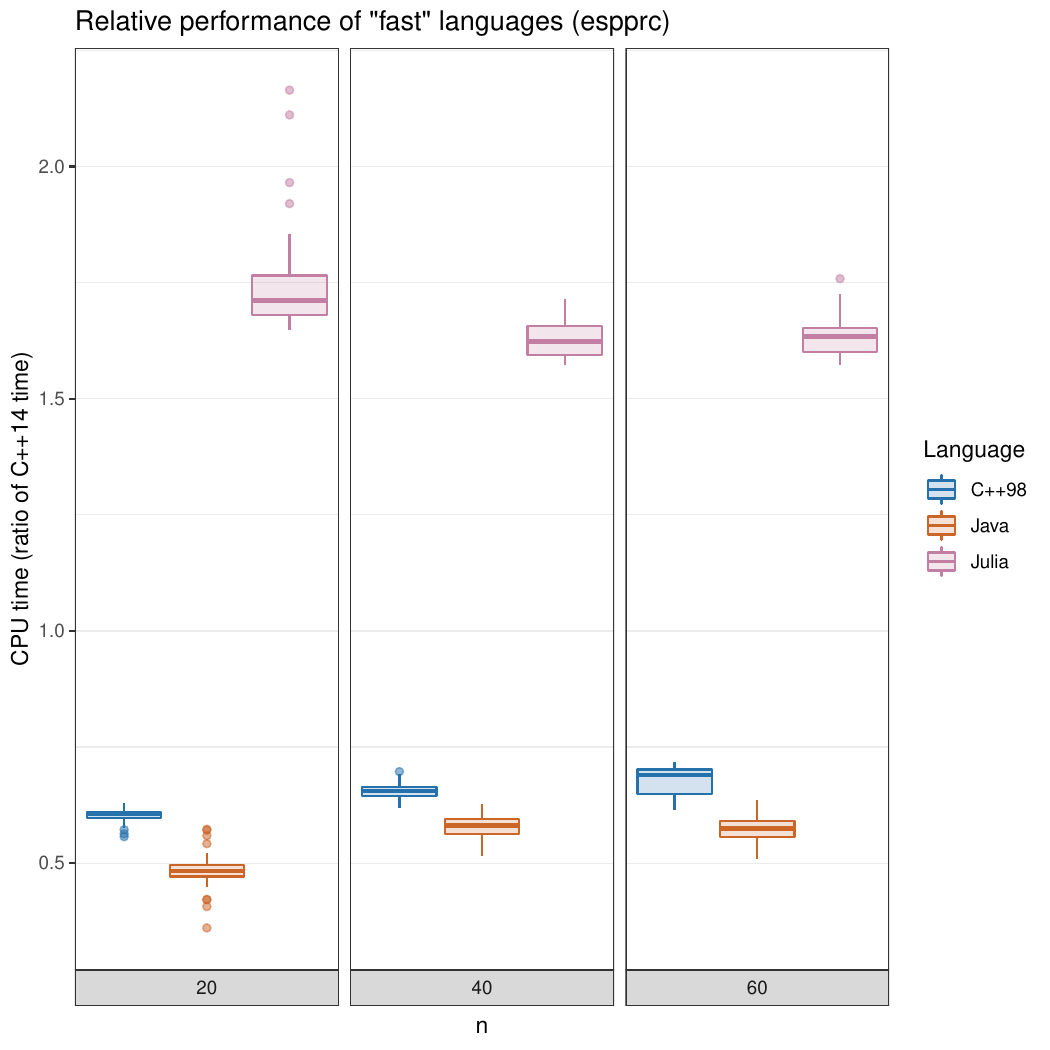}
    \includegraphics[scale=.22]{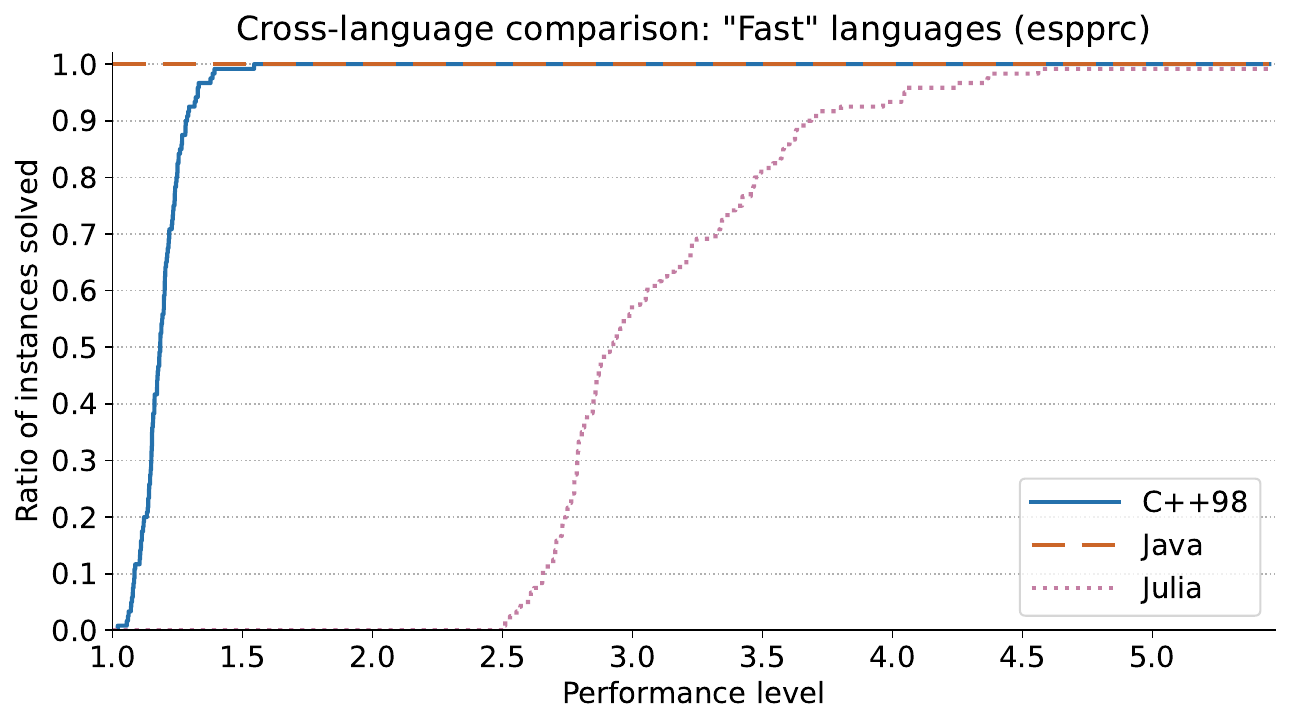}
    \caption{espprc benchmark}
  \end{subfigure}
  \begin{subfigure}{0.33\textwidth}
    \centering
    \includegraphics[scale=.28]{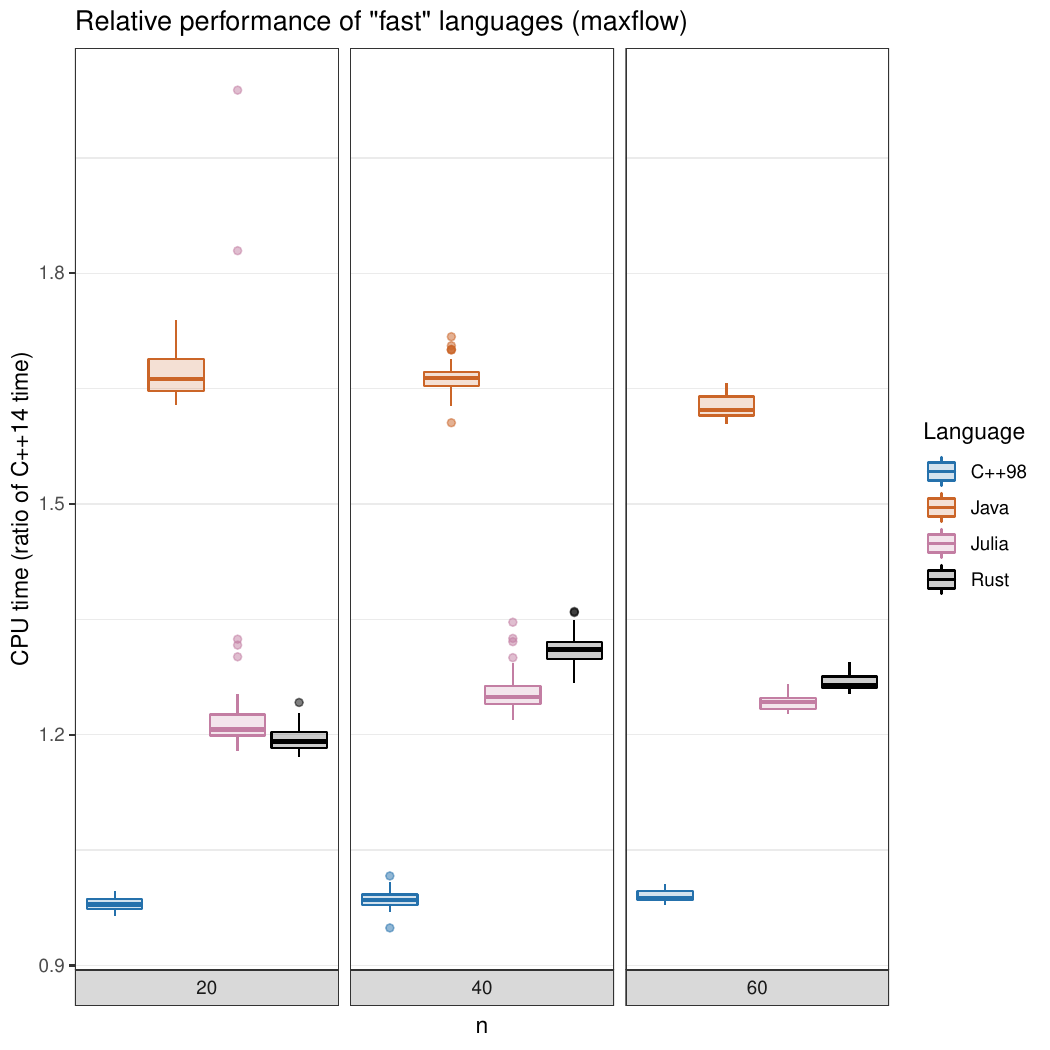}
    \includegraphics[scale=.22]{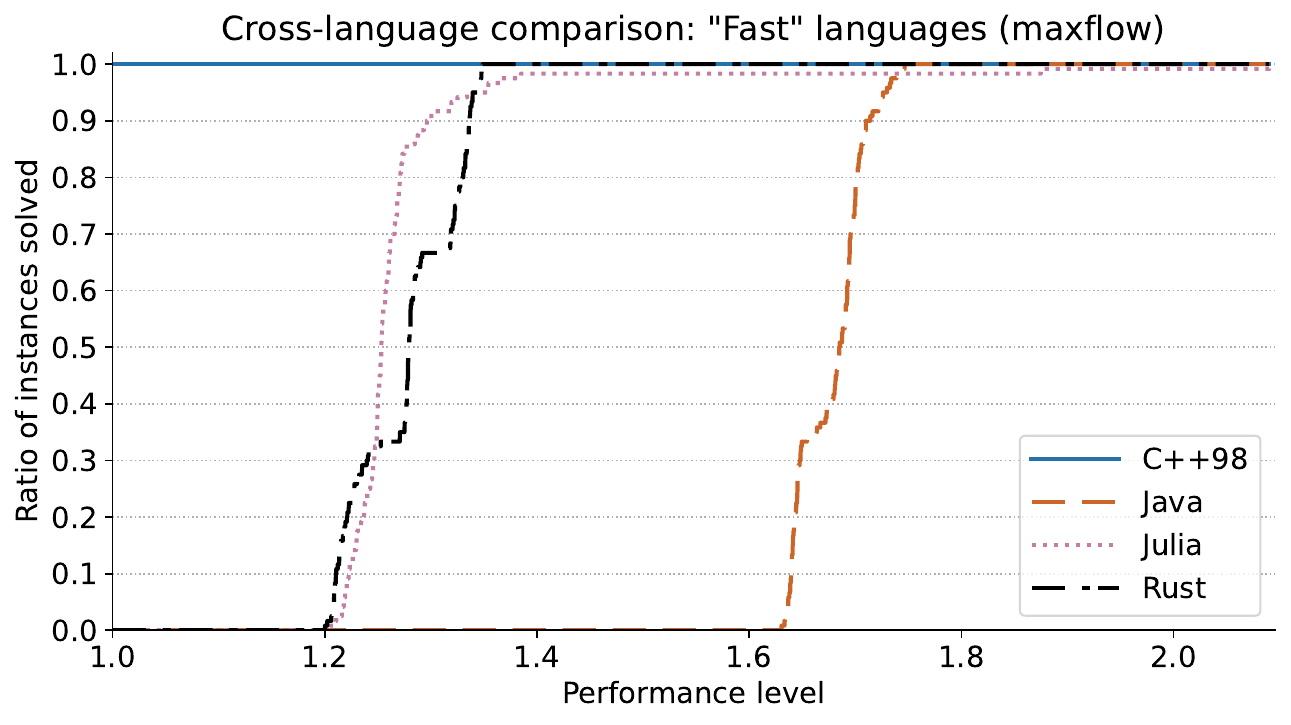}
    \caption{maxflow benchmark}
  \end{subfigure}
  \begin{subfigure}{0.33\textwidth}
    \centering
    \includegraphics[scale=.28]{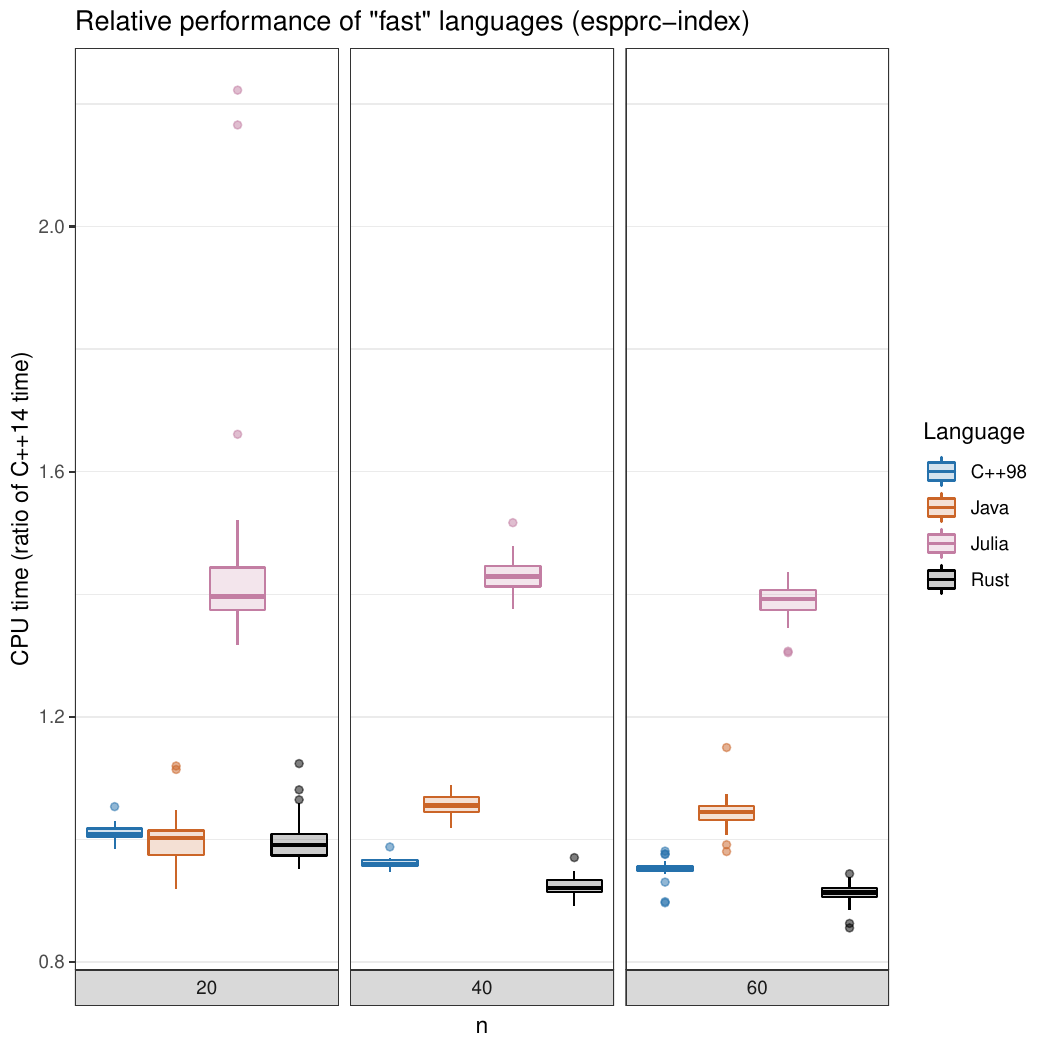}
    \includegraphics[scale=.22]{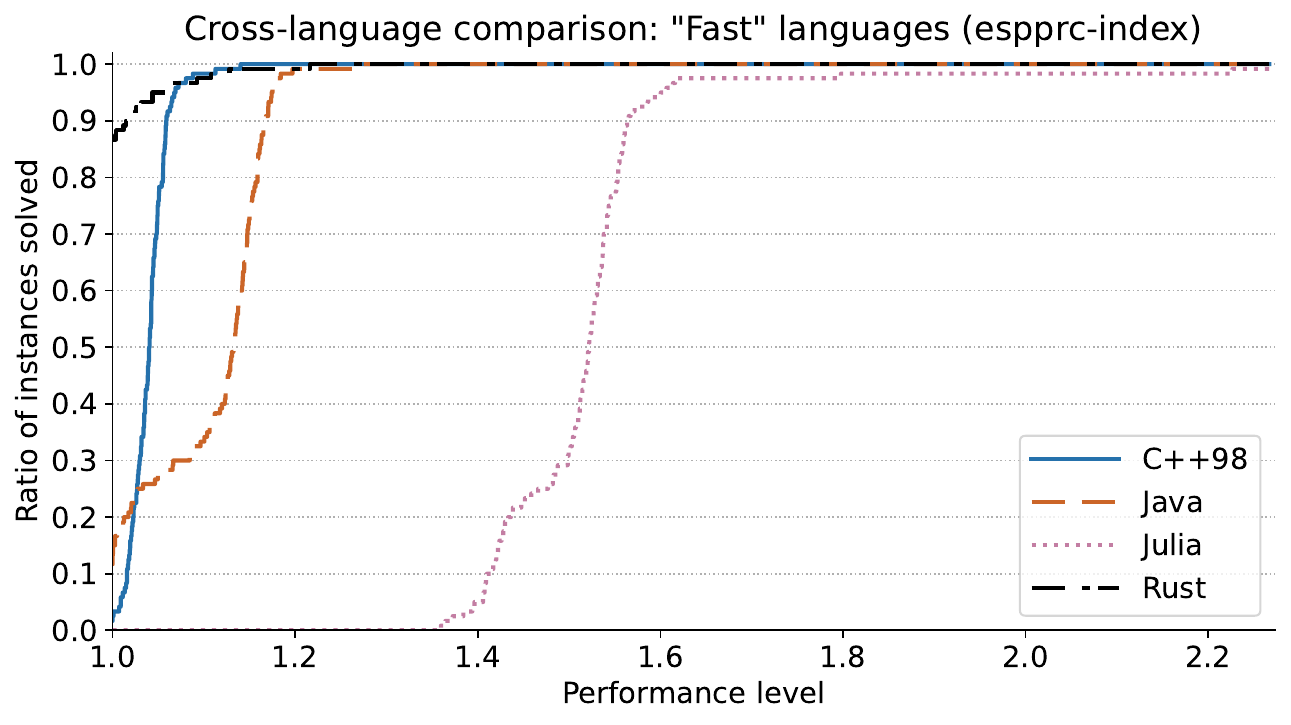}
    \caption{espprc-index benchmark}
  \end{subfigure}
  \caption{Performance of ``fast'' languages.}
  \label{fig:fast_languages}
\end{figure}

We compare the performance of fast languages in
Figure~\ref{fig:fast_languages}.

There is no clear winner: each language is the fastest for at least one
of the benchmarks. There are however some relatively large factors in
the execution speed in some cases, up to a factor 8 in the case of
Java for the Or-opt benchmark. However in many cases the factor
between two languages is never more than three. This is reassuring for
the researchers and students who want to use, in this setup, something
else than C++ and are worried about the performance hit.

\subsubsection{Cross-language comparison: ``interpreted''
  languages}
\begin{figure}
  \begin{subfigure}{0.33\textwidth}
    \centering
    \includegraphics[scale=.28]{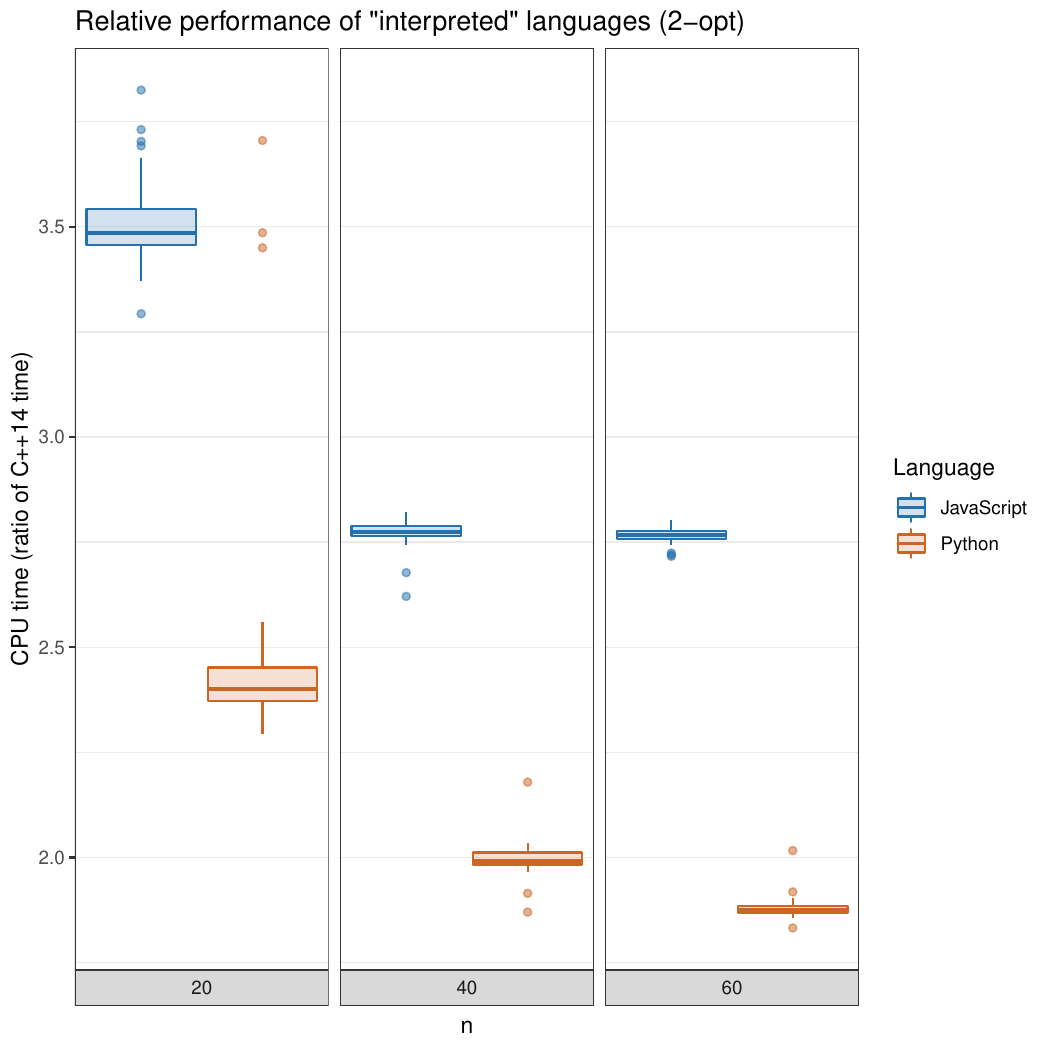}
    \includegraphics[scale=.22]{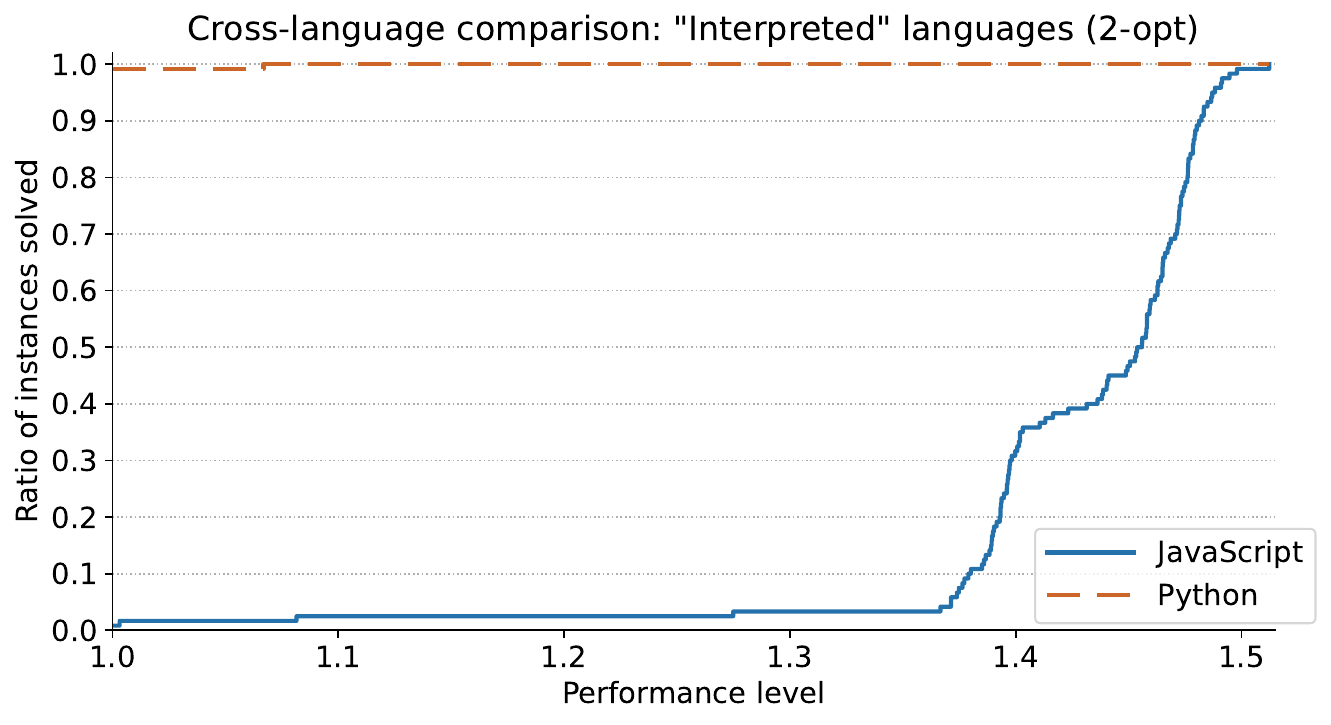}
    \caption{2-opt benchmark}
  \end{subfigure}
  \begin{subfigure}{0.33\textwidth}
    \centering
    \includegraphics[scale=.28]{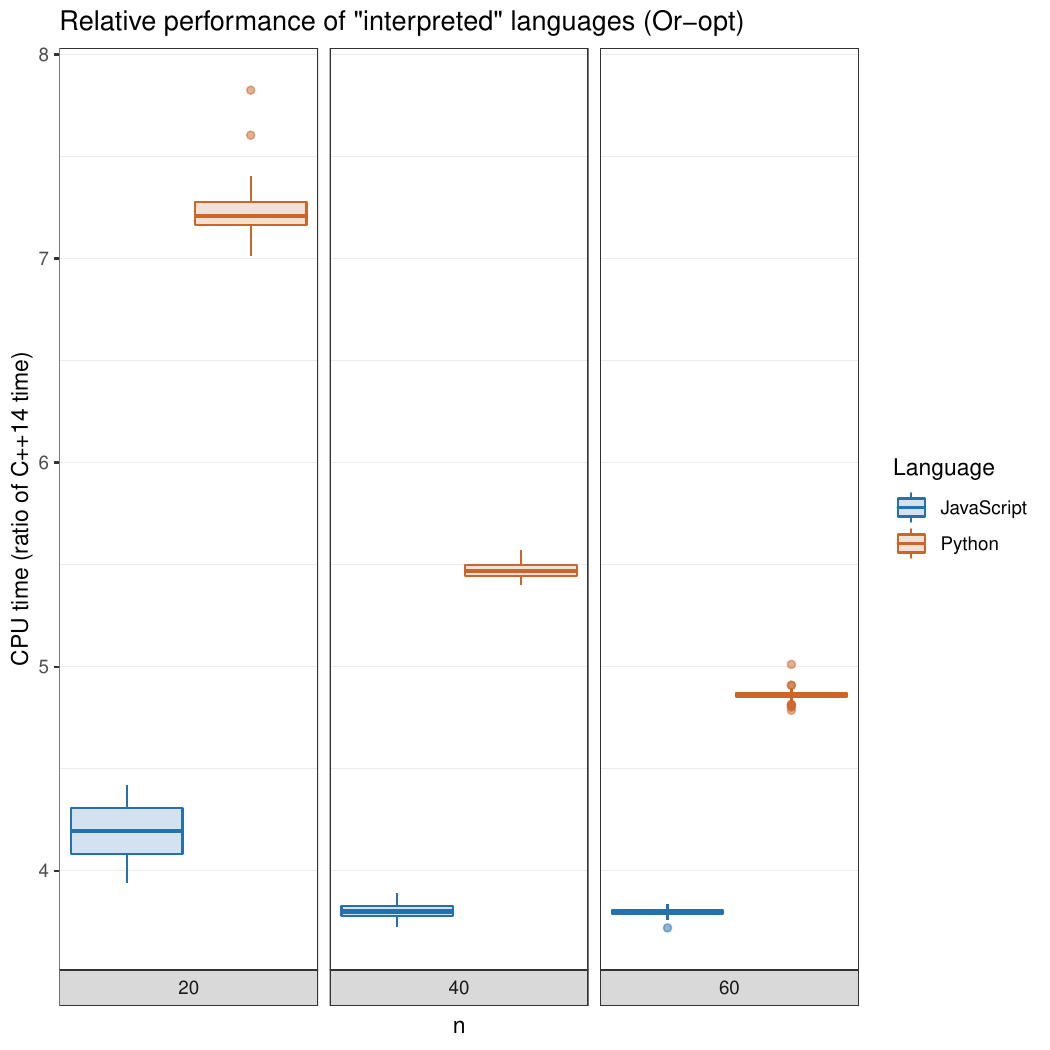}
    \includegraphics[scale=.22]{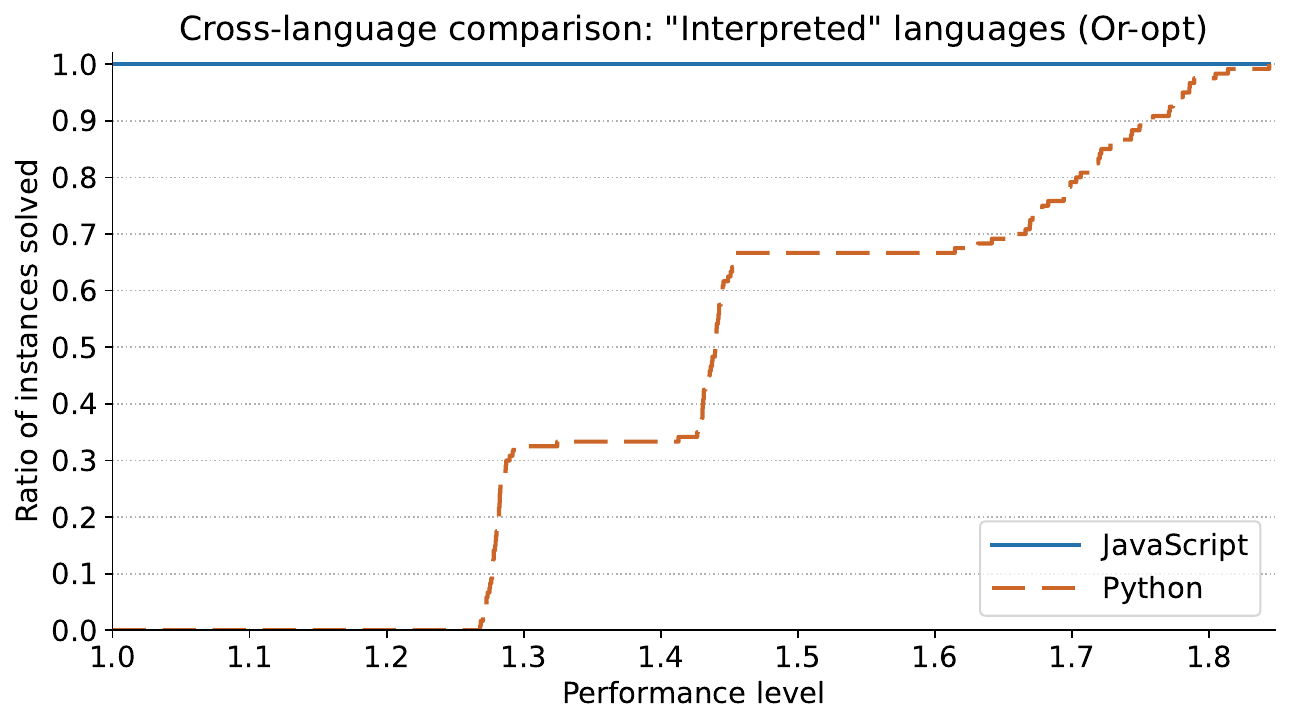}
    \caption{Or-opt benchmark}
  \end{subfigure}
  \begin{subfigure}{0.33\textwidth}
    \centering
    \includegraphics[scale=.28]{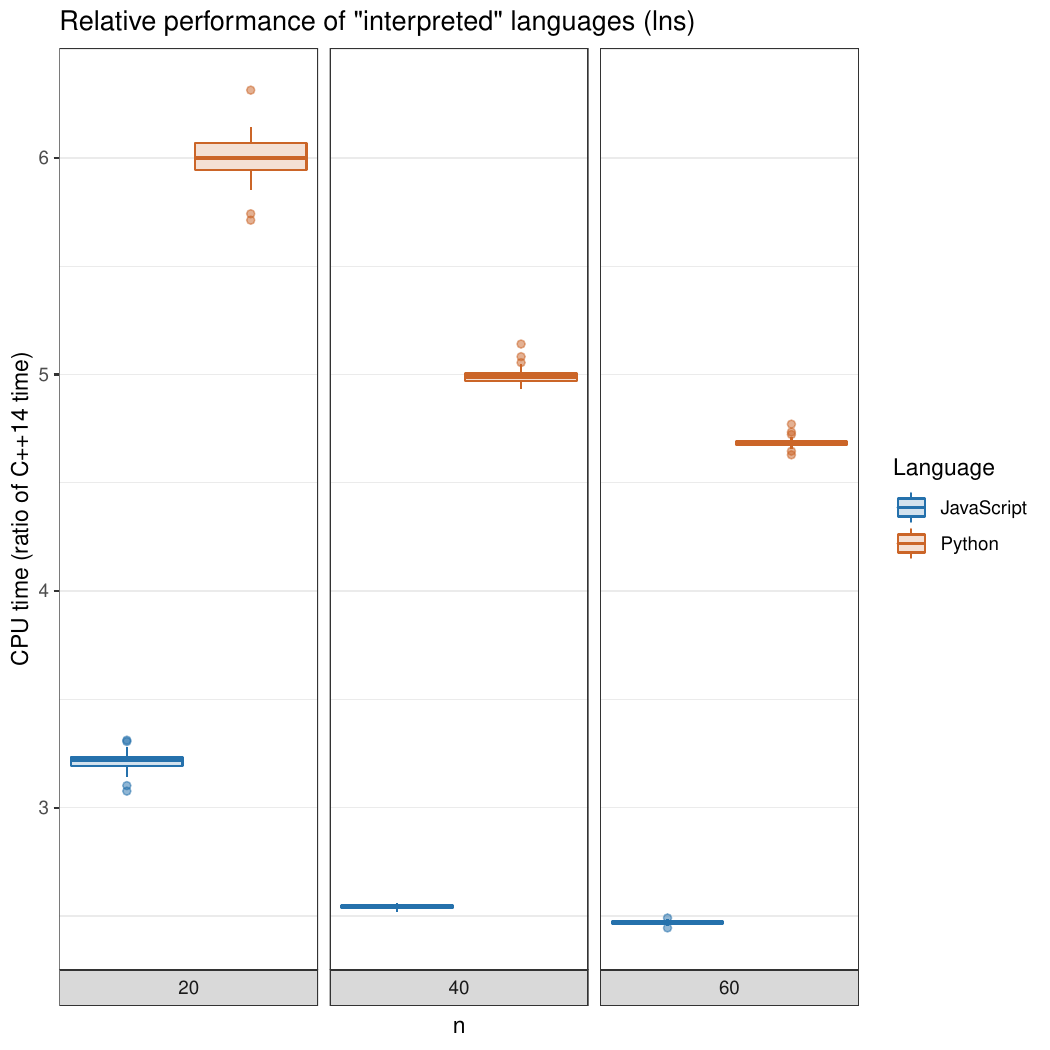}
    \includegraphics[scale=.22]{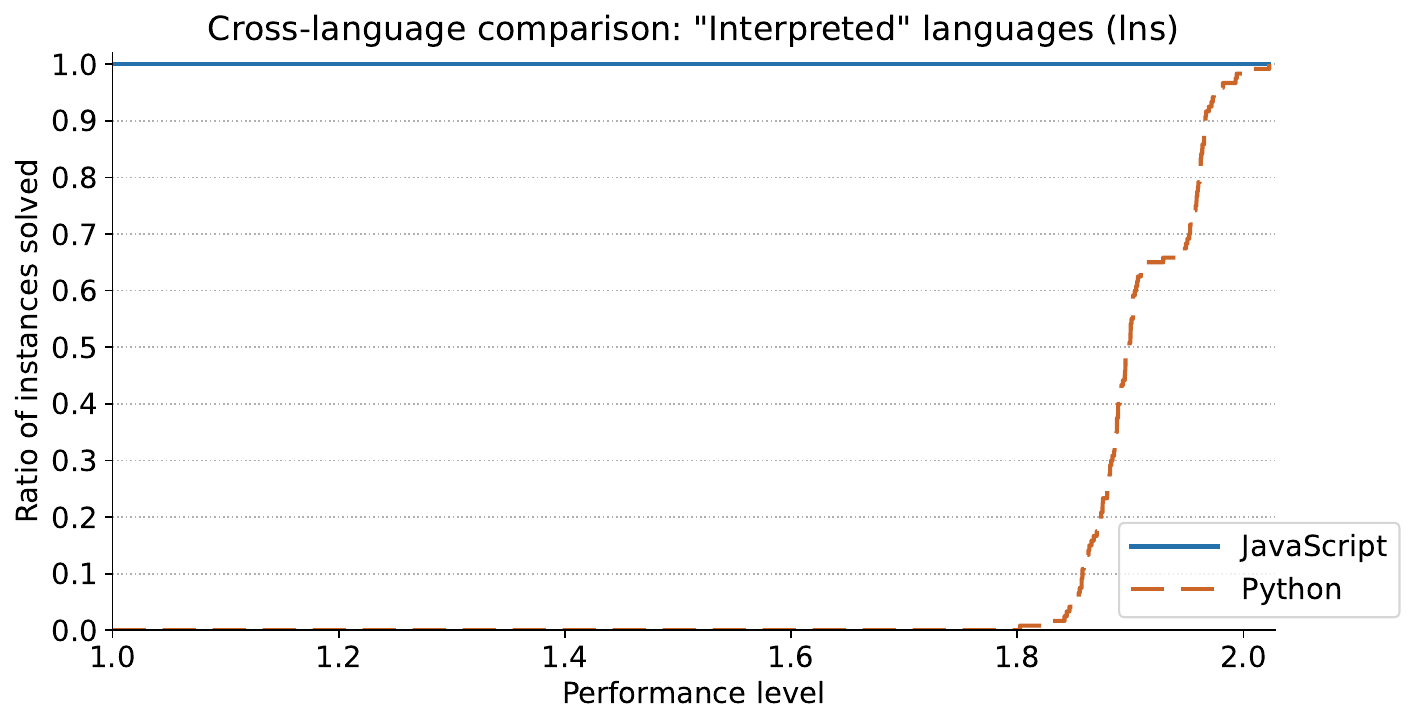}
    \caption{lns benchmark}
  \end{subfigure}
  \begin{subfigure}{0.33\textwidth}
    \centering
    \includegraphics[scale=.28]{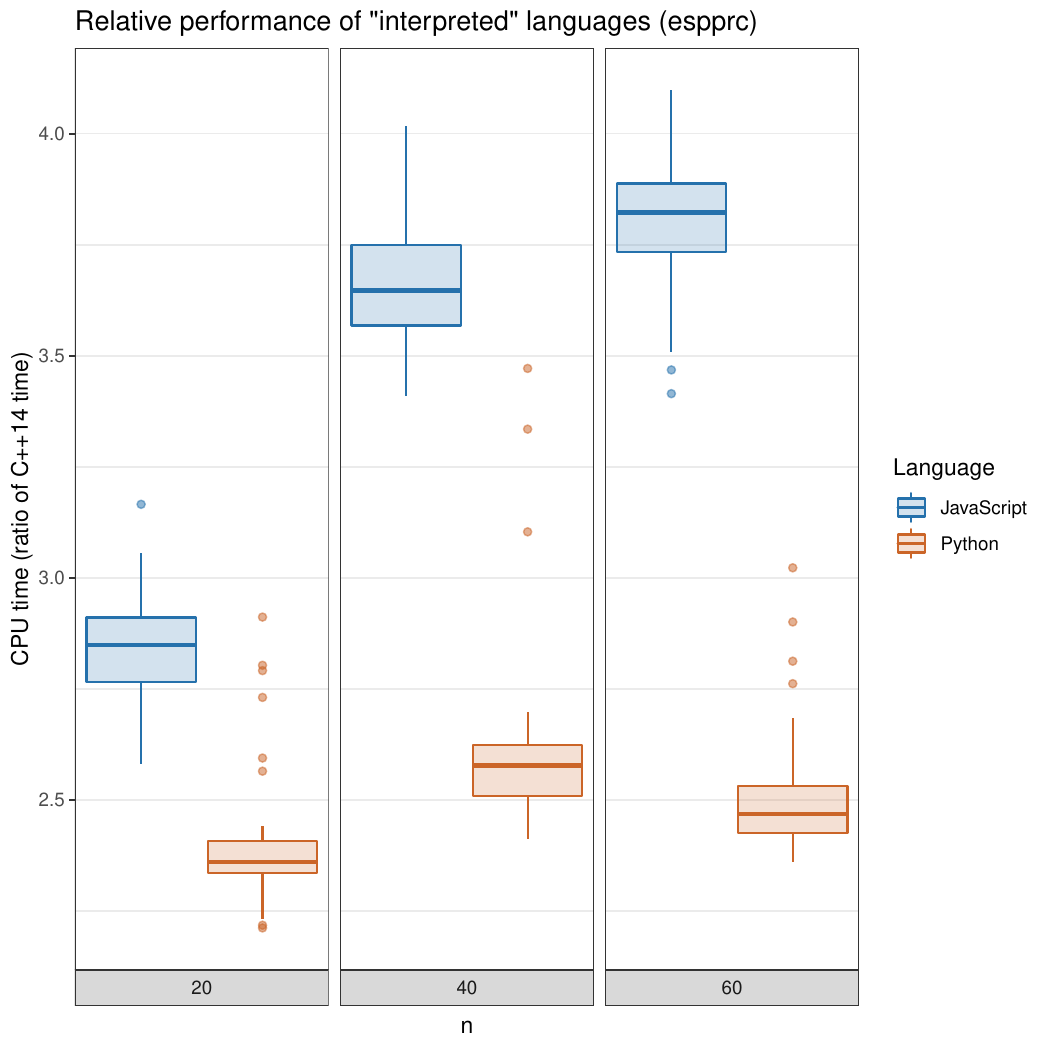}
    \includegraphics[scale=.22]{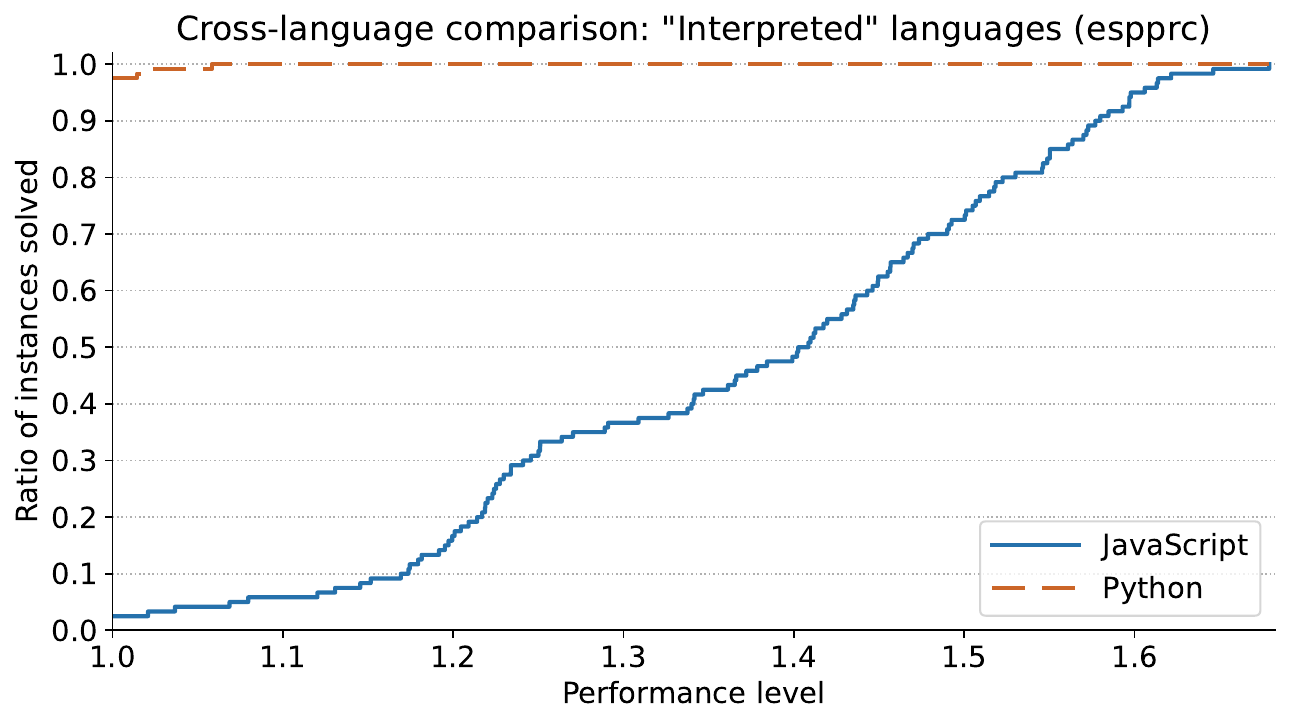}
    \caption{espprc benchmark}
  \end{subfigure}
  \begin{subfigure}{0.33\textwidth}
    \centering
    \includegraphics[scale=.28]{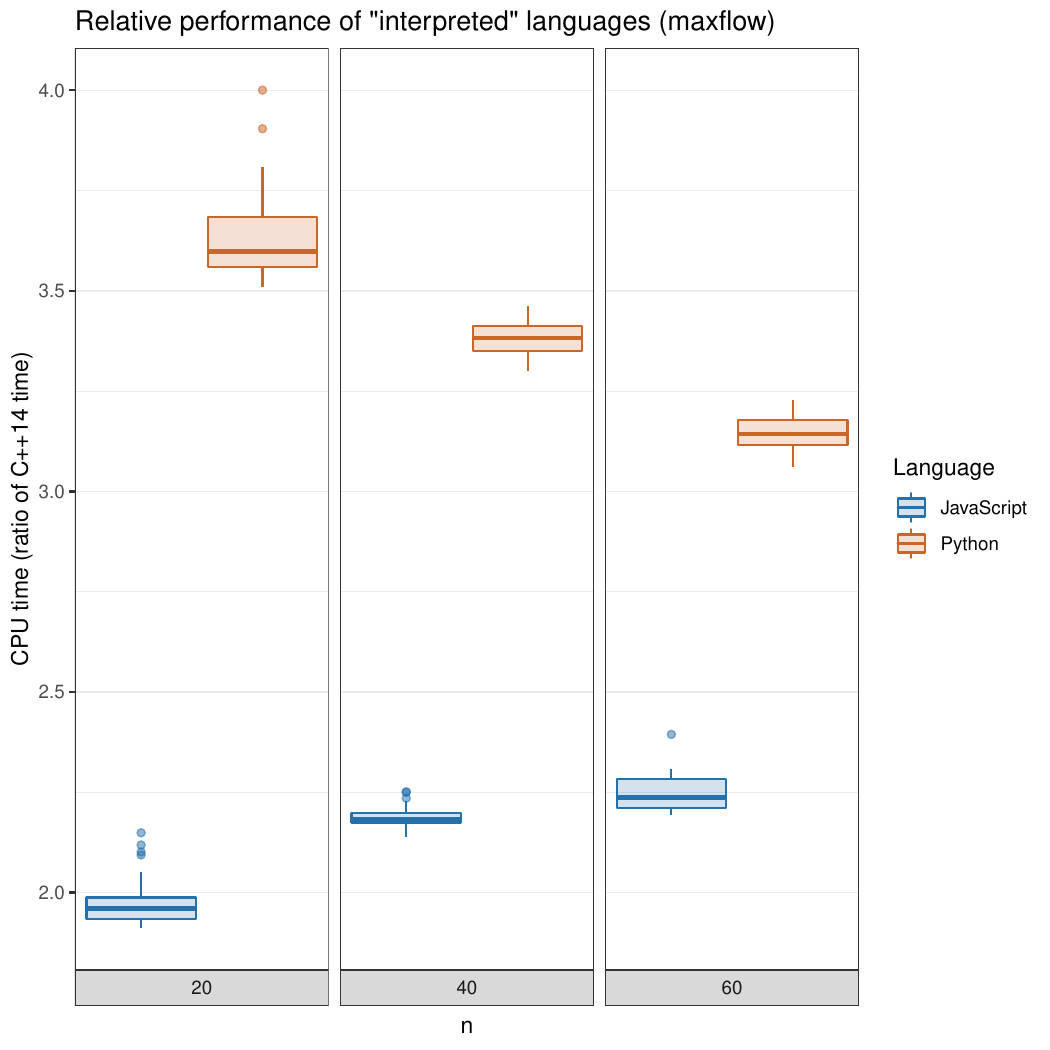}
    \includegraphics[scale=.22]{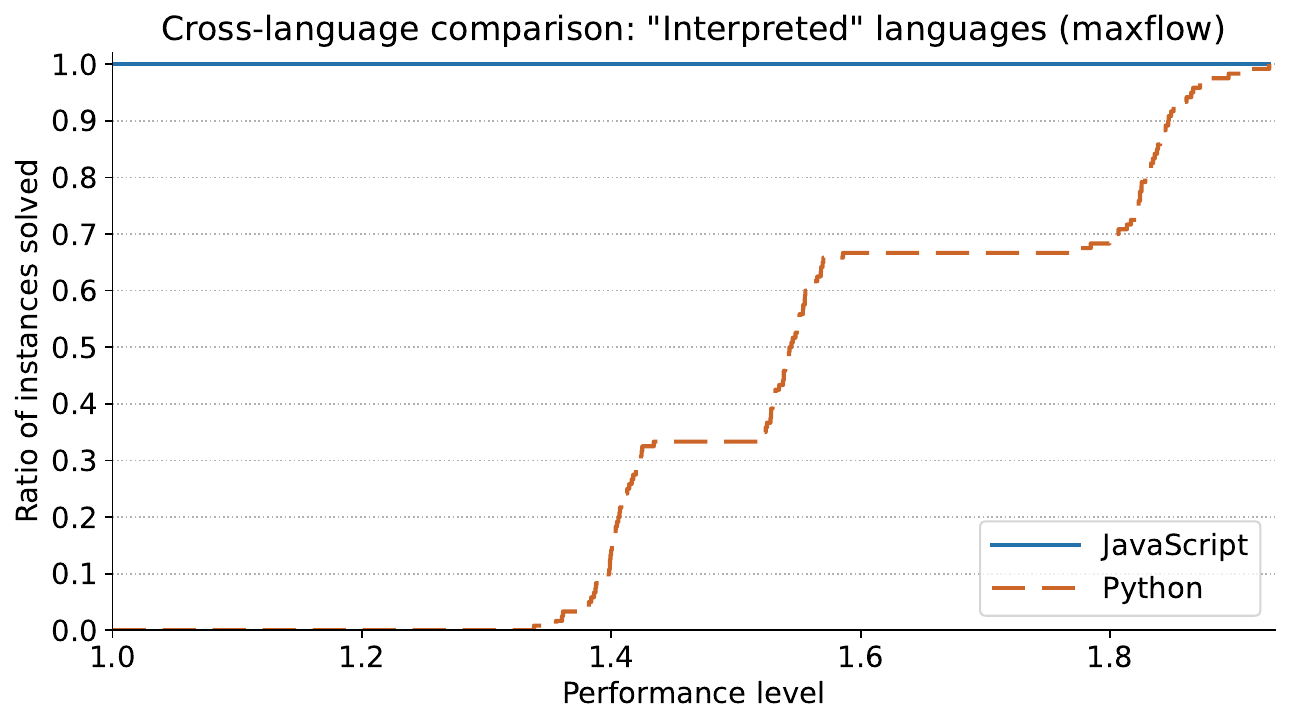}
    \caption{maxflow benchmark}
  \end{subfigure}
  \caption{Performance of ``interpreted'' languages.}
  \label{fig:interpreted_languages}
\end{figure}

We compare the performance of interpreted languages in
Figure~\ref{fig:interpreted_languages}.
Again there is no clear winner: both Python and JavaScript offer the
best performance for some of the benchmarks. More interestingly, these
implementations are never slower than ten times the effort required by
the baseline C++14 implementation (with the exception of one outlier
for the Python implementation of the Or-opt benchmark). In many cases,
they are less than five times slower even. This means that depending
on the use case, implementing routing optimization algorithms in
Python or JavaScript can be a viable (and convenient) option.

\subsection{Cross-platform comparison}
\begin{figure}
  \begin{subfigure}{0.5\textwidth}
    \centering
    \includegraphics[scale=.4]{pypy_vs_numba-2-opt-boxplot.pdf}
    \includegraphics[scale=.3]{pypy_vs_numba-2-opt-perfprof.pdf}
    \caption{Computing cluster: Intel Xeon E5-2650 v3, 20 cores, 3 GHz.}
  \end{subfigure}
  \begin{subfigure}{0.5\textwidth}
    \centering
    \includegraphics[scale=.4]{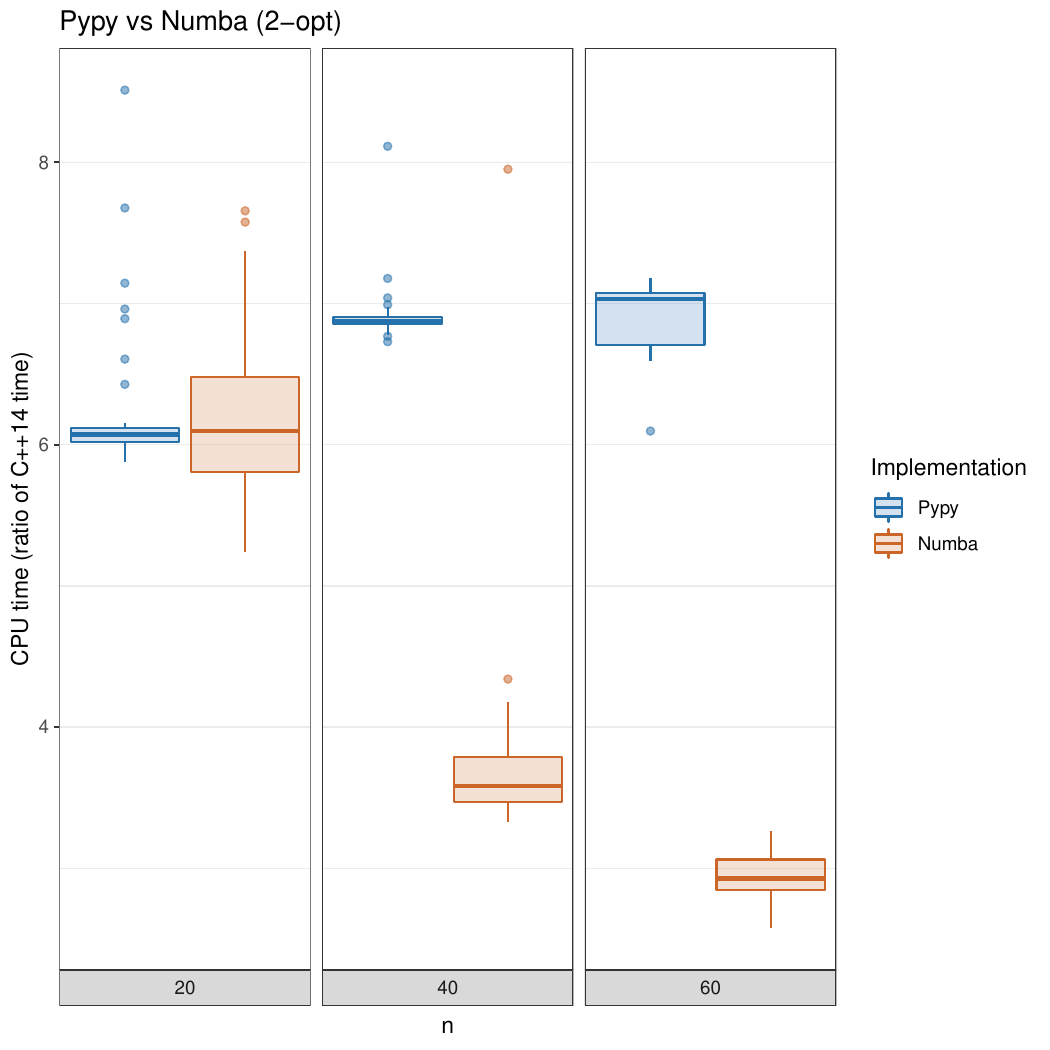}
    \includegraphics[scale=.3]{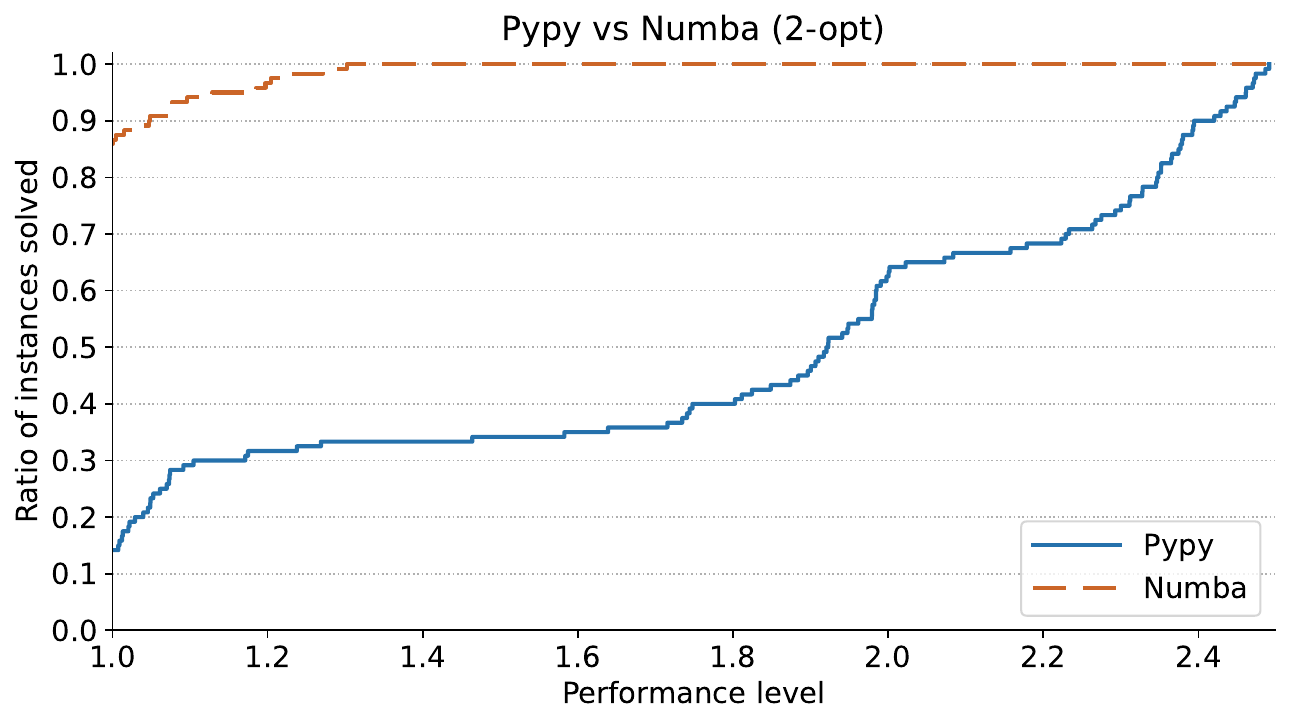}
    \caption{Raspberry Pi 4: ARMv7, 4 cores, 1.5 GHz.}
  \end{subfigure}
  \caption{Pypy versus Numba on 2-opt benchmark: different platforms.}
  \label{fig:pypy-vs-numba-platforms}
\end{figure}

Many findings presented above can be reproduced on different
platforms, however in some cases different platforms produce different
outcomes. We now illustrate this with an example, looking at the
comparison of Pypy and Numba on the 2-opt
benchmark. We run the 2-opt benchmark on the previously
mentioned Xeon CPU as well as on a Raspberry Pi 4 board with an ARMv7
CPU, 4 cores, 1.5 GHz. Pypy versions differ slightly: 7.0.0 on the
ARMv7 CPU versus 7.3.3 on the Xeon CPU. The Numba version is 0.42.0 on
both platforms.
Figure~\ref{fig:pypy-vs-numba-platforms} summarises and compares these
runs. Ratios are computed independently for each platform, e.g. the
CPU time on ARMv7 is presented as a ratio of the CPU time taken by
the C++14 implementation on ARMv7.
The trends are opposite on different platforms. This may be due to
several reasons, for instance certain JIT compilers working better with
certain CPUs, or version differences for Pypy and/or Numba. Ultimately, we
emphasize that it would be unwise to generalise the observations made
previously to different settings. The goal of this article is
not to provide final conclusions on what tool to use, rather to
introduce the BROUTE benchmark suite and encourage using it when
making early design decisions on implementation, using the appropriate
experimental platform to draw the appropriate conclusions.

\subsection{Experiments using standard VRP instances}
All instances used thus far have vertices whose coordinates are
generated using a uniform distribution. We now look at the same
experiments but using instances from the VRPLIB, available at
\url{http://or.dei.unibo.it/library/vrplib-vehicle-routing-problem-library},
which are more diverse in terms of vertex location generation. These results
are presented in Appendix~\ref{sec:vrplib}. They are largely
identical to the previously observed results.

\section{Discussion}
The question of what programming language to use when implementing
optimization algorithms is one that, by essence, does not have a final
answer. This being said, there is value in knowing how much
performance is gained or lost when using a certain
language on a certain architecture, or how much can be
gained by using different data structures. The benchmark suite
introduced in this paper is free software and can be used for that
purpose. We encourage students and researchers in
transport optimization to use it to support their implementation
decisions. Further experimental analyses, similar in flavor to the
ones conducted here, can also be good practice. For instance,
comparing the runtime performance of the same C++ code when compiled
with different compilers may provide a speedup for very little
effort.

Additionally, some insight can already be taken straight out of the
experiments conducted above. Using languages like Julia or Rust
does not incur a significant runtime penalty when compared to more
established languages like C++, and can actually be faster
sometimes. Interpreted languages like Python or JavaScript nowadays
have reasonably good runtime performance, usually less than five times
slower than pre-compiled languages, under the right conditions (typically,
using a JIT compiler). This means that the flexibility and ease of use
of these languages can be worth the performance hit, depending on the
application. Additionally, using JavaScript opens perspectives, as the
integration of optimization algorithms in web applications is easier
than it ever was, and these algorithms can now be run efficiently
client-side.

This is only the initial version of BROUTE; contributions are
encouraged, concerning both languages and algorithms. For example a
Matlab or C\# implementation of all benchmark algorithms would be
welcome as long as they can be run for free under the main operating
systems. Implementing a new benchmark in all currently supported
languages would also be welcome, or even re-implementing current
benchmark algorithms in currently supported languages but with
different data structures.

\section*{Acknowledgements}
The author is grateful to Sebastian Leitner for his valuable support
in relation to the Rust implementation, and to Belma Turan for her
valuable comments.

\section*{Declarations}
\subsection*{Conflicts of interest:}
On behalf of all authors, the corresponding author states that there
is no conflict of interest.
\subsection*{Availability of data and material:}
All data sets introduced in this paper are licensed under the GNU
General Public License version 3 and available at
\url{https://github.com/fa-bien/broute}.
\subsection*{Code availability:}
All code introduced in this work is licensed under the GNU General Public
License version 3 and available at
\url{https://github.com/fa-bien/broute}. This includes the code to
generate the charts.

\bibliographystyle{apalike}
\bibliography{paper_r1}

\newpage

\appendix

\section{Summary of languages and versions}
\label{app:lang}
\begin{table}[h]
  \begin{tabular}{l|ll}
    \toprule
    Language/tool & language or compiler version & options \\
    \midrule
    C++14 & g++ 8.3.0 & -Wall -ansi -pedantic -O3 -std=c++14 \\ 
    C++98 & g++ 8.3.0 & -Wall -ansi -pedantic -O3 -std=c++98 \\ 
    CPython & 3.7.3 & \\ 
    Pypy & 7.3.3-beta0 & \\ 
    Numpy & 1.16.2 & \\ 
    Numba & 0.42.0 & \\ 
    Java & OpenJDK version 11 & \\
    Julia & 1.6.0 & --check-bounds=no --inline=yes -O3 -t 1 \\
    Rust & 1.51 & cargo build --release \\
    JavaScript & Node v14.15.4 & \\
    \bottomrule   
  \end{tabular}
  \caption{Overview of languages and versions.}
\end{table}

\section{Pseudocode for each benchmark algorithm}
\label{app:abstract}
We use square braces to indicate array/vector indices, starting at index 0,
while $[]$ represents a new empty array/vector and $length(x)$ is the
number of elements in array $x$. Function $push(x, e)$ is used to
append value $e$ to array $x$. Function $delete(x, i)$ deletes, in
array $x$, the element at index $i$; function $insert(x, i, v)$
inserts value $v$ at index $i$ in array $x$, shifting values at index
$i$ and beyond to the right. Functions $delete()$ and $insert()$ have
linear worst-case time complexity.

In what follows, $n$ is the number of vertices in the distance matrix
and auxiliary graphs, $d_{ij}$ is the distance between
two vertices $i$ and $j$, while $s$ represents the \emph{seed}
solution. In algorithms performing local search (2-opt, Or-opt, lns),
this seed represents the initial solution. In other algorithms (espprc,
maxflow), it is used to generate the respective instance.

\subsection{2-opt}
The 2-opt benchmark is outlined in algorithm~\ref{alg:2-opt}. It
relies on function $first2eimprovement(s)$, which looks for an
improving 2-exchange in solution $s$. If it finds one, it performs it
(with a side effect on $s$)
and returns $true$; otherwise it returns $false$. The total number of
improvements performed is returned.

\begin{algorithm}[H]
  \caption{$2-opt(s)$}
  \begin{algorithmic}[1]
    \STATE $t \gets 0$
    \WHILE{$first2eimprovement(s)$}
    \STATE $t \gets t+1$
    \ENDWHILE
    \RETURN $t$
  \end{algorithmic}
  \label{alg:2-opt}
\end{algorithm}

\begin{algorithm}[H]
  \caption{$first2eimprovement(s)$}
  \begin{algorithmic}[1]
    \FOR{$p_1 \in \{1..|s| - 3\}$}
    \FOR{$p_2 \in \{p_1+1..|s| - 1\}$}
    \IF{$d_{s[p_1], s[p_1+1]} + d_{s[p_2], s[p_2+1]} > d_{s[p_1],
        s[p_2]} + d_{s[p_1+1], s[p_2+1]}$}
    \FOR{$i \in \{0..\lfloor\frac{p_2 - p_1 + 1}{2}\rfloor\}$}
    \STATE $s[p_1+1+i] \leftrightarrow s[p_2-i]$
    \ENDFOR
    \RETURN $true$
    \ENDIF
    \ENDFOR
    \ENDFOR
    \RETURN $false$
  \end{algorithmic}
  \label{alg:first2e}
\end{algorithm}

\subsection{Or-opt}
The Or-opt benchmark follows a similar pattern as the 2-opt
one. Algorithm~\ref{alg:Or-opt} outlines it.

\begin{algorithm}[H]
  \caption{$Or-opt(s)$}
  \begin{algorithmic}[1]
    \STATE $t \gets 0$
    \WHILE{$firstorimprovement(s)$}
    \STATE $t \gets t+1$
    \ENDWHILE
    \RETURN $t$
  \end{algorithmic}
  \label{alg:Or-opt}
\end{algorithm}

\begin{algorithm}[H]
  \caption{$firstorimprovement(s)$}
  \begin{algorithmic}[1]
    \FOR{$i \in \{1..|s| - 1\}$}
    \FOR{$l \in \{1..1+min(3, |s| - 1 - i\}$}
    \FOR{$p \in \{0..i-1\} \cup \{i+l..|s|-1\}$}
    \STATE $\delta \gets d_{s[i-1], s[i+l]} + d_{s[p], s[i]} + d_{s[i+l-1], s[p+1]} -
    d_{s[p], s[p+1]} - d_{s[i-1], s[i]} - d_{s[i+l-1], s[i+l]}$
    \IF{$\delta < 0$}
    \STATE $t \gets s[i..i+l]$
    \IF{$i < p$}
    \STATE $s[i..p-l+1] \gets s[i+l..p+1]$
    \STATE $s[p+1-l..p+1] \gets t$
    \ELSE
    \STATE $s[p+l..i-1+l] \gets s[p..i-1]$
    \STATE $s[p+1..p+1+l] \gets t$
    \ENDIF
    \RETURN $true$
    \ENDIF
    \ENDFOR
    \ENDFOR
    \ENDFOR
    \RETURN $false$
  \end{algorithmic}
  \label{alg:firstor}
\end{algorithm}

\subsection{lns}
The lns benchmark destroys then repairs the solution $niter$ times and
returns, as checksum value, the sum of the costs of all insertions
performed over its whole run. It is outlined in algorithm~\ref{alg:lns}.

\begin{algorithm}[H]
  \caption{$lns(s, niter)$}
  \begin{algorithmic}[1]
    \STATE $checksum \gets 0$
    \FOR{$iter \in \{1..niter\}$}
    \STATE $tmp \gets copy(s)$
    \STATE $unplanned \gets []$
    \STATE $where \gets 1$
    \WHILE{$where < length(tmp) - 1$}
    \STATE $push(unplanned, tmp[where])$
    \STATE $delete(tmp, where)$
    \STATE $where \gets where + 1$
    \ENDWHILE
    \WHILE{$length(unplanned) > 0$}
    \STATE $bc, bn, bf, bt \gets \infty, -1, -1, -1$
    \FOR{$p_1 \in \{0..unplanned\}$}
    \STATE $k \gets unplanned[p_1]$
    \FOR{$p_2 \in \{0..|tmp|-1\}$}
    \STATE $i, j \gets tmp[p], tmp[p+1]$
    \STATE $\delta \gets d_{ik} + d_{kj} - d_{ij}$
    \IF{$\delta < bc$}
    \STATE $bc \gets \delta$
    \STATE $bn, bf, bt \gets k, p_1, p_2$
    \ENDIF
    \ENDFOR
    \ENDFOR
    \STATE $insert(tmp, bt+1, bn)$
    \STATE $delete(unplanned, bf)$
    \STATE $checksum \gets checksum + bc$
    \ENDWHILE
    \STATE $s \gets tmp$
    \ENDFOR
    \RETURN $checksum$
  \end{algorithmic}
  \label{alg:lns}
\end{algorithm}

\subsection{maxflow}
For the sake of legibility, the generation of the maxflow instances is
not described here. In order to solve one maxflow instance, we need
two auxiliary graphs: one to store capacities and one to store flow
values. To represent these two graphs we use two arrays. These arrays
must be allocated in memory once when reading the input data, before the
timing of the benchmark starts, then reused for each separate
instance. Hence the memory allocation of auxiliary graphs is not
timed, however filling it with values for a given seed is included in
the time for the benchmark (but not represented below). Function
$edmondskarp(C, F, s, t)$ computes the maximum flow from source $s$
to sink $t$ in a graph of $n$ vertices where capacities are provided
in array $C$, and flows are stored in array $F$. $F$ is pre-allocated
but needs to be initialized to zero values. Both $C$ and $F$ are
nested arrays or multi-dimensional arrays depending on what the
language allows. $Q$ is a first-in-first-out (FIFO) queue. Function
$pop(Q)$ removes the first element in $Q$ and returns it.

\begin{algorithm}[H]
  \caption{$edmondskarp(C, F, s, t)$}
  \begin{algorithmic}[1]
    \STATE $totalflow \gets 0$
    \STATE $moreflow \gets true$
    \STATE $Q \gets \emptyset$
    \STATE $pred \gets $ array of size $n$
    \FOR{$i \in \{0..n-1\}$}
    \FOR{$j \in \{0..n-1\}$}
    \STATE $F[i,j] \gets 0$
    \ENDFOR
    \ENDFOR
    \WHILE{$moreflow$}
    \FOR{$i \in \{0..n-1\}$}
    \STATE $pred[i] \gets -1$
    \ENDFOR
    \STATE $push(Q, s)$
    \WHILE{$length(Q) > 0$}
    \STATE $c \gets pop(Q)$
    \FOR{$j \in \{0..n-1\}$}
    \IF{$j \neq c \land pred[j] = -1 \land j \neq s \land C[c,j] >
      F[c,j]$}
    \STATE $pred[j] \gets c$
    \STATE $push(Q, j)$
    \ENDIF
    \ENDFOR
    \ENDWHILE
    \IF{$pred[t] \neq -1$}
    \STATE $df \gets \infty$
    \STATE $i, j \gets pred[t], t$
    \WHILE{$i \neq -1$}
    \IF{$df > C[i,j] - F[i,j]$}
    \STATE $df \gets C[i,j] - F[i,j]$
    \ENDIF
    \STATE $i, j \gets pred[i], i$
    \ENDWHILE
    \STATE $i, j \gets pred[t], t$
    \WHILE{$i \neq -1$}
    \STATE $F[i,j] \gets F[i,j] + df$
    \STATE $i, j \gets pred[i], i$
    \ENDWHILE
    \STATE $totalflow += df$
    \ELSE
    \STATE $moreflow \gets false$
    \ENDIF
    \ENDWHILE
    \RETURN $totalflow$
  \end{algorithmic}
  \label{alg:maxflow}
\end{algorithm}

\subsection{espprc}
Similar to maxflow, we only provide the pseudocode to solve one
instance of espprc. Also similar to maxflow, the auxiliary graph, used
this time to store the reduced cost of each arc, is only allocated
once and this allocation is not timed. Let $rc$ be the auxiliary graph
(matrix) containing reduced costs, $nres$ the number of resources
considered, $C$ the capacity for each resource, $q_{ir}$ the resource
consumption for vertex $i$ and resource $r$ (0 or 1), and $M$ the
maximum length a tour may have. Function $esprrc(rc, nres, C, q, M)$ computes all
shortest paths from vertex 0 to itself which are elementary except for
the two visits at node 0, do not consume more resources than $C$ for
any resource, and do not exceed length $M$. Note that given the
definition of resource consumption provided in
Section~\ref{sec:espprc}, $q_{ir}$ can be inlined using bit
operations, and that is the case in every implementation. Function
$espprc()$ is outlined in Algorithm~\ref{alg:espprc}.
Similar to the maxflow algorithm, $Q$ is a FIFO queue. Function
$espprc()$ relies on \emph{labels}, each label containing the following
attributes for a partial path:
\begin{itemize}
\item $cost$: total cost according to reduced cost matrix $rc$.
\item $length$: total length according to distance matrix $d$.
\item $visited$: set of nodes that are already visited.
\item $predecesor$: predecessor label (for path reconstruction).
\item $successors$: collections of successor labels (for recursive marking in
  case this label is found to be dominated).
\item $q$ : resource consumption for each resource (index with $r$).
\item $marked$: boolean flag determining whether a label is marked to be
    ignored.
\end{itemize}
In the following we consider that these attributes can be accessed
using the dot operator, e.g. $l.length$ is the length of label $l$.
We perform the following operations with such labels:
\begin{itemize}
\item $emptylabel()$: create an empty label.
\item $extend(l, v)$: extend label $l$ to vertex $v$.
\item $updatedominance(l, nl)$: compare a new label $nl$ against a
  collection $l$ of labels, all representing a path to the same
  vertex. If $nl$ is dominated by at least one element of $l$, return
  $false$. Otherwise, add $nl$ to $l$, marking all elements of $l$
  that are dominated by $nl$, and return $true$. No label is deleted
  during the algorithm, instead they are \emph{marked} to be
  ignored. Memory is only freed after the algorithm has converged.
\end{itemize}
For the sake of legibility, we do not detail these label operations
here. Moreover, they are typically heavily language-dependent.
\begin{algorithm}
  \caption{$espprc(rc, nres, C, q, M$}
  \begin{algorithmic}[1]
    \STATE $Q \gets \empty$
    \STATE $labels \gets$ array of empty arrays, one per vertex
    \STATE $push(labels[0], emptylabel())$
    \WHILE{$Q \neq \emptyset$}
    \STATE $c \gets pop(Q)$
    \FOR{$l \in labels[c]$}
    \IF{$\lnot l.marked$}
    \FOR{$succ \in \{0..n-1\}$}
    \IF{$succ \in l.visited \lor succ = c \lor l.length + d_{n, succ}
      + d_{succ, 0} > M$}
    \STATE continue (skip remainder of this iteration of the loop)
    \ENDIF
    \STATE $rfeas \gets true$
    \FOR{$r \in \{1..nres\}$}
    \IF{$q_{succ, r} > 0 \land l.q_r + 1 > C$}
    \STATE $rfeas \gets false$
    \STATE break (exit this loop)
    \ENDIF
    \ENDFOR
    \IF{$\lnot rfeas$}
    \STATE continue
    \ENDIF
    \STATE $nl \gets extend(l, succ)$
    \STATE $added \gets updatedominance(labels[succ], nl)$
    \IF{$added$}
    \STATE $push(l.successors, nl)$
    \IF{$succ \notin Q \land succ \neq 0$}
    \STATE $push(Q, succ)$
    \ENDIF
    \ENDIF
    \ENDFOR
    \ENDIF
    \ENDFOR
    \ENDWHILE
    \RETURN minimum cost among all labels at vertex $0$
  \end{algorithmic}
  \label{alg:espprc}
\end{algorithm}

\section{Experimental results on VRPLIB instances}
\label{sec:vrplib}
We conduct experiments on instances from the VRPLIB, which can be
found at
\url{http://or.dei.unibo.it/library/vrplib-vehicle-routing-problem-library}. We
use the symmetric CVRP instances, ignoring vehicle capacity. These
instances use Euclidean distance but vertex coordinates are not
randomly generated with a uniform distribution like our instances.
Some of these
instances use the same distance matrix; we consider a maximal subset
of VRPLIB instances with unique distance matrices. They are
summarized in Table~\ref{tab:vrplib}.

\begin{table}
  \centering
  \begin{tabular}{lc}
    \toprule
    Instance & \# vertices \\
    \midrule
    E016-03m & 16 \\
    E021-04m & 21 \\
    E022-04g & 22 \\
    E023-03g & 23 \\
    E026-08m & 26 \\
    E030-03g & 30 \\
    E031-09h & 31 \\
    E033-04g & 33 \\
    E036-11h & 36 \\
    E041-14h & 41 \\
    E045-04f & 45 \\
    E048-04y & 48 \\
    E051-05e & 51 \\
    E072-04f & 72 \\
    E076-07s & 76 \\
    E076A10r & 76 \\
    E076B09r & 76 \\
    E076C09r & 76 \\
    E076D09r & 76 \\
    \bottomrule
  \end{tabular}
  \caption{VRPLIB instances}
  \label{tab:vrplib}
\end{table}

\begin{figure}
  \begin{subfigure}{0.5\textwidth}
    \centering
    \includegraphics[scale=.3]{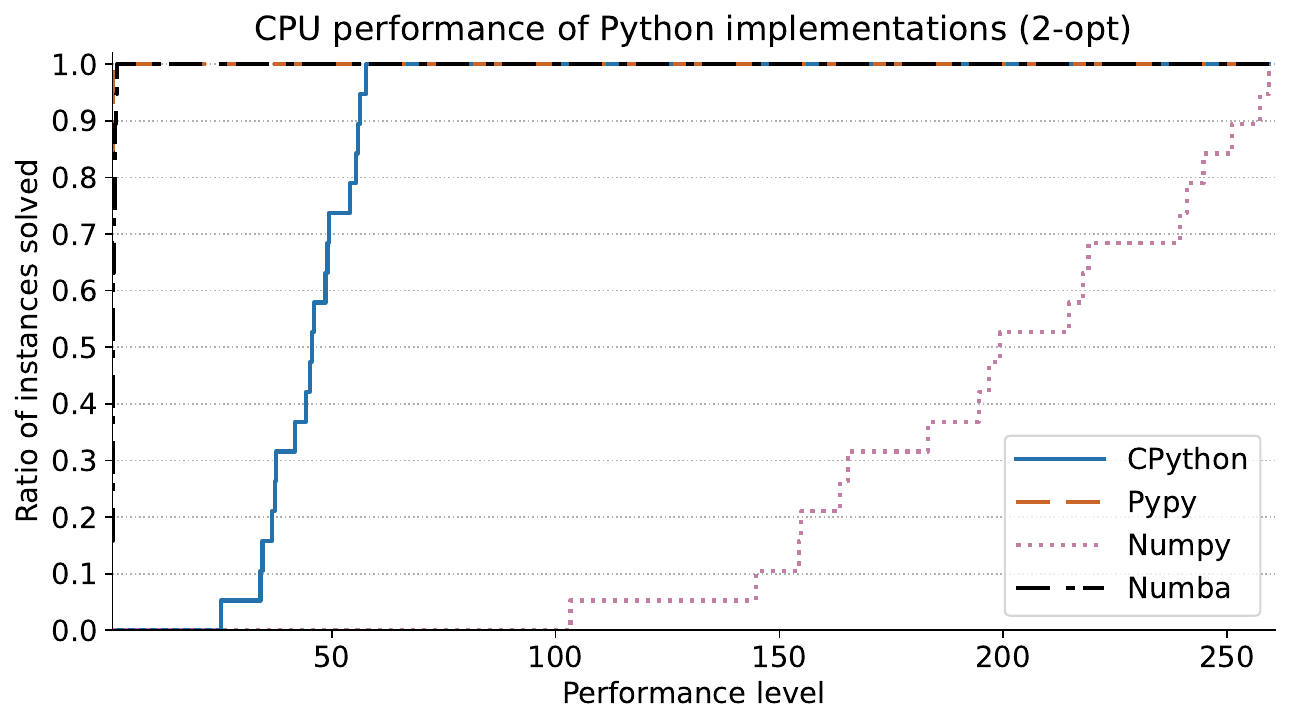}
    \caption{2-opt benchmark}
  \end{subfigure}
  \begin{subfigure}{0.5\textwidth}
    \centering
    \includegraphics[scale=.3]{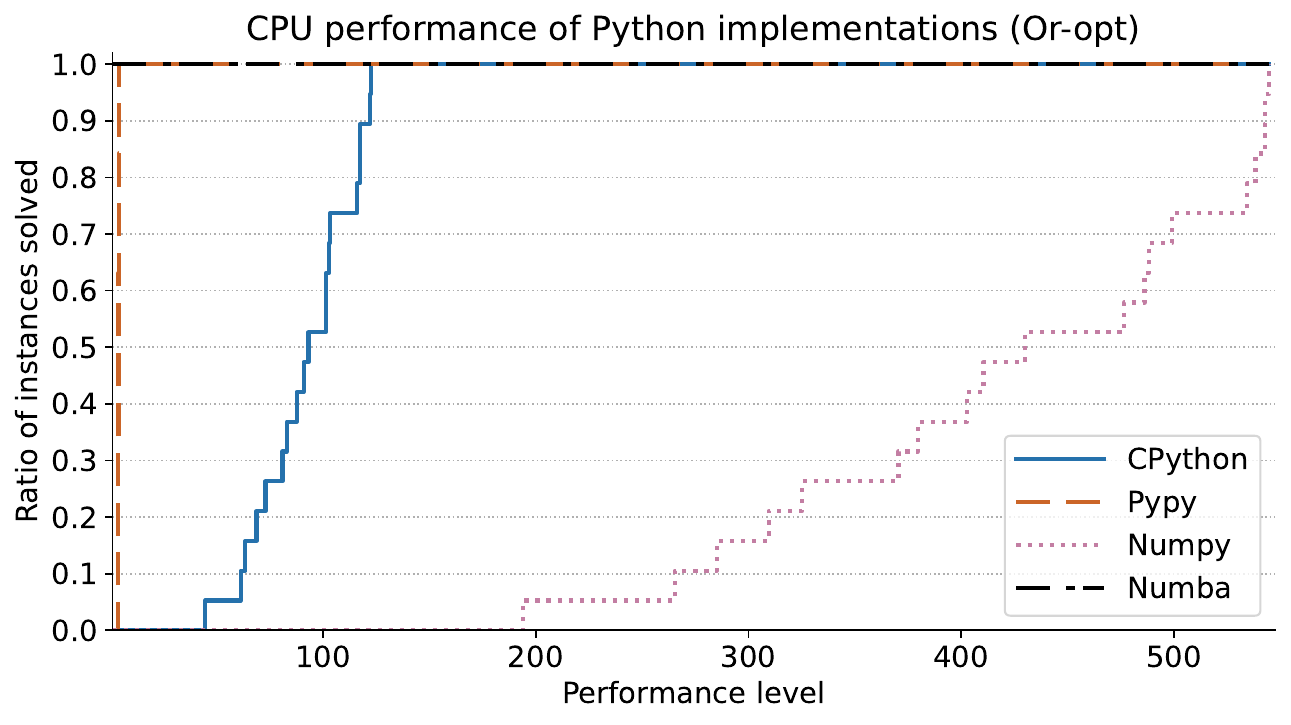}
    \caption{Or-opt benchmark}
  \end{subfigure}
  \caption{Python implementations: performance profiles (VRPLIB)}
  \label{fig:python-implementations-vrplib}
\end{figure}

\begin{figure}
  \begin{subfigure}{0.33\textwidth}
    \centering
    \includegraphics[scale=.22]{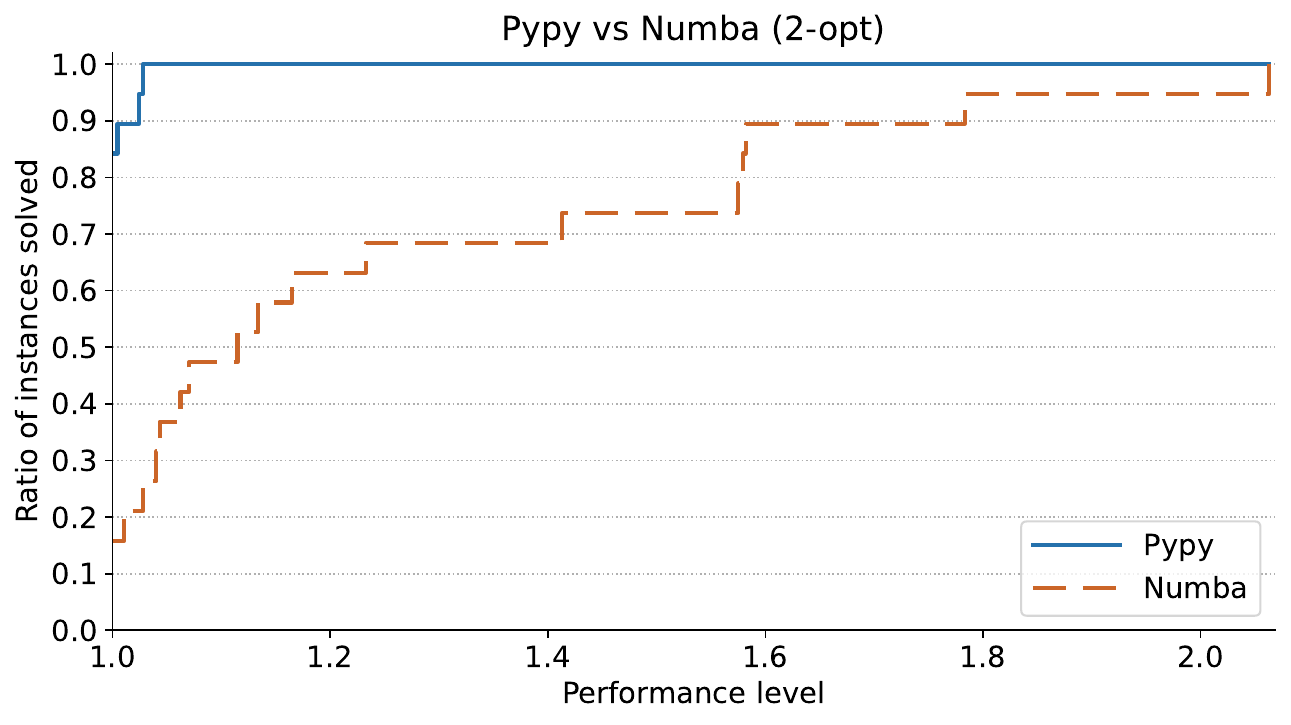}
    \caption{2-opt benchmark}
  \end{subfigure}
  \begin{subfigure}{0.33\textwidth}
    \centering
    \includegraphics[scale=.22]{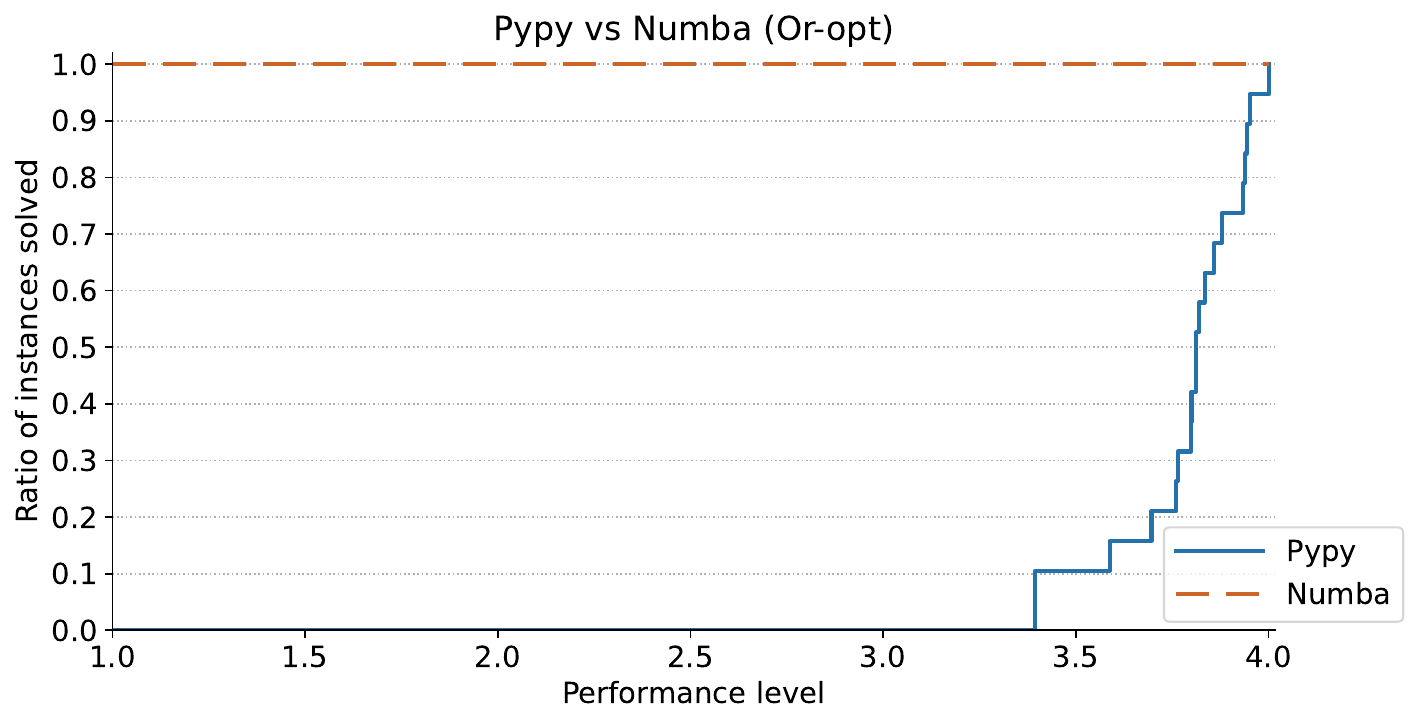}
    \caption{Or-opt benchmark}
  \end{subfigure}
  \begin{subfigure}{0.33\textwidth}
    \centering
    \includegraphics[scale=.22]{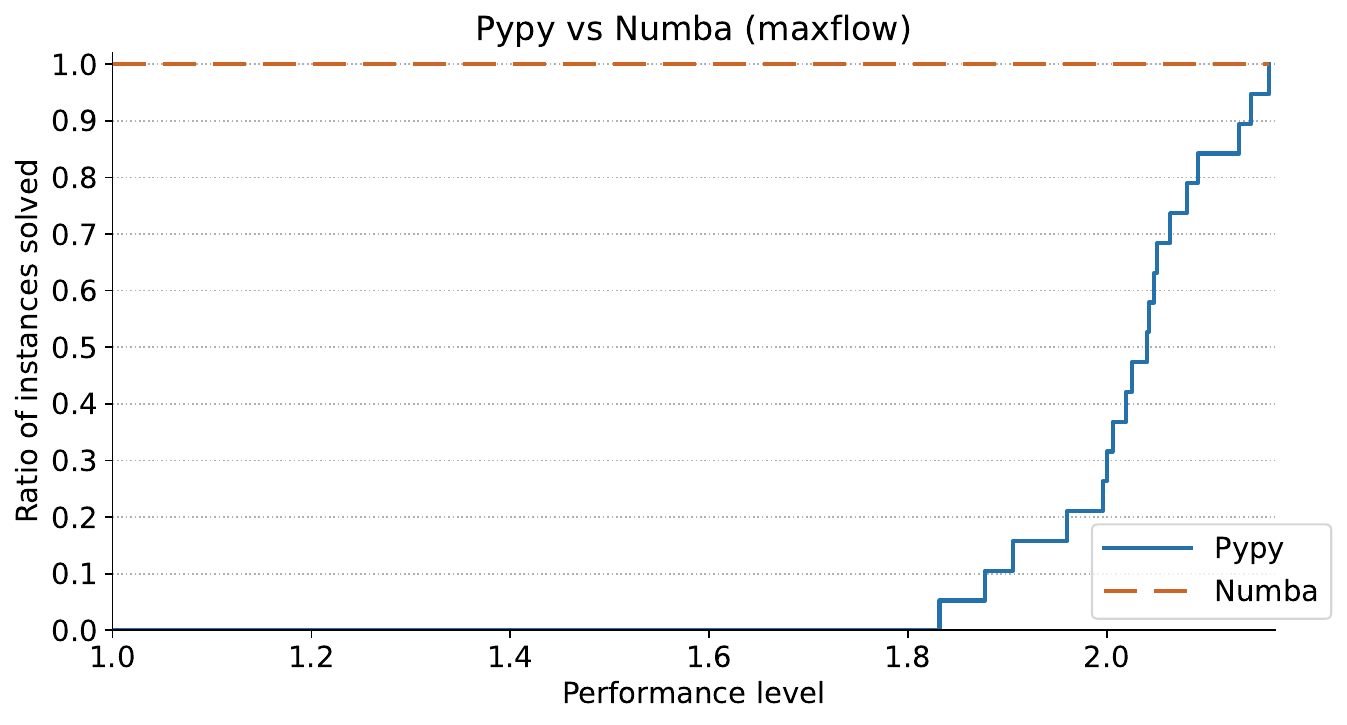}
    \caption{maxflow benchmark}
  \end{subfigure}
  \caption{Pypy vs Numba: performance profiles (VRPLIB)}
  \label{fig:pypy-vs-numba-vrplib}
\end{figure}

\begin{figure}
  \begin{subfigure}{0.33\textwidth}
    \centering
    \includegraphics[scale=.22]{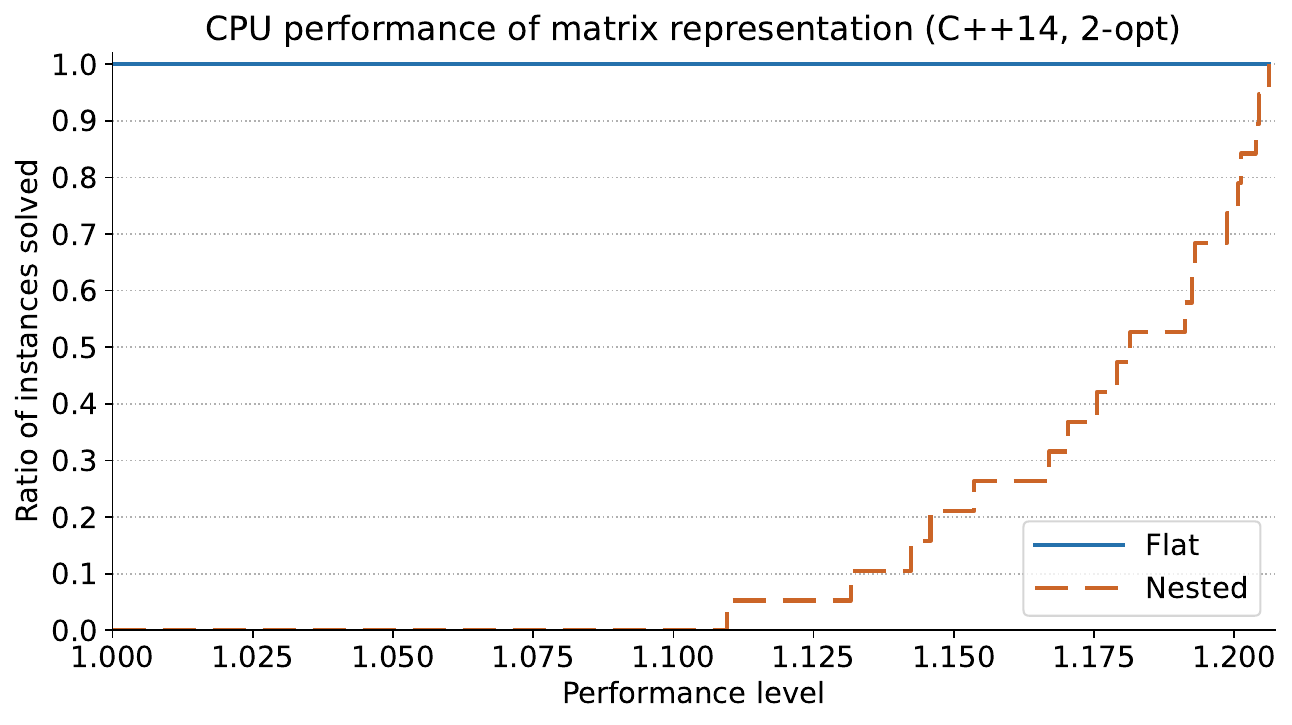}
    \caption{2-opt benchmark}
  \end{subfigure}
  \begin{subfigure}{0.33\textwidth}
    \centering
    \includegraphics[scale=.22]{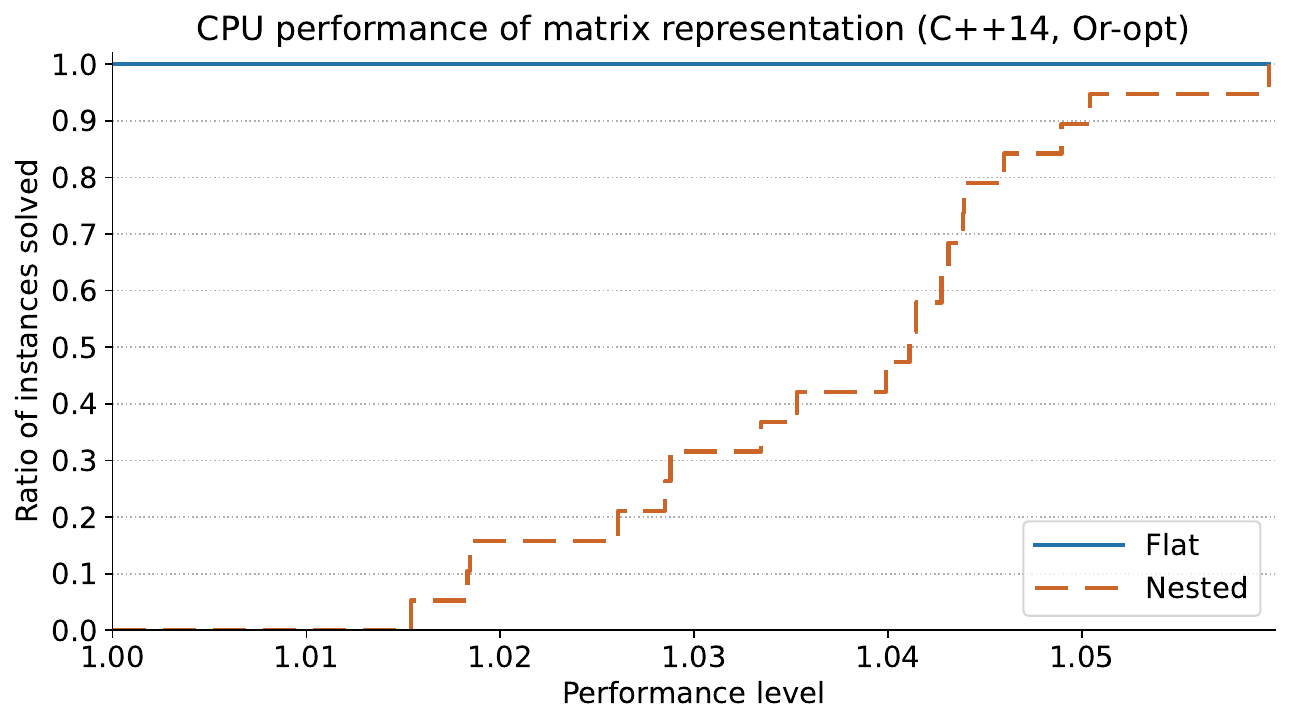}
    \caption{Or-opt benchmark}
  \end{subfigure}
  \begin{subfigure}{0.33\textwidth}
    \centering
    \includegraphics[scale=.22]{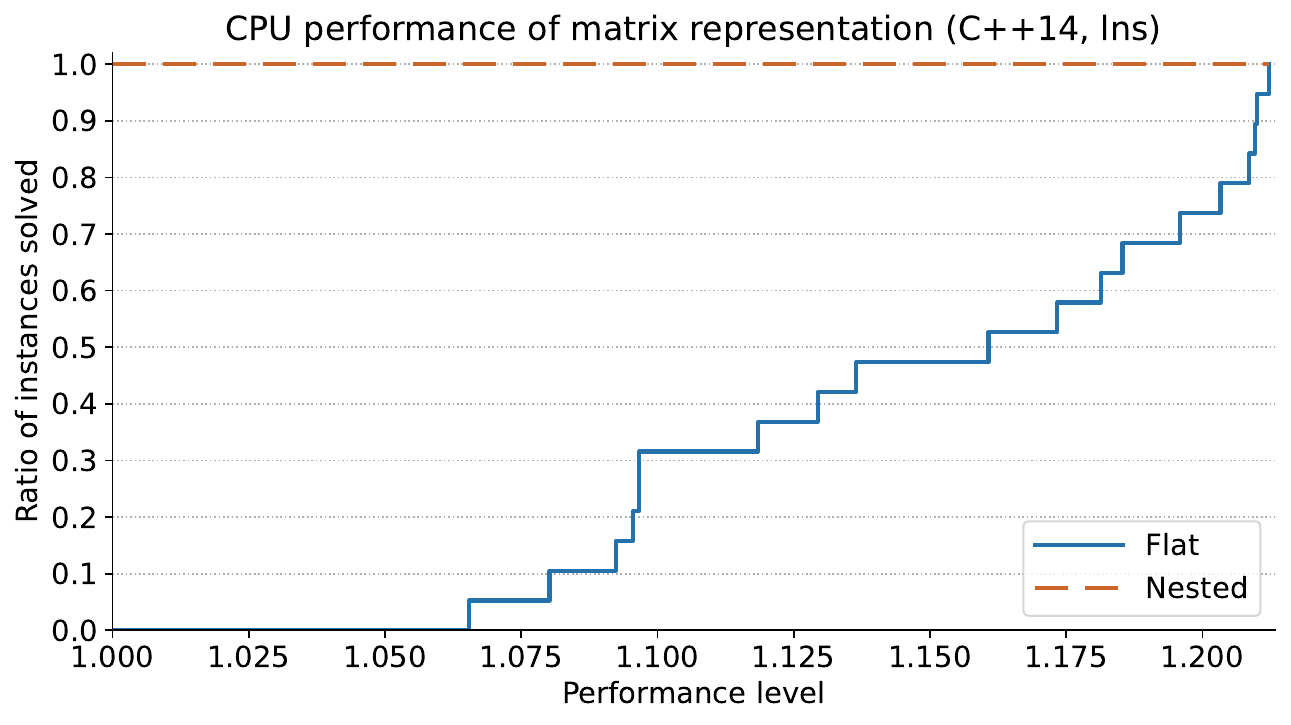}
    \caption{lns benchmark}
  \end{subfigure}
  \caption{C++14: comparison of matrix representations (VRPLIB)}
  \label{fig:c++14-matrix-vrplib}
\end{figure}

\begin{figure}
  \begin{subfigure}{0.33\textwidth}
    \centering
    \includegraphics[scale=.22]{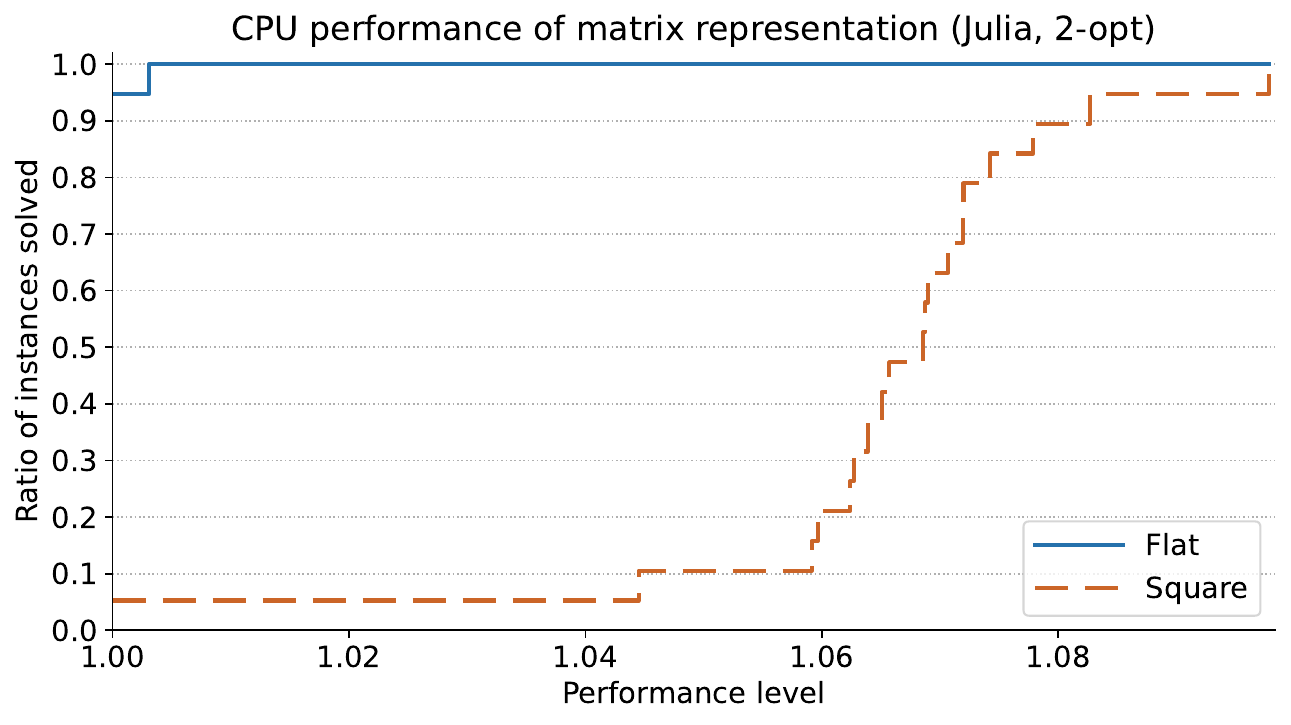}
    \caption{2-opt benchmark}
  \end{subfigure}
  \begin{subfigure}{0.33\textwidth}
    \centering
    \includegraphics[scale=.22]{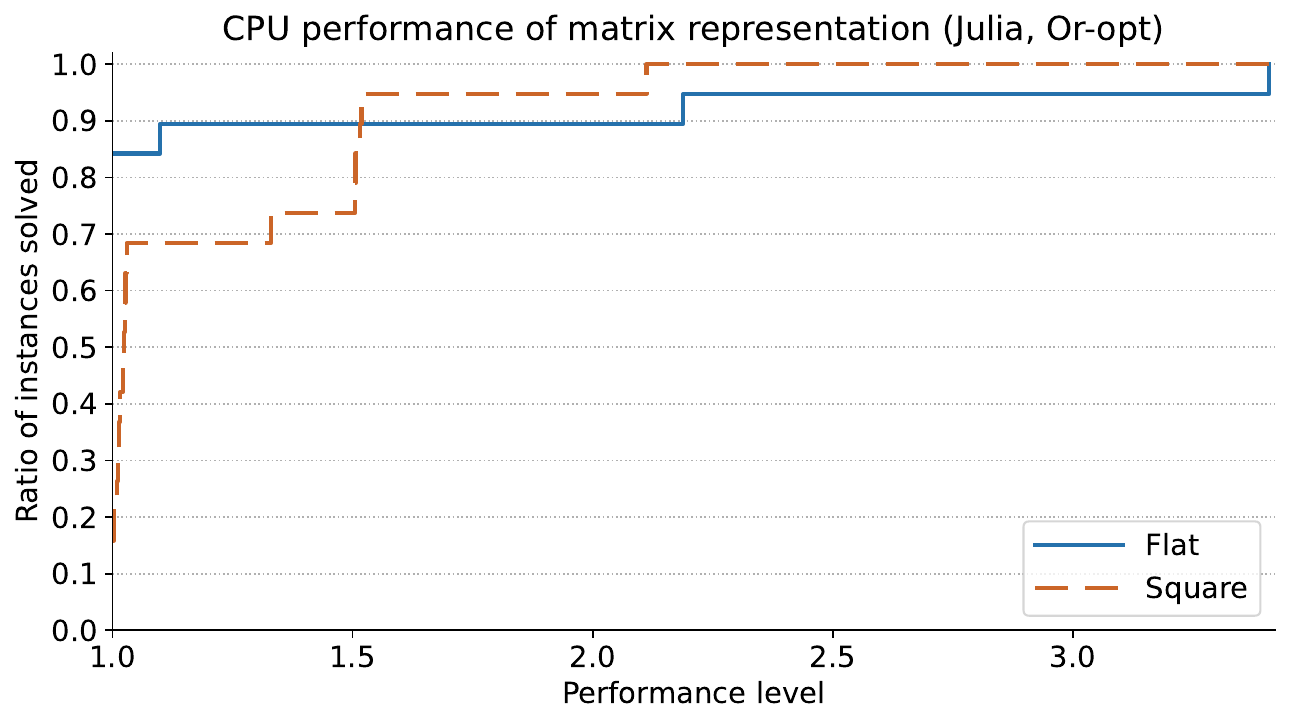}
    \caption{Or-opt benchmark}
  \end{subfigure}
  \begin{subfigure}{0.33\textwidth}
    \centering
    \includegraphics[scale=.22]{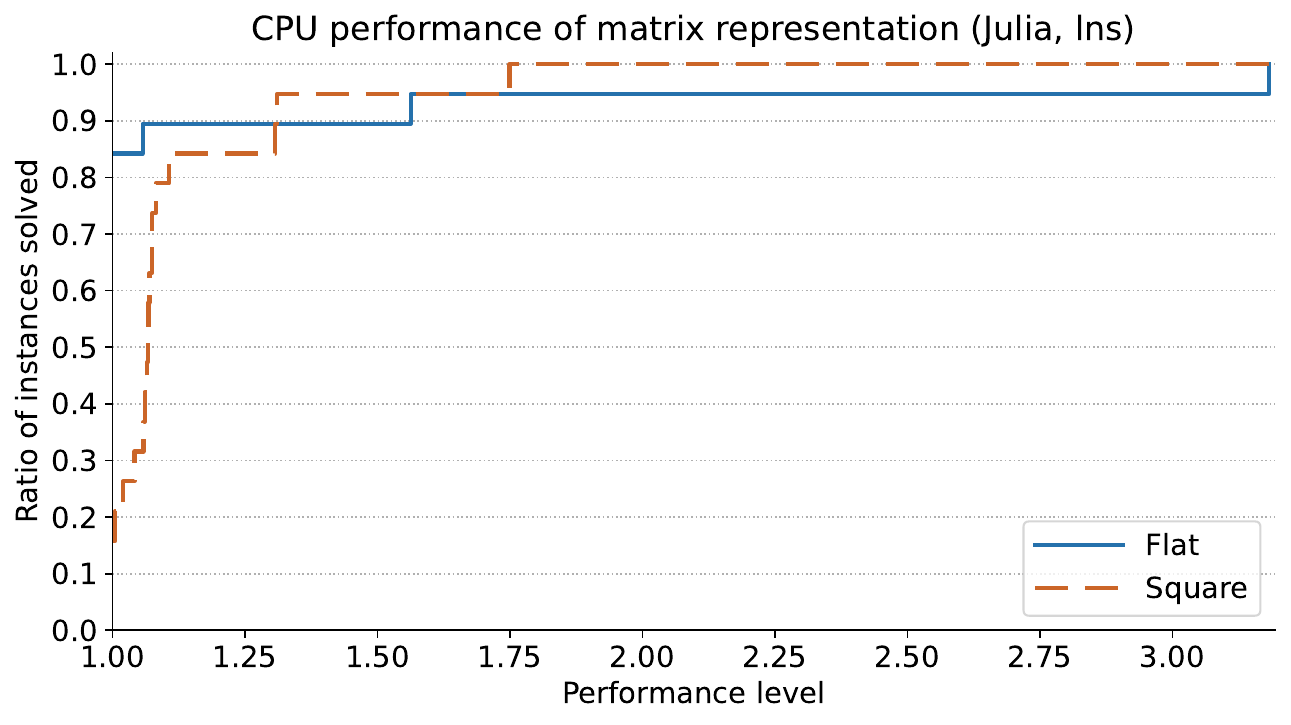}
    \caption{lns benchmark}
  \end{subfigure}
  \caption{Julia: comparison of matrix representations (VRPLIB)}
  \label{fig:julia-matrix-vrplib}
\end{figure}

\begin{figure}
  \begin{subfigure}{0.5\textwidth}
    \centering
    \includegraphics[scale=.3]{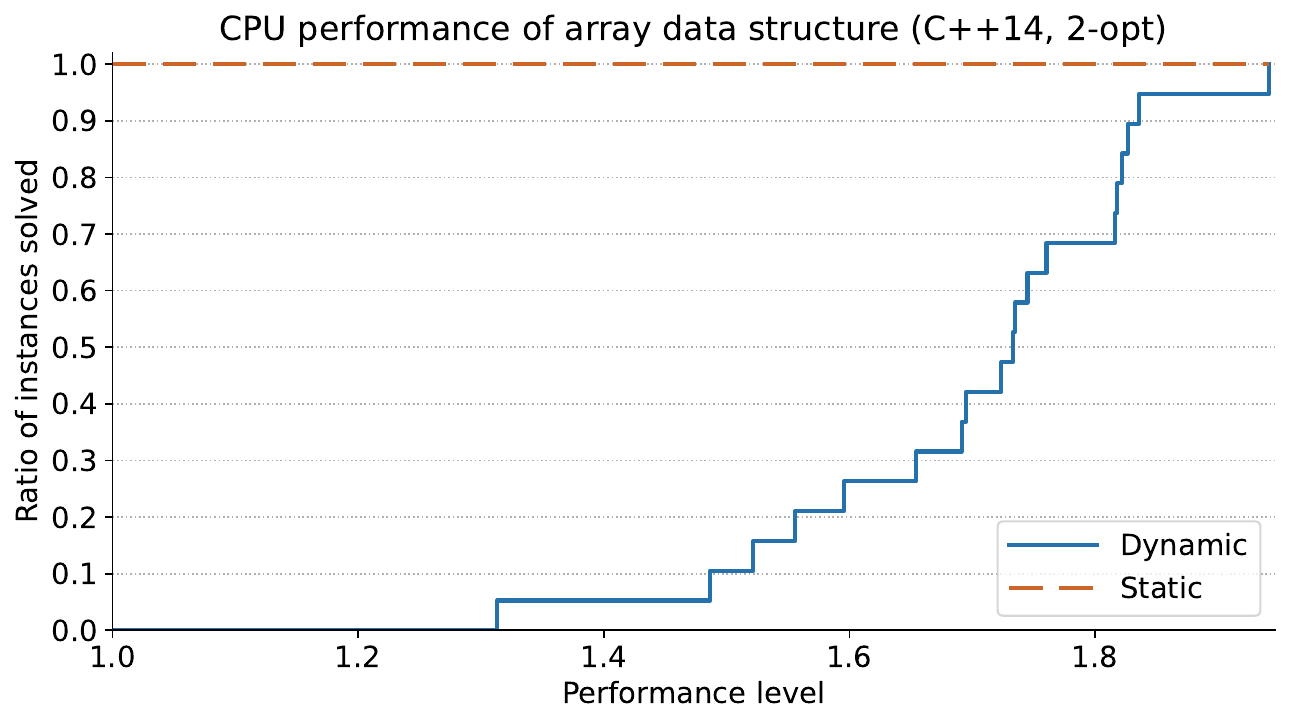}
    \caption{C++14, 2-opt benchmark}
  \end{subfigure}
  \begin{subfigure}{0.5\textwidth}
    \centering
    \includegraphics[scale=.3]{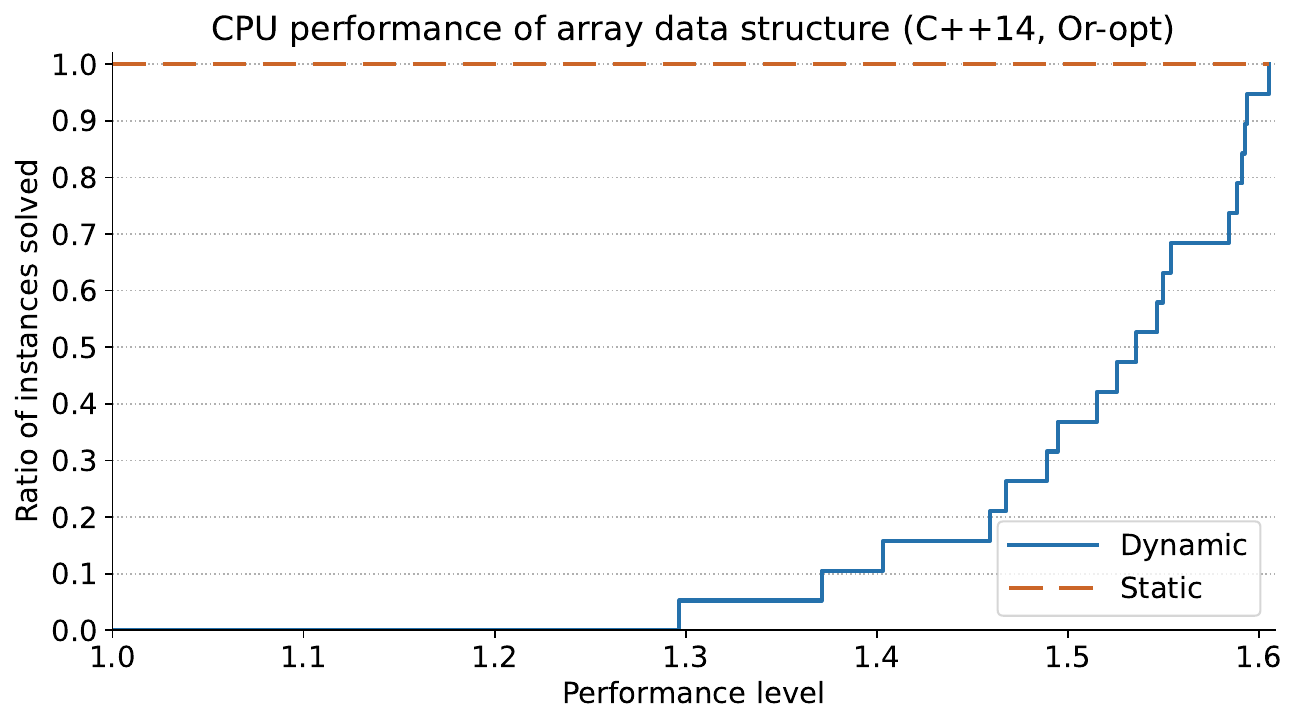}
    \caption{C++14, Or-opt benchmark}
  \end{subfigure}
  \begin{subfigure}{0.5\textwidth}
    \centering
    \includegraphics[scale=.3]{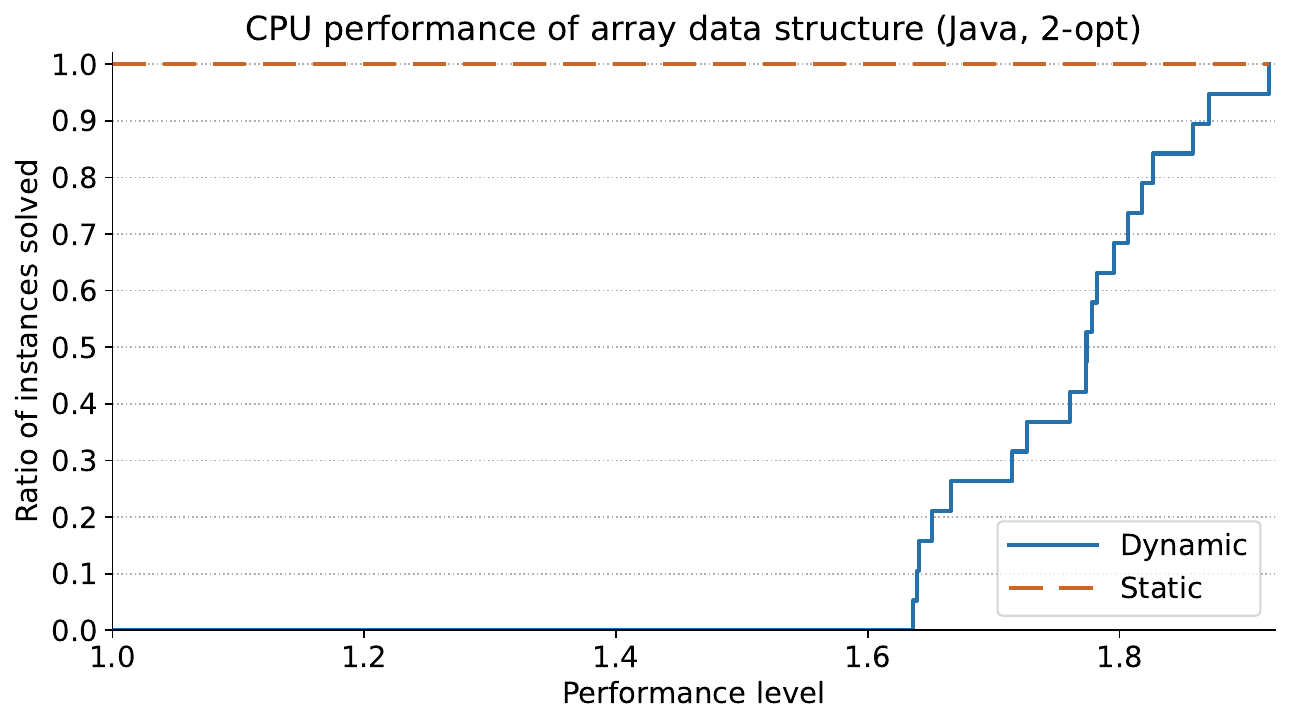}
    \caption{Java, 2-opt benchmark}
  \end{subfigure}
  \begin{subfigure}{0.5\textwidth}
    \centering
    \includegraphics[scale=.3]{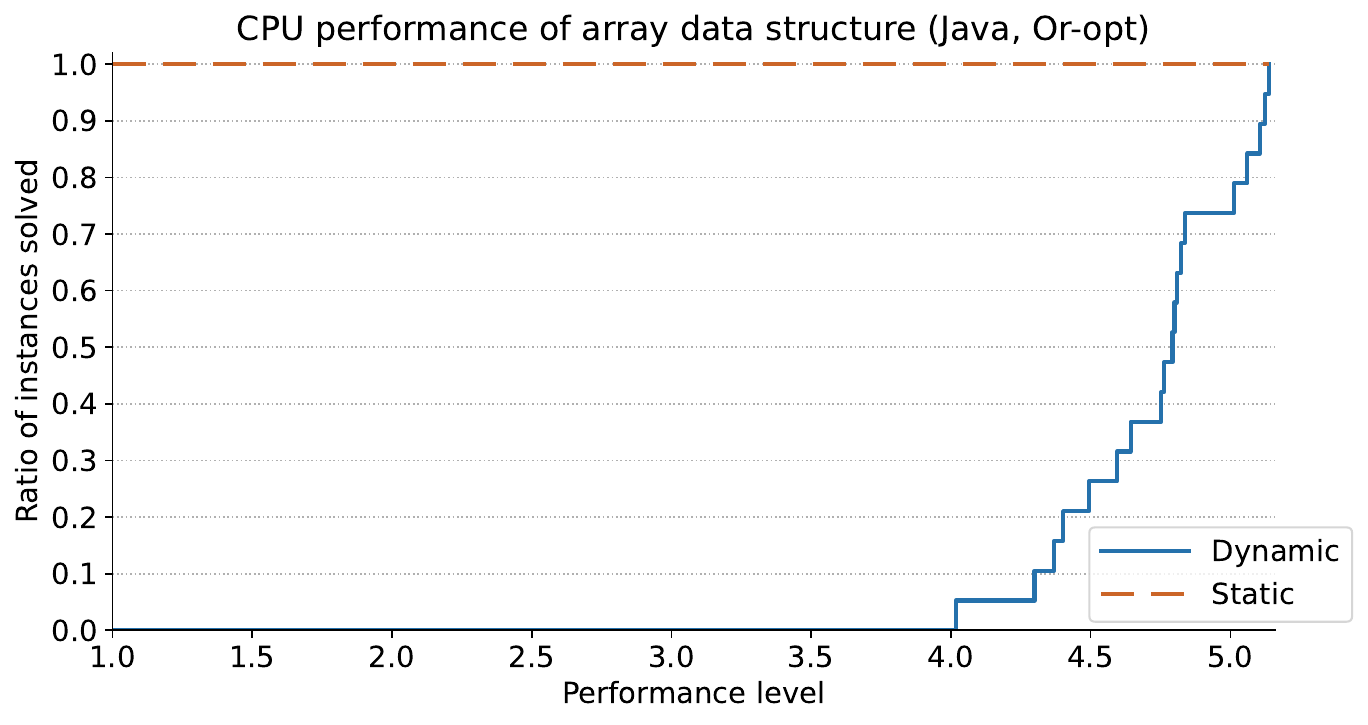}
    \caption{Java, Or-opt benchmark}
  \end{subfigure}
  \caption{Performance of static arrays versus variable-size vectors (VRPLIB)}
  \label{fig:static-arrays-vrplib}
\end{figure}

\begin{figure}
  \begin{subfigure}{0.33\textwidth}
    \centering
    \includegraphics[scale=.22]{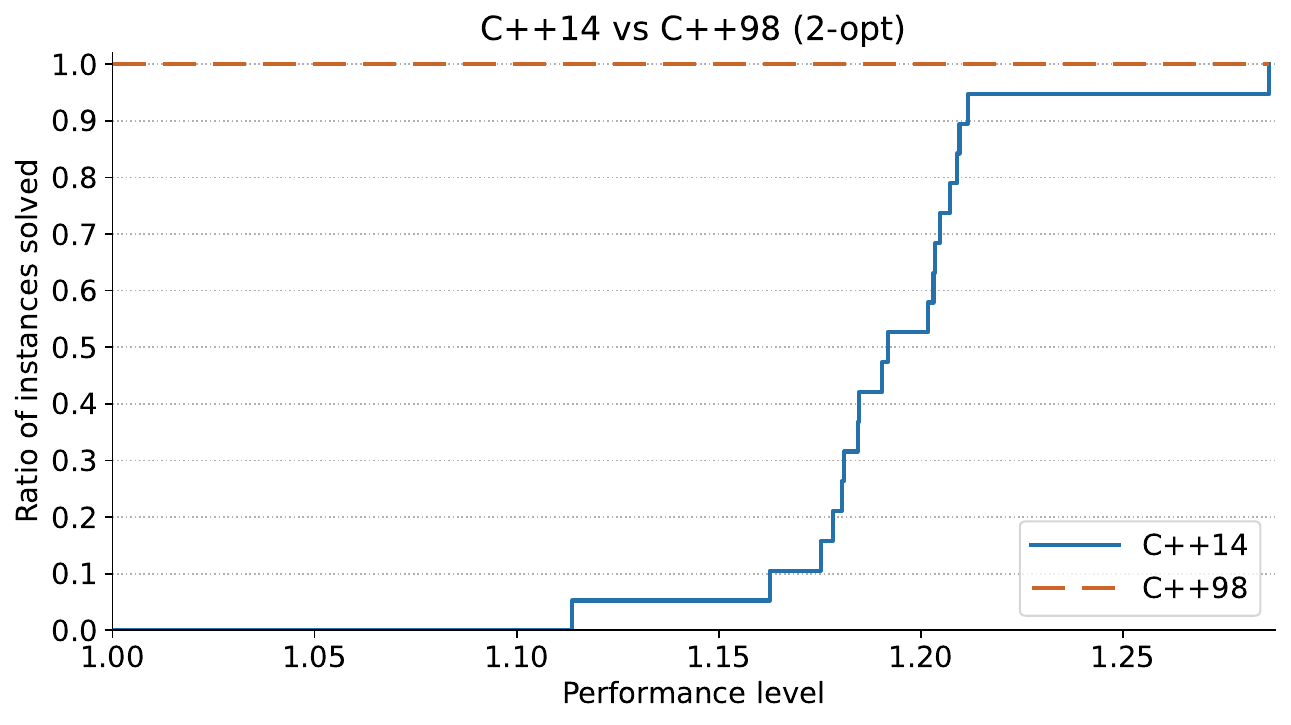}
    \caption{2-opt benchmark}
  \end{subfigure}
  \begin{subfigure}{0.33\textwidth}
    \centering
    \includegraphics[scale=.22]{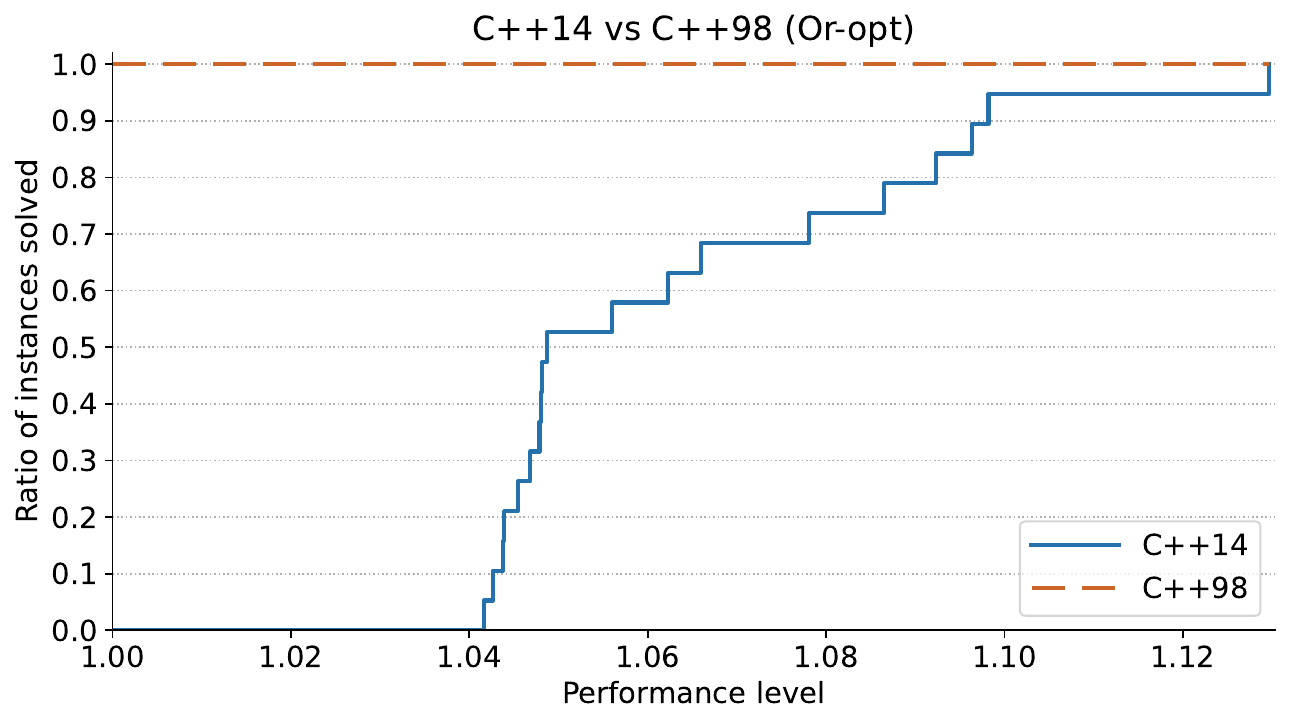}
    \caption{Or-opt benchmark}
  \end{subfigure}
  \begin{subfigure}{0.33\textwidth}
    \centering
    \includegraphics[scale=.22]{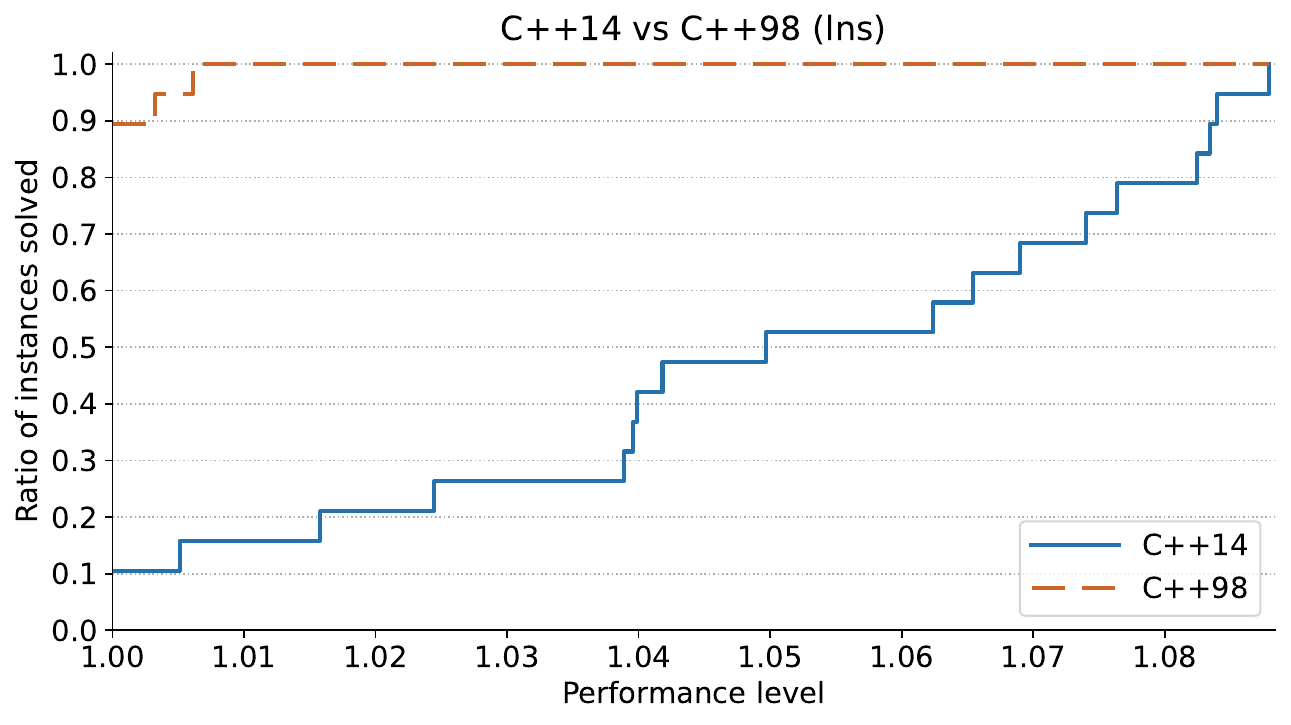}
    \caption{lns benchmark}
  \end{subfigure}
  \begin{subfigure}{0.33\textwidth}
    \centering
    \includegraphics[scale=.22]{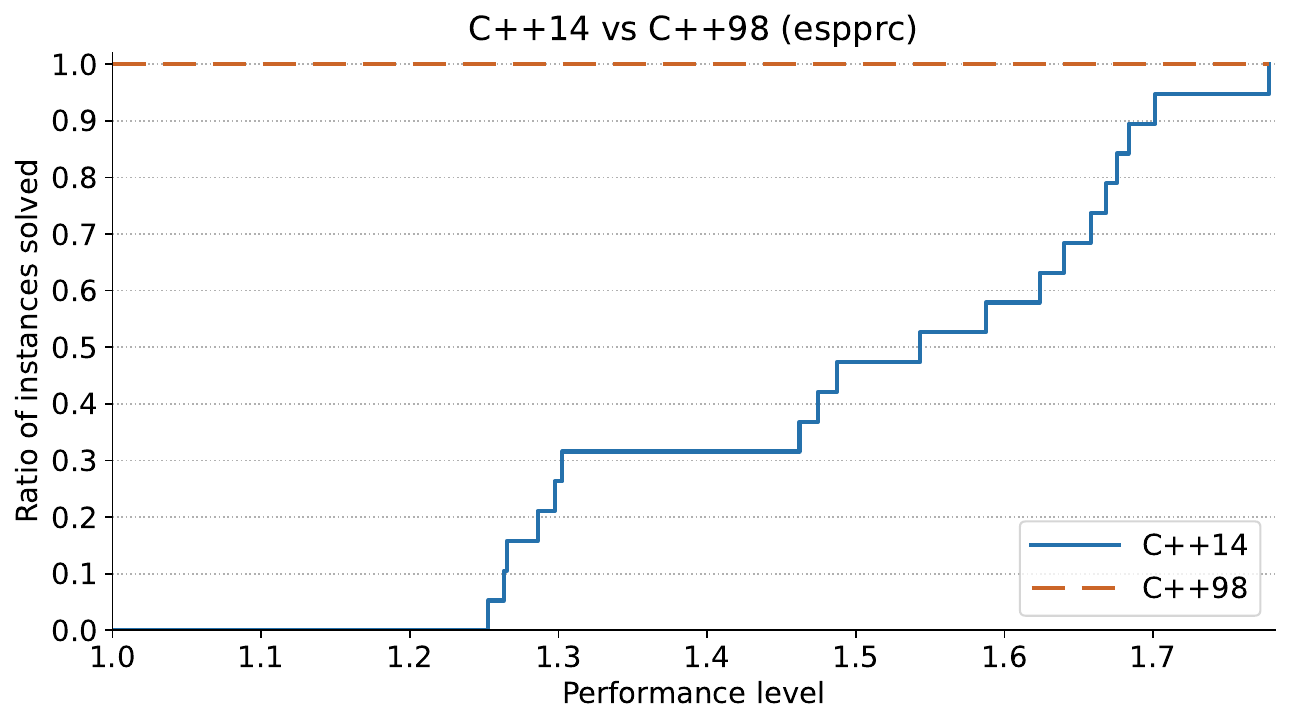}
    \caption{espprc benchmark}
  \end{subfigure}
  \begin{subfigure}{0.33\textwidth}
    \centering
    \includegraphics[scale=.22]{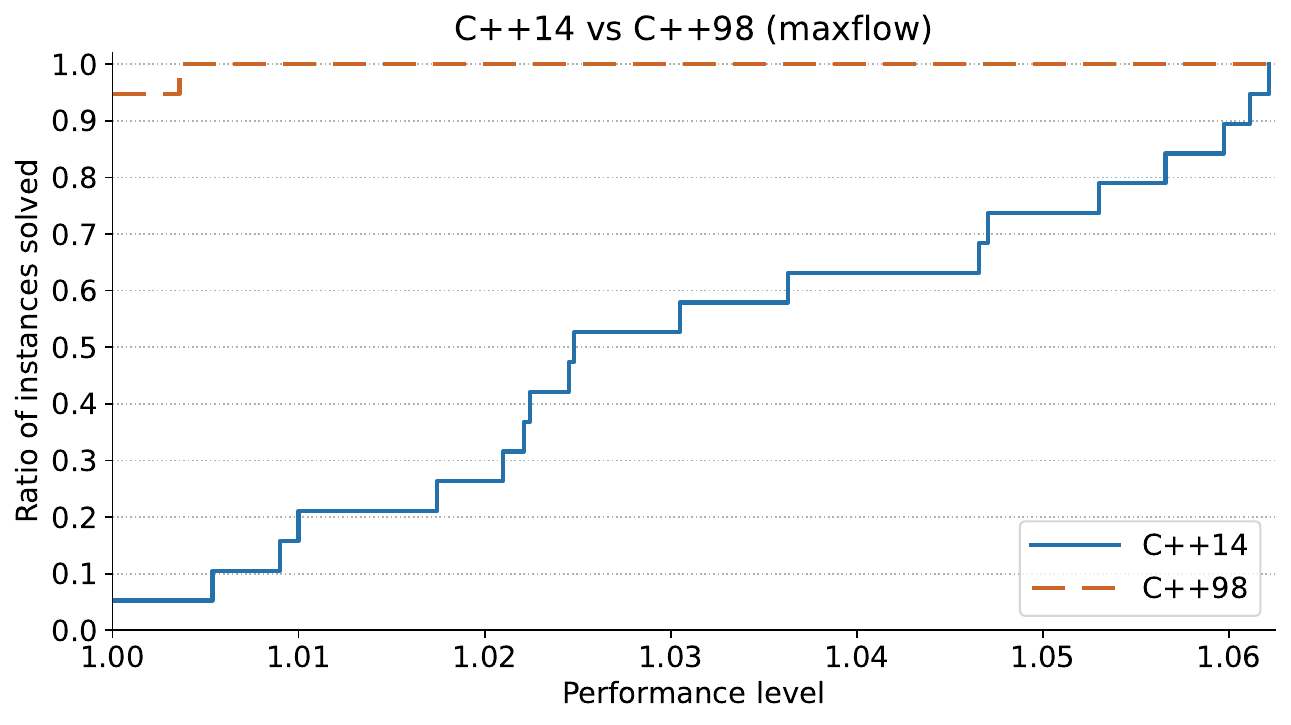}
    \caption{maxflow benchmark}
  \end{subfigure}
  \caption{Performance of C++14 versus C++98 (VRPLIB)}
  \label{fig:c++-implementations-vrplib}
\end{figure}

\begin{figure}
  \begin{subfigure}{0.33\textwidth}
    \centering
    \includegraphics[scale=.22]{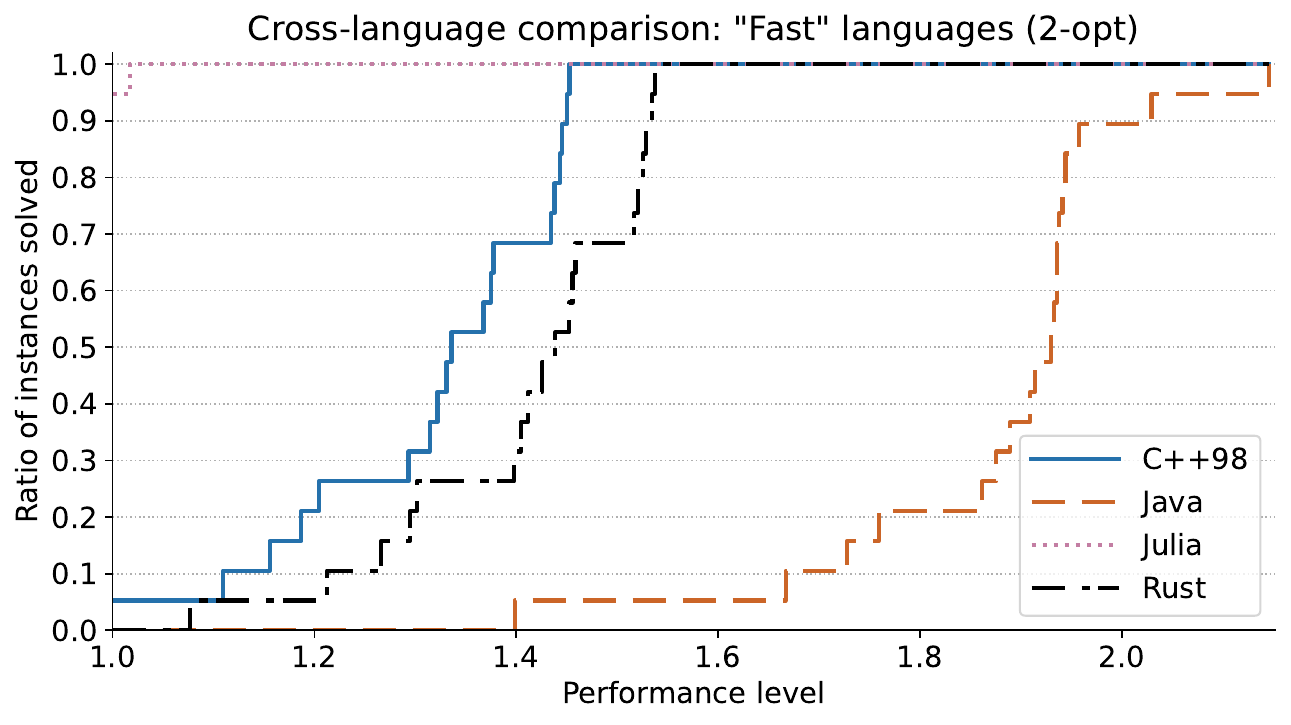}
    \caption{2-opt benchmark}
  \end{subfigure}
  \begin{subfigure}{0.33\textwidth}
    \centering
    \includegraphics[scale=.22]{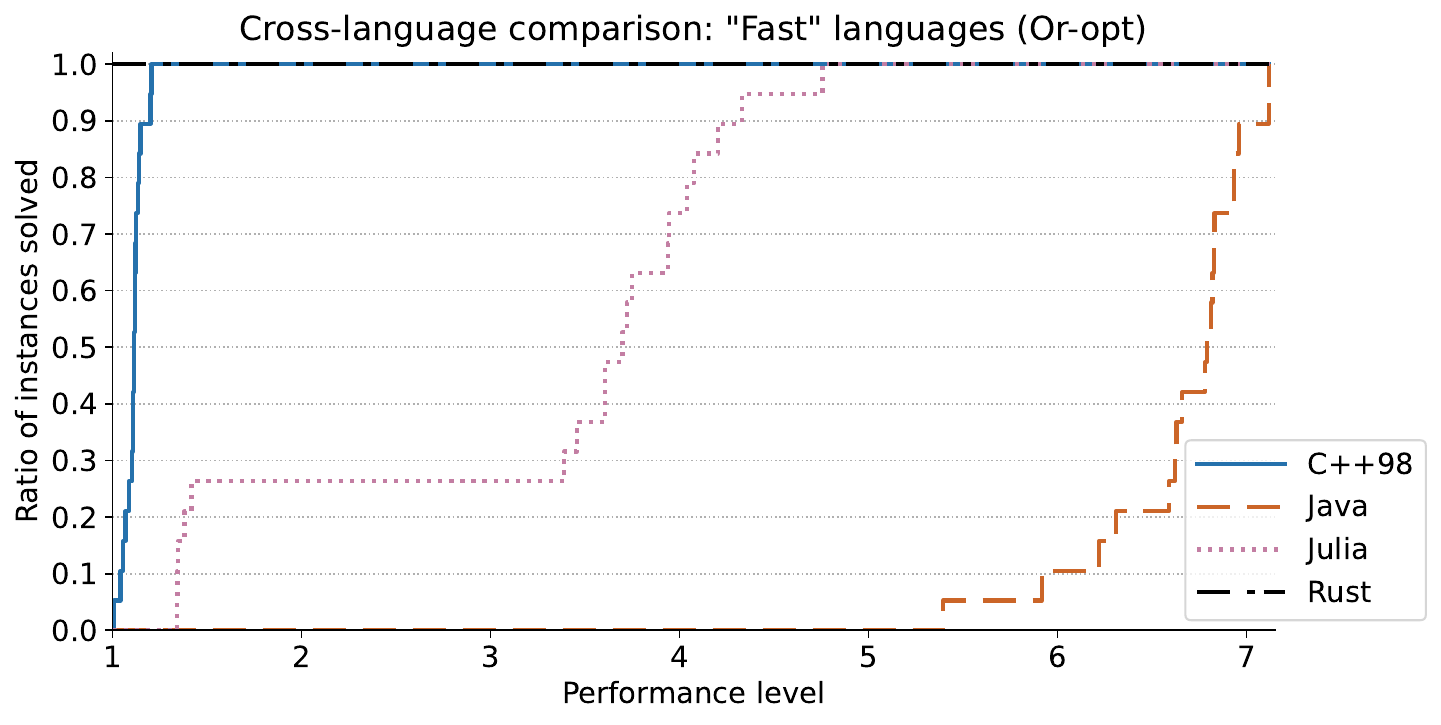}
    \caption{Or-opt benchmark}
  \end{subfigure}
  \begin{subfigure}{0.33\textwidth}
    \centering
    \includegraphics[scale=.22]{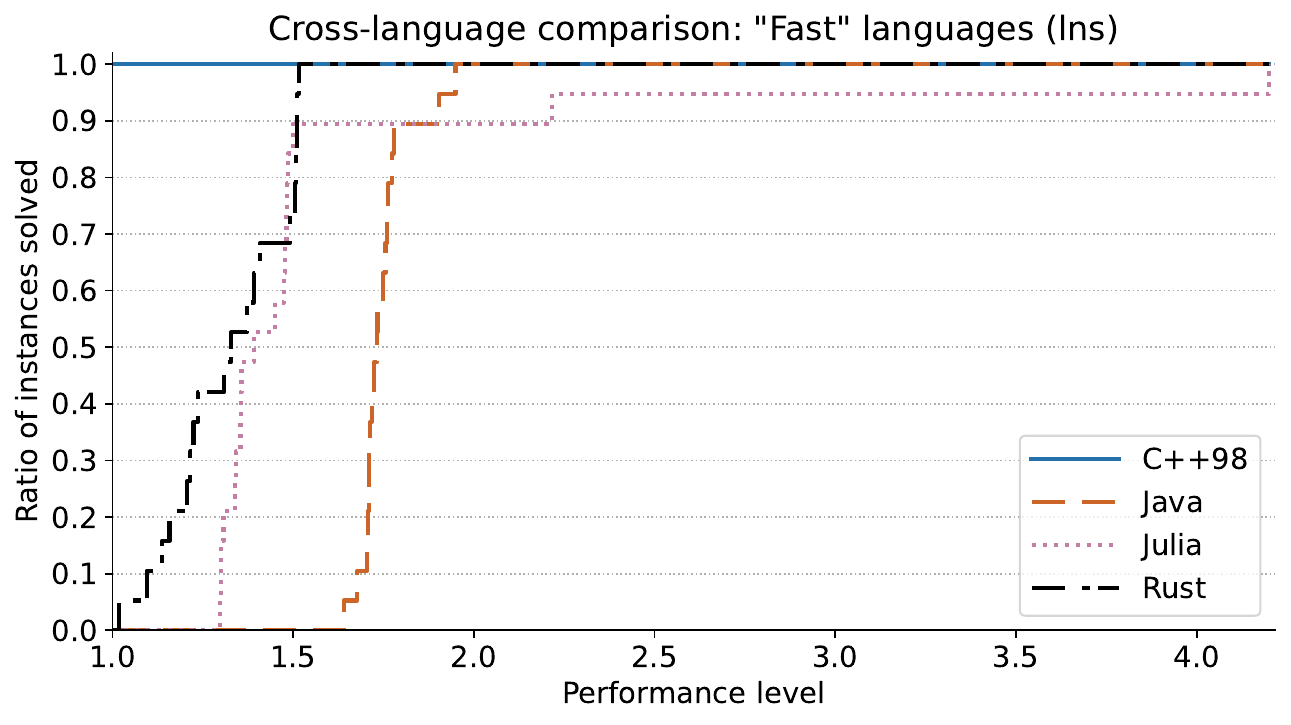}
    \caption{lns benchmark}
  \end{subfigure}
  \begin{subfigure}{0.33\textwidth}
    \centering
    \includegraphics[scale=.22]{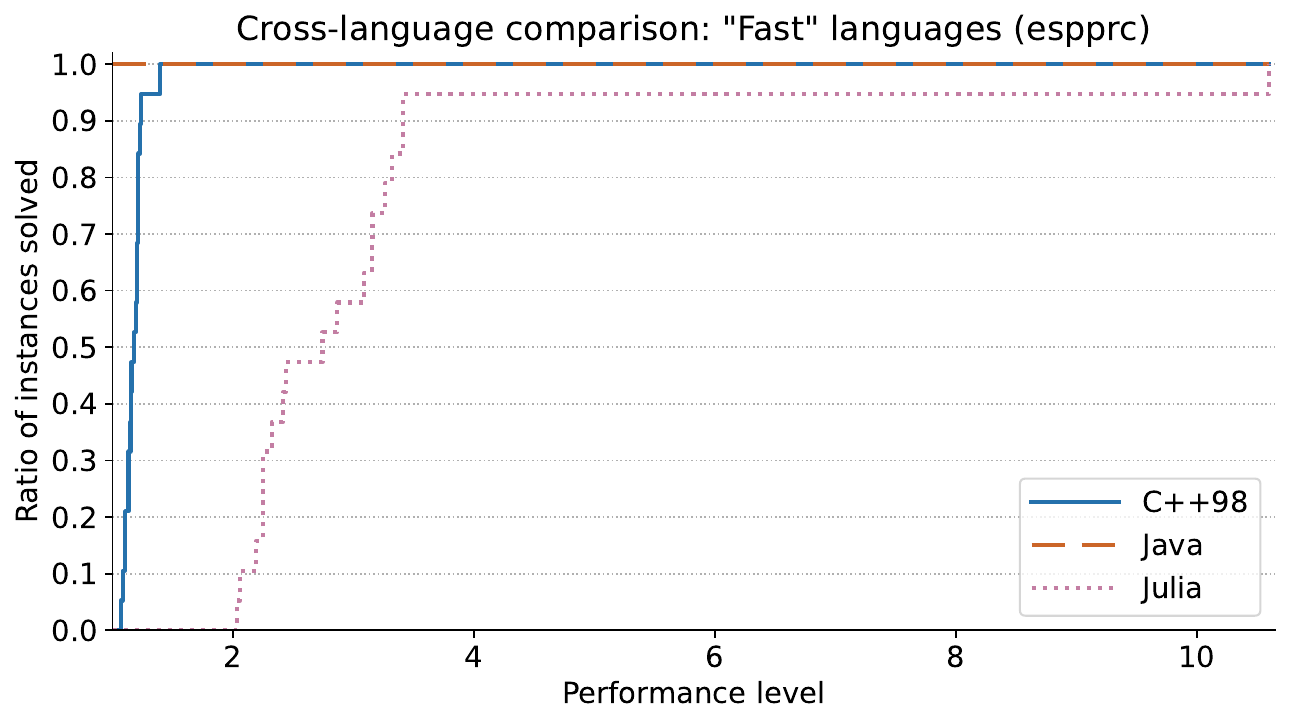}
    \caption{espprc benchmark}
  \end{subfigure}
  \begin{subfigure}{0.33\textwidth}
    \centering
    \includegraphics[scale=.22]{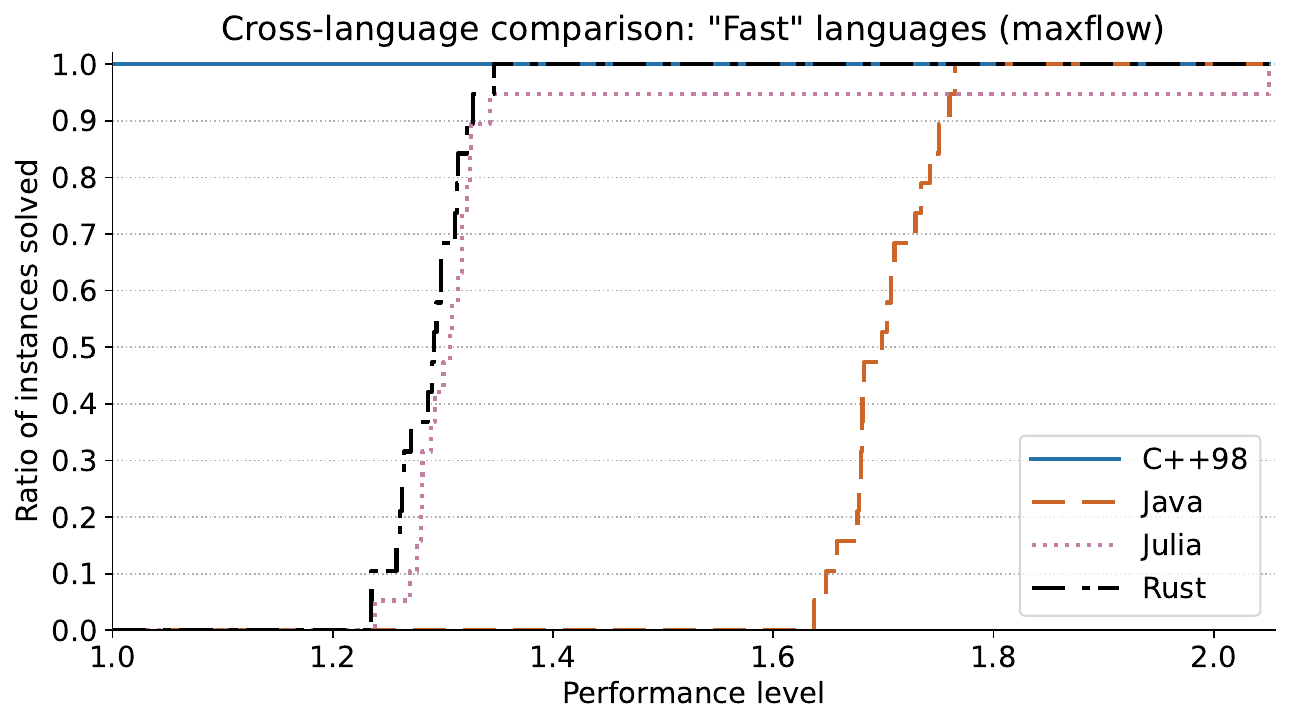}
    \caption{maxflow benchmark}
  \end{subfigure}
  \begin{subfigure}{0.33\textwidth}
    \centering
    \includegraphics[scale=.22]{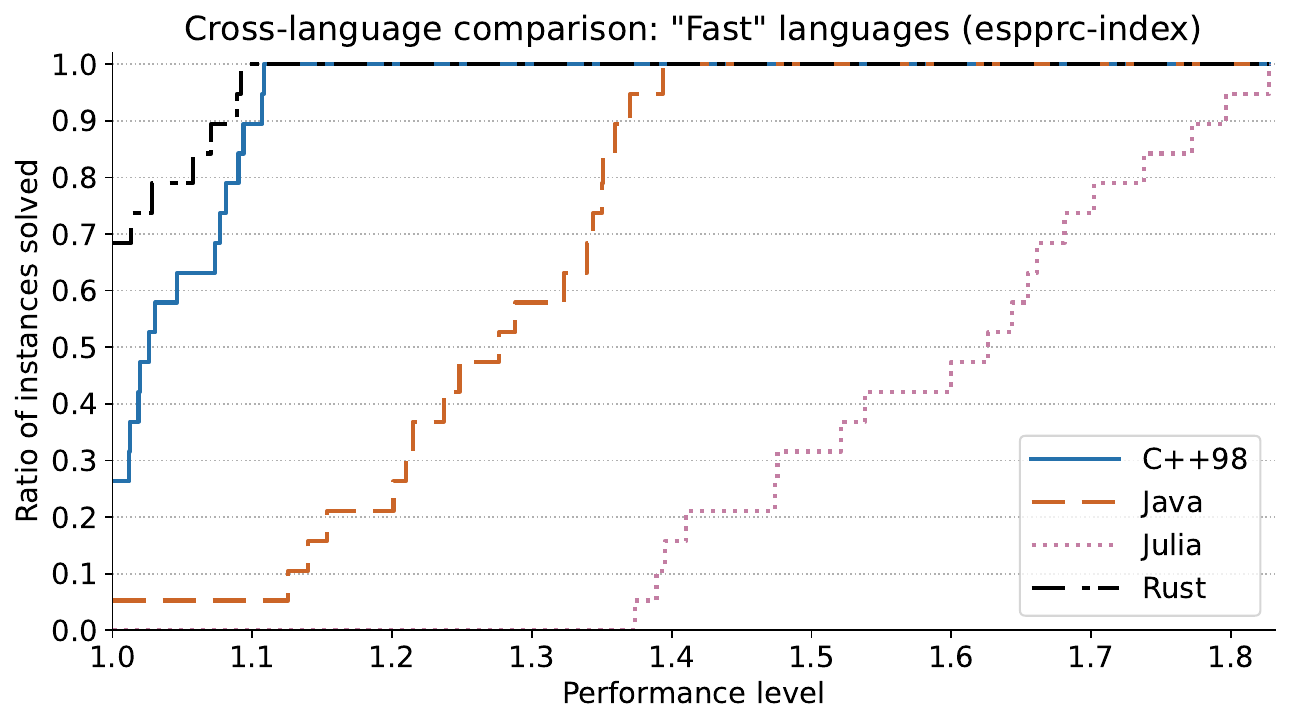}
    \caption{espprc-index benchmark}
  \end{subfigure}
  \caption{Performance of ``fast'' languages (VRPLIB)}
  \label{fig:fast_languages-vrplib}
\end{figure}

\begin{figure}
  \begin{subfigure}{0.33\textwidth}
    \centering
    \includegraphics[scale=.22]{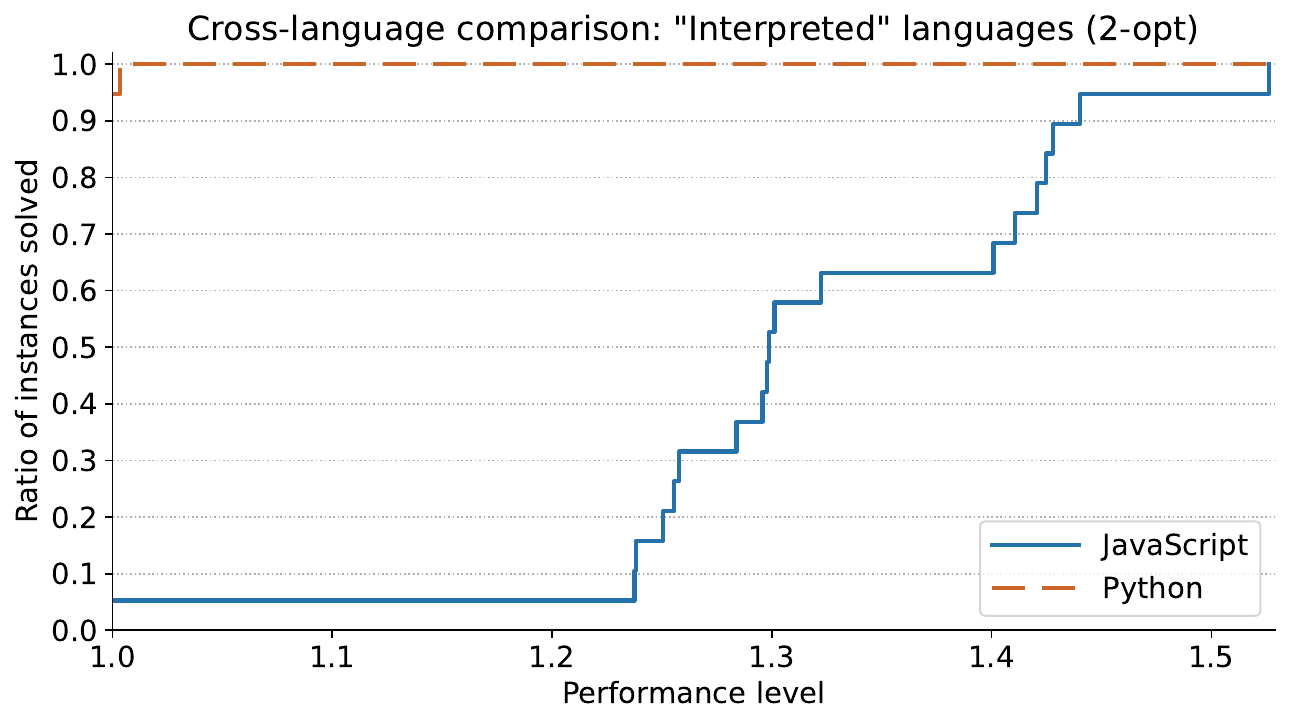}
    \caption{2-opt benchmark}
  \end{subfigure}
  \begin{subfigure}{0.33\textwidth}
    \centering
    \includegraphics[scale=.22]{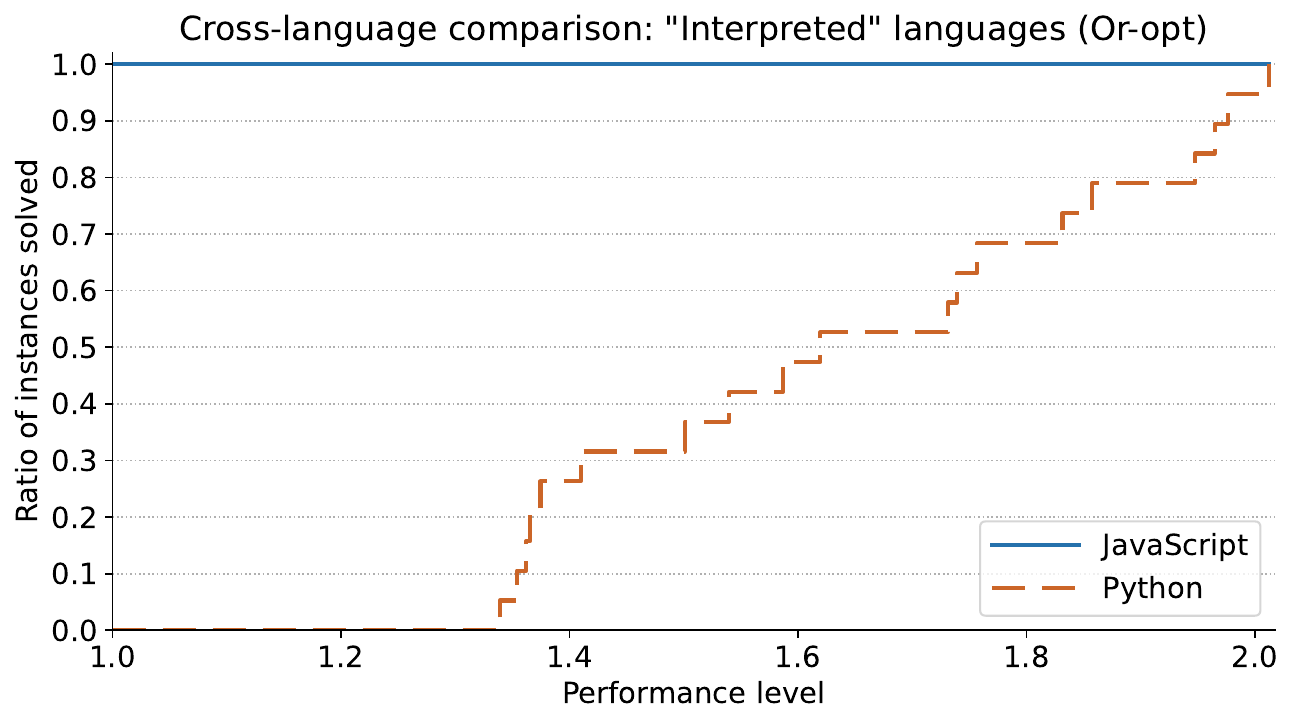}
    \caption{Or-opt benchmark}
  \end{subfigure}
  \begin{subfigure}{0.33\textwidth}
    \centering
    \includegraphics[scale=.22]{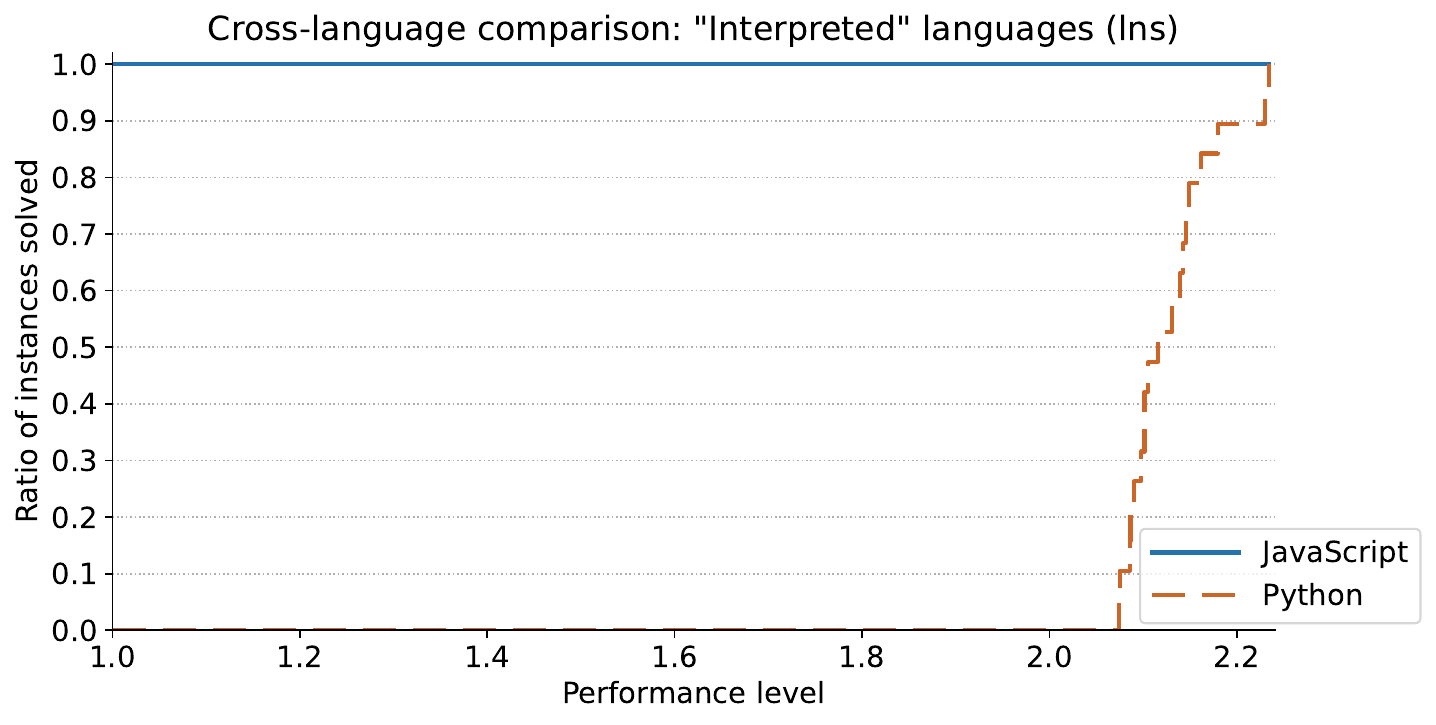}
    \caption{lns benchmark}
  \end{subfigure}
  \begin{subfigure}{0.33\textwidth}
    \centering
    \includegraphics[scale=.22]{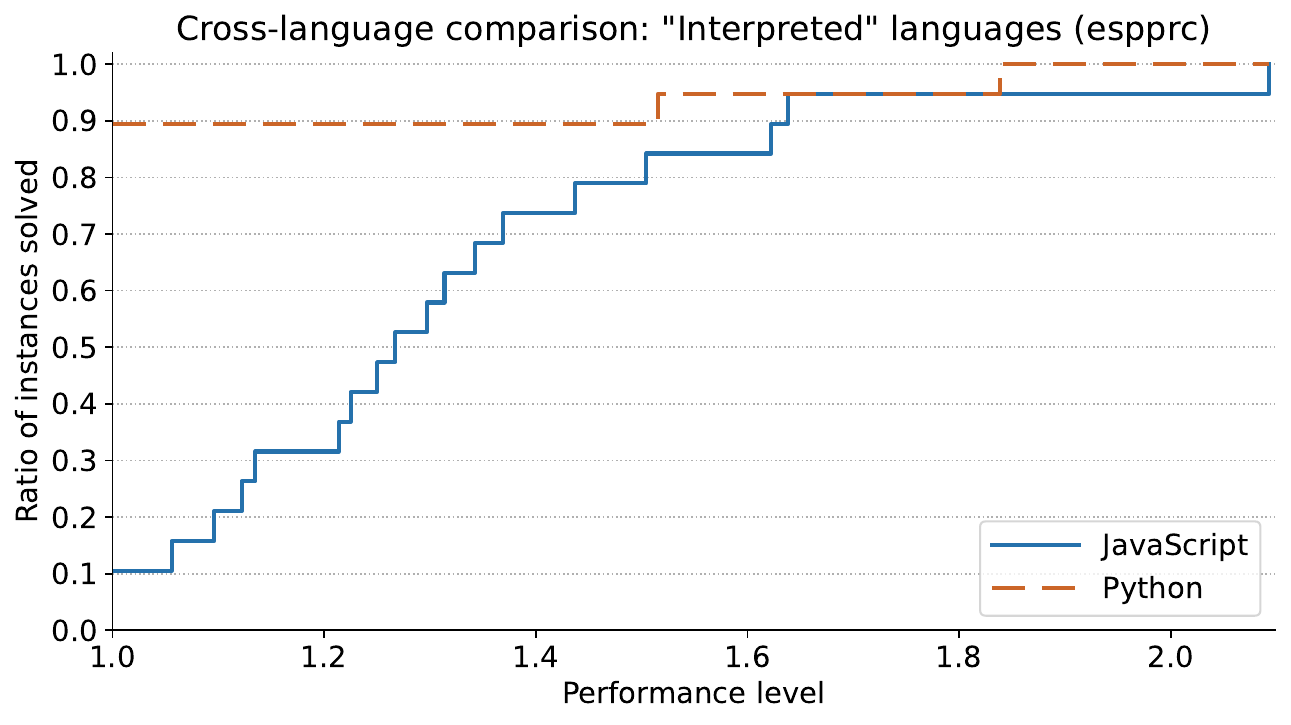}
    \caption{espprc benchmark}
  \end{subfigure}
  \begin{subfigure}{0.33\textwidth}
    \centering
    \includegraphics[scale=.22]{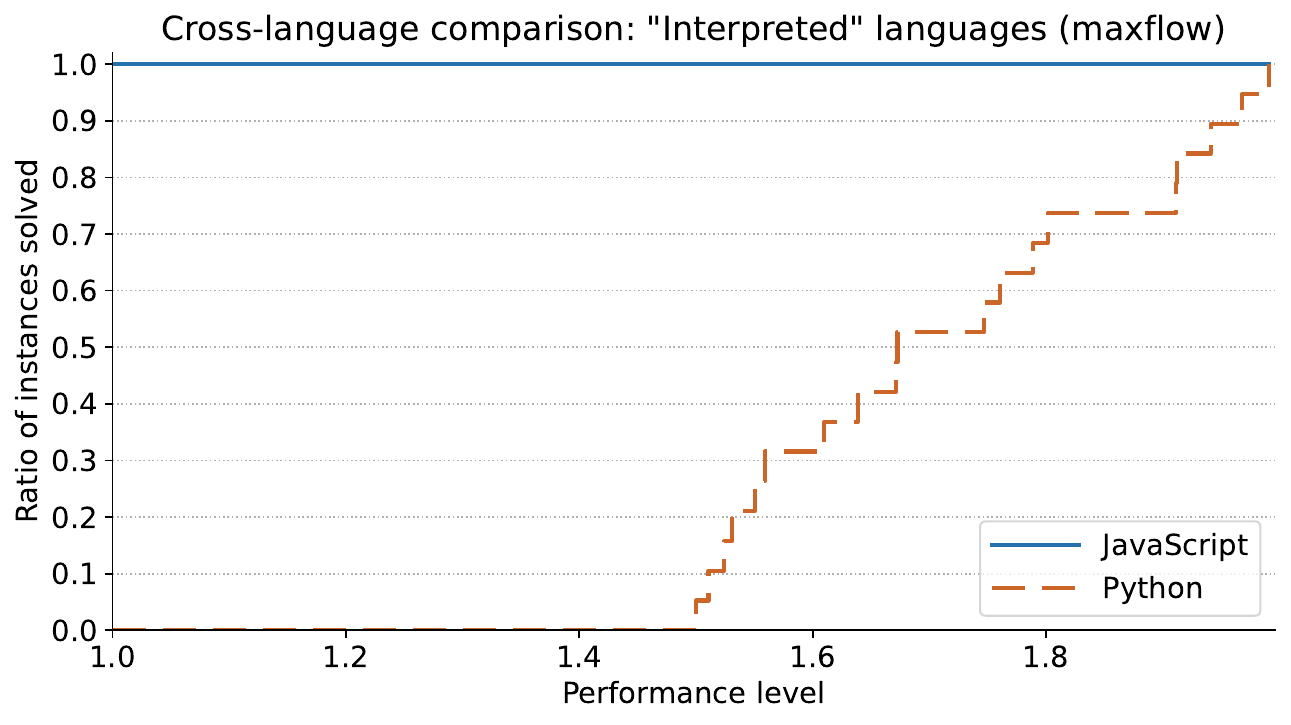}
    \caption{maxflow benchmark}
  \end{subfigure}
  \caption{Performance of ``interpreted'' languages (VRPLIB)}
  \label{fig:interpreted_languages-vrplib}
\end{figure}

For each instance we generate 1000 starting seeds randomly, as
previously. The variations in instance size do not allow to group them
by size conveniently, which is what we used for box 
plots. Therefore we only report performance profiles here. We produce
the same figures as \ref{fig:python-implementations} --
\ref{fig:interpreted_languages} from Section~\ref{sec:exp}, in the same
order, in figures \ref{fig:python-implementations-vrplib} --
\ref{fig:interpreted_languages-vrplib}. We omit the cross-platform
comparison as it is of limited interest here.

Since there are less instances as in the previous experiments, the
steps in performance profiles are more noticeable: there are 19
instances here, so each profile has at most 19 steps.
The trends are the same as before. The comparison of fast
languages on the 
espprc benchmark looks a bit different but is in fact very
similar, except it is stretched horizontally due to one unusually
longer Julia run.
Overall, it appears that the way the distance matrix is generated does
not impact the relative runtime budget required by the different
implementations.

\end{document}